\documentclass[sigconf]{acmart}

\AtBeginDocument{%
  }

\copyrightyear{2024} 
\acmYear{2024} 
\setcopyright{rightsretained} 
\acmConference[CCS '24]{Proceedings of the 2024 ACM SIGSAC Conference on
Computer and Communications Security}{October 14--18, 2024}{Salt Lake City,
UT, USA}
\acmBooktitle{Proceedings of the 2024 ACM SIGSAC Conference on Computer
and Communications Security (CCS '24), October 14--18, 2024, Salt Lake City,
UT, USA}
\acmDOI{10.1145/3658644.3690275}
\acmISBN{979-8-4007-0636-3/24/10}

\newif\ifSUPPLEMENT
\SUPPLEMENTtrue

\usepackage{acronym}
\usepackage{xspace}
\usepackage{listings}

\usepackage{enumitem}
\usepackage{graphicx}
\usepackage{subcaption}
\usepackage{multirow}   
\usepackage{fontawesome}
\usepackage{color, colortbl}
\usepackage{setspace}

\newfloat{tcolorboxfloat}{htbp}{lop}
\definecolor{bittersweet}{rgb}{1.0, 0.44, 0.37}
\definecolor{darkorange}{rgb}{1.0, 0.55, 0.0}
\definecolor{forestgreen}{RGB}{34, 139, 34}

\newcommand{\aravind}[1]{\textcolor{green}{Aravind: #1}}
\newcommand{\machiry}[1]{\textcolor{green}{Aravind: #1}}
\newcommand{\shank}[1]{\textcolor{blue}{Shank: #1}}
\newcommand{\ayushi}[1]{\textcolor{red}{Ayushi: #1}}
\newcommand{\santiago}[1]{\textcolor{olive}{Santiago: #1}}

\newcommand{\todo}[1]{TODO: {#1}}
\newcommand{\blue}[1]{\textcolor{blue}{#1}}

\acrodef{EMSW}{Embedded Software}
\acrodef{EMS}{Embedded System}
\acrodef{CFI}{Control Flow Integrity}
\acrodef{MCU}{Microcontroller Unit}
\acrodefplural{EMS}[EMSes]{Embedded Systems}
\acrodef{DEP}{Data Execution Prevention}
\acrodef{MMU}{Memory Management Unit}
\acrodef{ASLR}{Address Space Layout Randomization}
\acrodef{MPU}{Memory Protection Unit}
\acrodef{ISA}{Instruction Set Architecture}
\acrodef{ABI}{Application Binary Interface}
\acrodef{RISC}{Reduced Instruction Set Computer}
\acrodef{MMIO}{Memory Mapped Input Output}
\acrodef{SoC}{System on Chip}
\acrodef{FFI}{Foreign Function Interface}
\acrodef{SAST}{Static Application Security Testing}
\acrodef{AST}{Abstract Syntax Tree}
\acrodef{MIR}{Mid-level IR}
\acrodef{MPK}{Memory Protection Key}

\acrodef{RTOS}{Real Time Operating System}
\acrodefplural{RTOS}[RTOSes]{Real Time Operating Systems}

\acrodef{OS}{Operating System}
\acrodefplural{OS}[OSes]{Operating Systems}

\newcommand{\etal}{\textit{et al.,}\xspace}
\newcommand{\ie}{\textit{i.e.,}\xspace}
\newcommand{\eg}{\textit{e.g.,}\xspace}
\newcommand{\rustc}{{\tt rustc}\xspace}
\newcommand{\rust}{{\sc Rust}\xspace}

\newcommand{\conrust}{{\sc CoR}\xspace}
\newcommand{\rustonc}{{\sc RoC}\xspace}
\newcommand{\rustwc}{{\sc RwC}\xspace}
\newcommand{\mirchecker}{{\sc MirChecker}\xspace}

\newcommand{\freertos}{{\sc FreeRTOS}\xspace}
\newcommand{\freertosrs}{{\sc FreeRTOS.rs}\xspace}
\newcommand{\lilos}{{\sc lilos}\xspace}

\newcommand{\lockbud}{{\sc Lockbud}\xspace}
\newcommand{\rudra}{{\sc Rudra}\xspace}
\newcommand{\safedrop}{{\sc SafeDrop}\xspace}
\newcommand{\yuga}{{\sc Yuga}\xspace}
\newcommand{\rcanary}{{\sc rCanary}\xspace}
\newcommand{\ffichecker}{{\sc FFIChecker}\xspace}
\newcommand{\qrates}{{\sc Qrates}\xspace}
\newcommand{\clametrics}{{\sc cla-metrics}\xspace}
\newcommand{\crusts}{{\sc CRustS}\xspace}
\newcommand{\ctorust}{{\sc c2rust}\xspace}
\newcommand{\laertes}{{\sc laertes}\xspace}
\newcommand{\clang}{{\sc clang}\xspace}

\newcommand{\easyeffort}{\textcolor{green}{\faSmileO}}
\newcommand{\mediumeffort}{\textcolor{orange}{\faMehO}}
\newcommand{\hardeffort}{\textcolor{red}{\faFrownO}}
\newcommand{\greencheck}{{\textcolor{forestgreen}\faCheck}}
\newcommand{\redcross}{{\textcolor{red}\faTimes}}

\newcommand{\tbl}[1]{Tbl.~\ref{#1}}
\newcommand{\sect}[1]{$\S$\ref{#1}}
\newcommand{\fig}[1]{Fig.~\ref{#1}}
\newcommand{\lst}[1]{Lis.~\ref{#1}}

\ifSUPPLEMENT
 \newcommand{\apdx}[1]{Appendix~\ref{#1}}
\else
 \newcommand{\extendedreport}{Our Extended Report~\cite{extendedreport}\xspace}
 \newcommand{\apdx}[1]{\extendedreport}
\fi

\newcommand{\myparagraph}[1]{\noindent\textbf{#1}.\xspace}

 \newcommand{\code}[1]{%
  \lstinline[language=C++, basicstyle=\small\ttfamily]{#1}%
}

\lstdefinelanguage{Rust}{
 sensitive%
, morecomment=[l]{//}%
, morecomment=[s]{/*}{*/}%
, moredelim=[s][{\itshape\color[rgb]{0,0,0.75}}]{\#[}{]}%
, morestring=[b]{"}%
, alsodigit={}%
, alsoother={}%
, alsoletter={!}%
, morekeywords={break, continue, else, for, if, in, loop, match, return, while}  
, morekeywords={as, const, let, move, mut, ref, static}  
, morekeywords={dyn, enum, fn, impl, Self, self, struct, trait, type, union, use, where}  
, morekeywords={crate, extern, mod, pub, super}  
, morekeywords={unsafe}  
, morekeywords={abstract, alignof, become, box, do, final, macro, offsetof, override, priv, proc, pure, sizeof, typeof, unsized, virtual, yield}  
, morekeywords=[2]{Add, AddAssign, Any, AsciiExt, AsInner, AsInnerMut, AsMut, AsRawFd, AsRawHandle, AsRawSocket, AsRef, Binary, BitAnd, BitAndAssign, Bitor, BitOr, BitOrAssign, BitXor, BitXorAssign, Borrow, BorrowMut, Boxed, BoxPlace, BufRead, BuildHasher, CastInto, CharExt, Clone, CoerceUnsized, CommandExt, Copy, Debug, DecodableFloat, Default, Deref, DerefMut, DirBuilderExt, DirEntryExt, Display, Div, DivAssign, DoubleEndedIterator, DoubleEndedSearcher, Drop, EnvKey, Eq, Error, ExactSizeIterator, ExitStatusExt, Extend, FileExt, FileTypeExt, Float, Fn, FnBox, FnMut, FnOnce, Freeze, From, FromInner, FromIterator, FromRawFd, FromRawHandle, FromRawSocket, FromStr, FullOps, FusedIterator, Generator, Hash, Hasher, Index, IndexMut, InPlace, Int, Into, IntoCow, IntoInner, IntoIterator, IntoRawFd, IntoRawHandle, IntoRawSocket, IsMinusOne, IsZero, Iterator, JoinHandleExt, LargeInt, LowerExp, LowerHex, MetadataExt, Mul, MulAssign, Neg, Not, Octal, OpenOptionsExt, Ord, OsStrExt, OsStringExt, Packet, PartialEq, PartialOrd, Pattern, PermissionsExt, Place, Placer, Pointer, Product, Put, RangeArgument, RawFloat, Read, Rem, RemAssign, Seek, Shl, ShlAssign, Shr, ShrAssign, Sized, SliceConcatExt, SliceExt, SliceIndex, Stats, Step, StrExt, Sub, SubAssign, Sum, Sync, TDynBenchFn, Terminal, Termination, ToOwned, ToSocketAddrs, ToString, Try, TryFrom, TryInto, UnicodeStr, Unsize, UpperExp, UpperHex, WideInt, Write}
, morekeywords=[2]{Send}  
, morekeywords=[3]{bool, char, f32, f64, i8, i16, i32, i64, isize, str, u8, u16, u32, u64, unit, usize, i128, u128}  
, morekeywords=[4]{Err, false, None, Ok, Some, true}  
, morekeywords=[3]{AccessError, Adddf3, AddI128, AddoI128, AddoU128, ADDRESS, ADDRESS64, addrinfo, ADDRINFOA, AddrParseError, Addsf3, AddU128, advice, aiocb, Alignment, AllocErr, AnonPipe, Answer, Arc, Args, ArgsInnerDebug, ArgsOs, Argument, Arguments, ArgumentV1, Ashldi3, Ashlti3, Ashrdi3, Ashrti3, AssertParamIsClone, AssertParamIsCopy, AssertParamIsEq, AssertUnwindSafe, AtomicBool, AtomicPtr, Attr, auxtype, auxv, BackPlace, BacktraceContext, Barrier, BarrierWaitResult, Bencher, BenchMode, BenchSamples, BinaryHeap, BinaryHeapPlace, blkcnt, blkcnt64, blksize, BOOL, boolean, BOOLEAN, BoolTrie, BorrowError, BorrowMutError, Bound, Box, bpf, BTreeMap, BTreeSet, Bucket, BucketState, Buf, BufReader, BufWriter, Builder, BuildHasherDefault, BY, BYTE, Bytes, CannotReallocInPlace, cc, Cell, Chain, CHAR, CharIndices, CharPredicateSearcher, Chars, CharSearcher, CharsError, CharSliceSearcher, CharTryFromError, Child, ChildPipes, ChildStderr, ChildStdin, ChildStdio, ChildStdout, Chunks, ChunksMut, ciovec, clock, clockid, Cloned, cmsgcred, cmsghdr, CodePoint, Color, ColorConfig, Command, CommandEnv, Component, Components, CONDITION, condvar, Condvar, CONSOLE, CONTEXT, Count, Cow, cpu, CRITICAL, CStr, CString, CStringArray, Cursor, Cycle, CycleIter, daddr, DebugList, DebugMap, DebugSet, DebugStruct, DebugTuple, Decimal, Decoded, DecodeUtf16, DecodeUtf16Error, DecodeUtf8, DefaultEnvKey, DefaultHasher, dev, device, Difference, Digit32, DIR, DirBuilder, dircookie, dirent, dirent64, DirEntry, Discriminant, DISPATCHER, Display, Divdf3, Divdi3, Divmoddi4, Divmodsi4, Divsf3, Divsi3, Divti3, dl, Dl, Dlmalloc, Dns, DnsAnswer, DnsQuery, dqblk, Drain, DrainFilter, Dtor, Duration, DwarfReader, DWORD, DWORDLONG, DynamicLibrary, Edge, EHAction, EHContext, Elf32, Elf64, Empty, EmptyBucket, EncodeUtf16, EncodeWide, Entry, EntryPlace, Enumerate, Env, epoll, errno, Error, ErrorKind, EscapeDebug, EscapeDefault, EscapeUnicode, event, Event, eventrwflags, eventtype, ExactChunks, ExactChunksMut, EXCEPTION, Excess, ExchangeHeapSingleton, exit, exitcode, ExitStatus, Failure, fd, fdflags, fdsflags, fdstat, ff, fflags, File, FILE, FileAttr, filedelta, FileDesc, FilePermissions, filesize, filestat, FILETIME, filetype, FileType, Filter, FilterMap, Fixdfdi, Fixdfsi, Fixdfti, Fixsfdi, Fixsfsi, Fixsfti, Fixunsdfdi, Fixunsdfsi, Fixunsdfti, Fixunssfdi, Fixunssfsi, Fixunssfti, Flag, FlatMap, Floatdidf, FLOATING, Floatsidf, Floatsisf, Floattidf, Floattisf, Floatundidf, Floatunsidf, Floatunsisf, Floatuntidf, Floatuntisf, flock, ForceResult, FormatSpec, Formatted, Formatter, Fp, FpCategory, fpos, fpos64, fpreg, fpregset, FPUControlWord, Frame, FromBytesWithNulError, FromUtf16Error, FromUtf8Error, FrontPlace, fsblkcnt, fsfilcnt, fsflags, fsid, fstore, fsword, FullBucket, FullBucketMut, FullDecoded, Fuse, GapThenFull, GeneratorState, gid, glob, glob64, GlobalDlmalloc, greg, group, GROUP, Guard, GUID, Handle, HANDLE, Handler, HashMap, HashSet, Heap, HINSTANCE, HMODULE, hostent, HRESULT, id, idtype, if, ifaddrs, IMAGEHLP, Immut, in, in6, Incoming, Infallible, Initializer, ino, ino64, inode, input, InsertResult, Inspect, Instant, int16, int32, int64, int8, integer, IntermediateBox, Internal, Intersection, intmax, IntoInnerError, IntoIter, IntoStringError, intptr, InvalidSequence, iovec, ip, IpAddr, ipc, Ipv4Addr, ipv6, Ipv6Addr, Ipv6MulticastScope, Iter, IterMut, itimerspec, itimerval, jail, JoinHandle, JoinPathsError, KDHELP64, kevent, kevent64, key, Key, Keys, KV, l4, LARGE, lastlog, launchpad, Layout, Lazy, lconv, Leaf, LeafOrInternal, Lines, LinesAny, LineWriter, linger, linkcount, LinkedList, load, locale, LocalKey, LocalKeyState, Location, lock, LockResult, loff, LONG, lookup, lookupflags, LookupHost, LPBOOL, LPBY, LPBYTE, LPCSTR, LPCVOID, LPCWSTR, LPDWORD, LPFILETIME, LPHANDLE, LPOVERLAPPED, LPPROCESS, LPPROGRESS, LPSECURITY, LPSTARTUPINFO, LPSTR, LPVOID, LPWCH, LPWIN32, LPWSADATA, LPWSAPROTOCOL, LPWSTR, Lshrdi3, Lshrti3, lwpid, M128A, mach, major, Map, mcontext, Metadata, Metric, MetricMap, mflags, minor, mmsghdr, Moddi3, mode, Modsi3, Modti3, MonitorMsg, MOUNT, mprot, mq, mqd, msflags, msghdr, msginfo, msglen, msgqnum, msqid, Muldf3, Mulodi4, Mulosi4, Muloti4, Mulsf3, Multi3, Mut, Mutex, MutexGuard, MyCollection, n16, NamePadding, NativeLibBoilerplate, nfds, nl, nlink, NodeRef, NoneError, NonNull, NonZero, nthreads, NulError, OccupiedEntry, off, off64, oflags, Once, OnceState, OpenOptions, Option, Options, OptRes, Ordering, OsStr, OsString, Output, OVERLAPPED, Owned, Packet, PanicInfo, Param, ParseBoolError, ParseCharError, ParseError, ParseFloatError, ParseIntError, ParseResult, Part, passwd, Path, PathBuf, PCONDITION, PCONSOLE, Peekable, PeekMut, Permissions, PhantomData, pid, Pipes, PlaceBack, PlaceFront, PLARGE, PoisonError, pollfd, PopResult, port, Position, Powidf2, Powisf2, Prefix, PrefixComponent, PrintFormat, proc, Process, PROCESS, processentry, protoent, PSRWLOCK, pthread, ptr, ptrdiff, PVECTORED, Queue, radvisory, RandomState, Range, RangeFrom, RangeFull, RangeInclusive, RangeMut, RangeTo, RangeToInclusive, RawBucket, RawFd, RawHandle, RawPthread, RawSocket, RawTable, RawVec, Rc, ReadDir, Receiver, recv, RecvError, RecvTimeoutError, ReentrantMutex, ReentrantMutexGuard, Ref, RefCell, RefMut, REPARSE, Repeat, Result, Rev, Reverse, riflags, rights, rlim, rlim64, rlimit, rlimit64, roflags, Root, RSplit, RSplitMut, RSplitN, RSplitNMut, RUNTIME, rusage, RwLock, RWLock, RwLockReadGuard, RwLockWriteGuard, sa, SafeHash, Scan, sched, scope, sdflags, SearchResult, SearchStep, SECURITY, SeekFrom, segment, Select, SelectionResult, sem, sembuf, send, Sender, SendError, servent, sf, Shared, shmatt, shmid, ShortReader, ShouldPanic, Shutdown, siflags, sigaction, SigAction, sigevent, sighandler, siginfo, Sign, signal, signalfd, SignalToken, sigset, sigval, Sink, SipHasher, SipHasher13, SipHasher24, size, SIZE, Skip, SkipWhile, Slice, SmallBoolTrie, sockaddr, SOCKADDR, sockcred, Socket, SOCKET, SocketAddr, SocketAddrV4, SocketAddrV6, socklen, speed, Splice, Split, SplitMut, SplitN, SplitNMut, SplitPaths, SplitWhitespace, spwd, SRWLOCK, ssize, stack, STACKFRAME64, StartResult, STARTUPINFO, stat, Stat, stat64, statfs, statfs64, StaticKey, statvfs, StatVfs, statvfs64, Stderr, StderrLock, StderrTerminal, Stdin, StdinLock, Stdio, StdioPipes, Stdout, StdoutLock, StdoutTerminal, StepBy, String, StripPrefixError, StrSearcher, subclockflags, Subdf3, SubI128, SuboI128, SuboU128, subrwflags, subscription, Subsf3, SubU128, Summary, suseconds, SYMBOL, SYMBOLIC, SymmetricDifference, SyncSender, sysinfo, System, SystemTime, SystemTimeError, Take, TakeWhile, tcb, tcflag, TcpListener, TcpStream, TempDir, TermInfo, TerminfoTerminal, termios, termios2, TestDesc, TestDescAndFn, TestEvent, TestFn, TestName, TestOpts, TestResult, Thread, threadattr, threadentry, ThreadId, tid, time, time64, timespec, TimeSpec, timestamp, timeval, timeval32, timezone, tm, tms, ToLowercase, ToUppercase, TraitObject, TryFromIntError, TryFromSliceError, TryIter, TryLockError, TryLockResult, TryRecvError, TrySendError, TypeId, U64x2, ucontext, ucred, Udivdi3, Udivmoddi4, Udivmodsi4, Udivmodti4, Udivsi3, Udivti3, UdpSocket, uid, UINT, uint16, uint32, uint64, uint8, uintmax, uintptr, ulflags, ULONG, ULONGLONG, Umoddi3, Umodsi3, Umodti3, UnicodeVersion, Union, Unique, UnixDatagram, UnixListener, UnixStream, Unpacked, UnsafeCell, UNWIND, UpgradeResult, useconds, user, userdata, USHORT, Utf16Encoder, Utf8Error, Utf8Lossy, Utf8LossyChunk, Utf8LossyChunksIter, utimbuf, utmp, utmpx, utsname, uuid, VacantEntry, Values, ValuesMut, VarError, Variables, Vars, VarsOs, Vec, VecDeque, vm, Void, WaitTimeoutResult, WaitToken, wchar, WCHAR, Weak, whence, WIN32, WinConsole, Windows, WindowsEnvKey, winsize, WORD, Wrapping, wrlen, WSADATA, WSAPROTOCOL, WSAPROTOCOLCHAIN, Wtf8, Wtf8Buf, Wtf8CodePoints, xsw, xucred, Zip, zx}
, morekeywords=[5]{assert!, assert_eq!, assert_ne!, cfg!, column!, compile_error!, concat!, concat_idents!, debug_assert!, debug_assert_eq!, debug_assert_ne!, env!, eprint!, eprintln!, file!, format!, format_args!, include!, include_bytes!, include_str!, line!, module_path!, option_env!, panic!, print!, println!, select!, stringify!, thread_local!, try!, unimplemented!, unreachable!, vec!, write!, writeln!}  
}

\lstdefinestyle{colouredRust}%
{ basicstyle=\ttfamily%
, identifierstyle=%
, commentstyle=\color[rgb]{0, 0.5, 0}
, stringstyle=\color[rgb]{0, 0, 0.5}%
, keywordstyle=\bfseries
, keywordstyle=[2]\color[rgb]{0.75, 0, 0}
, keywordstyle=[3]\color[rgb]{0, 0.5, 0}
, keywordstyle=[4]\color[rgb]{0, 0.5, 0}
, keywordstyle=[5]\color[rgb]{0, 0, 0.75}
, columns=spaceflexible%
, keepspaces=true%
, showspaces=false%
, showtabs=false%
, showstringspaces=true%
}%

\lstdefinestyle{boxed}{
  style=colouredRust%
, numbers=left%
, firstnumber=auto%
, numberblanklines=true%
, frame=trbL%
, numberstyle=\tiny%
, frame=leftline%
, numbersep=7pt%
, framesep=5pt%
, framerule=10pt%
, xleftmargin=15pt%
, backgroundcolor=\color[gray]{0.97}%
, rulecolor=\color[gray]{0.90}%
}

\lstdefinelanguage{Markdown}{
  basicstyle=\ttfamily,
  commentstyle=\color{gray}\ttfamily\itshape, 
  stringstyle=\color{red}, 
  morecomment=[l]{<!--}, 
  morecomment=[s]{<!--}{-->}, 
  literate=
    {\#}{{\textcolor{blue}{\#}}}1   
    {*}{{\textcolor{blue}{*}}}1    
    {`}{\textasciigrave}1          
    {-}{{\textcolor{blue}{-}}}1    
    {[}{{\textcolor{cyan}{[}}}1    
    {]}{{\textcolor{cyan}{]}}}1    
    {(}{{\textcolor{cyan}{(}}}1    
    {)}{{\textcolor{cyan}{)}}}1    
}

\lstdefinelanguage{JSON}{
    basicstyle=\ttfamily,              
    stringstyle=\color{brown},         
    numberstyle=\color{blue},          
    keywordstyle=\color{red}\bfseries, 
    commentstyle=\color{gray},         
    showstringspaces=false,            
    morestring=[b]",                   
    morestring=[d]{:},                 
    literate=
      *{0}{{{\color{blue}0}}}{1}       
       {1}{{{\color{blue}1}}}{1}
       {2}{{{\color{blue}2}}}{1}
       {3}{{{\color{blue}3}}}{1}
       {4}{{{\color{blue}4}}}{1}
       {5}{{{\color{blue}5}}}{1}
       {6}{{{\color{blue}6}}}{1}
       {7}{{{\color{blue}7}}}{1}
       {8}{{{\color{blue}8}}}{1}
       {9}{{{\color{blue}9}}}{1}
       {:}{{{\color{black}:}}}{1}      
       {,}{{{\color{black},}}}{1}      
       {true}{{{\color{red}\bfseries true}}}{4}   
       {false}{{{\color{red}\bfseries false}}}{5} 
       {null}{{{\color{red}\bfseries null}}}{4}   
}

\newcommand{\rustcode}[1]{%
  \lstinline[language=Rust]{#1}%
}

 \newcommand{\shell}[1]{%
  \lstinline[language=bash, basicstyle=\footnotesize\ttfamily]{#1}%
}

\newcommand*\encircle[1]{%
 \raisebox{.5pt}{\textcircled{\raisebox{-.9pt}{#1}}}%
}

\newcommand{\abbrrtoscr}{{\sc Rtos}\xspace}
\newcommand{\abbrdrivercr}{{\sc Dr}\xspace}
\newcommand{\abbrhalcr}{{\sc Hal}\xspace}
\newcommand{\abbrbspcr}{{\sc Bsp}\xspace}
\newcommand{\abbrpaccr}{{\sc Pac}\xspace}
\newcommand{\abbrarchcr}{{\sc Arch}\xspace}
\newcommand{\abbrutilcr}{{\sc Util}\xspace}
\newcommand{\abbruncatcr}{{\sc Uncat}\xspace}

\def\innerradius{0.8cm}
\def\outerradius{1.2cm}

\newcommand{\wheelchart}[1]{
    \pgfmathsetmacro{\totalnum}{0}
    \foreach \value/\colour/\name in {#1} {
        \pgfmathparse{\value+\totalnum}
        \global\let\totalnum=\pgfmathresult
    }

    \begin{tikzpicture}

      \pgfmathsetmacro{\wheelwidth}{\outerradius-\innerradius}
      \pgfmathsetmacro{\midradius}{(\outerradius+\innerradius)/2}

      \begin{scope}[rotate=90]

      \pgfmathsetmacro{\cumnum}{0}
      \foreach \value/\colour/\name in {#1} {
            \pgfmathsetmacro{\newcumnum}{\cumnum + \value/\totalnum*360}

            \pgfmathsetmacro{\percentage}{\value/\totalnum*100}
            \pgfmathsetmacro{\midangle}{-(\cumnum+\newcumnum)/2}

            \pgfmathparse{
               (-\midangle<180?"west":"east")
            } \edef\textanchor{\pgfmathresult}
            \pgfmathsetmacro\labelshiftdir{1-2*(-\midangle>180)}

            \fill[\colour] (-\cumnum:\outerradius) arc (-\cumnum:-(\newcumnum):\outerradius) --
            (-\newcumnum:\innerradius) arc (-\newcumnum:-(\cumnum):\innerradius) -- cycle;

            \draw  [-,thin] node [append after command={(\midangle:\midradius pt) -- (\midangle:\outerradius + 1ex) -- (\tikzlastnode)}] at (\midangle:\outerradius + 1ex) [xshift=\labelshiftdir*0.5cm,inner sep=0pt, outer sep=0pt, anchor=\textanchor]{\name: \pgfmathprintnumber{\percentage}\%};

            \global\let\cumnum=\newcumnum
        }

      \end{scope}
    \end{tikzpicture}
}

\newcommand{\rustembcrates}{$Cr_{e}$\xspace}
\newcommand{\rustnonembcrates}{$Cr_{n}$\xspace}

\newcommand{\nonnativerustrots}{$RRT_{nr}$\xspace}
\newcommand{\nativerustrots}{$RRT_{r}$\xspace}
\newcommand{\nonrustrots}{$CRT$\xspace}
\newcommand{\totalyesrust}{149}
\newcommand{\totalyesrustperc}{66.2\%}

\newcommand{\totalnorust}{76}

\newcommand{\totalnorustperc}{33.8\%}

\newcommand{\didnotconsiderrust}{7}

\newcommand{\didnotconsiderrustperc}{9\%}

\newcommand{\considerrust}{69}

\newcommand{\considerrustperc}{91\%}

\newcommand{\crawldate}{Feb 2024} 
\newcommand{\totalnostdunique}{{11,002}\xspace}

\newcommand{\totalnostdsuccess}{{8,148}\xspace}
\newcommand{\totalnostdsuccessperc}{{90.77\%}\xspace}

\newcommand{\totalnostdfailureperc}{{9.23\%}\xspace}

\newcommand{\totalembeddedtargets}{{23}\xspace}
\newcommand{\totaltargets}{{85}\xspace}
\newcommand{\cratesnighgltydepend}{{2,025}\xspace}
\newcommand{\cratesnighgltydependperc}{{18.4\%}\xspace}
\newcommand{\totalnonembeddednostd}{2,569}

\newcommand{\numembcrates}{2,836}
\newcommand{\numnonembcrates}{11,000}
\newcommand{\numnonnativertos}{6}
\newcommand{\numnativertos}{10}
\newcommand{\numcrtos}{16}

\newcommand{\numfindings}{16\xspace}
\newcommand{\numopenproblems}{8\xspace}
\newcommand{\ruststableversion}{1.77.2\xspace}

\newcommand{\totalstablebuildable}{8,977\xspace}

\newcommand{\nummcufamilies}{43}

\newcommand{\numrtoscrates}{6}
\newcommand{\numdrivercrates}{466}
\newcommand{\numhalcrates}{57}
\newcommand{\numbspcrates}{114}
\newcommand{\numpaccrates}{565}
\newcommand{\numarchcrates}{15}
\newcommand{\numutilcrates}{4,764}
\newcommand{\numuncategorizedcrates}{421}
\newcommand{\totalnumcrates}{6,408}

\newcommand{\numrtoscrwrappers}{2}
\newcommand{\numdrivercrwrappers}{31}    
\newcommand{\numhalcrwrappers}{29} 
\newcommand{\numbspcrwrappers}{100}       
\newcommand{\numpaccrwrappers}{439}     
\newcommand{\numarchcrwrappers}{9}   
\newcommand{\numutilcrwrappers}{173}
\newcommand{\numuncatcrwrappers}{30}
\newcommand{\numtotalwrappers}{813}

\newcommand{\perrtoscrwrappers}{33.33\%}
\newcommand{\perdrivercrwrappers}{6.65\%}
\newcommand{\perhalcrwrappers}{50.88\%}
\newcommand{\perbspcrwrappers}{87.72\%}
\newcommand{\perpaccrwrappers}{77.70\%}
\newcommand{\perarchcrwrappers}{60.00\%}
\newcommand{\perutilcrwrappers}{3.63\%}
\newcommand{\peruncatcrwrappers}{7.18\%}
\newcommand{\numtotalwrappersperc}{12.69\%}

\newcommand{\numrtoscompiled}{3}
\newcommand{\perrtoscompiled}{50.00\%}
\newcommand{\numrtosub}{3}
\newcommand{\perrtosub}{100.00\%}
\newcommand{\numrtosuf}{2}
\newcommand{\perrtosuf}{66.67\%}
\newcommand{\numrtosut}{0}
\newcommand{\perrtosut}{0.00\%}
\newcommand{\numrtosuti}{2}
\newcommand{\perrtosuti}{66.67\%}
\newcommand{\numrtosuatone}{3}
\newcommand{\perrtosuatone}{100.00\%}

\newcommand{\numdrivercompiled}{423}
\newcommand{\perdrivercompiled}{90.77\%}
\newcommand{\numdriverub}{70}
\newcommand{\perdriverub}{16.55\%}
\newcommand{\numdriveruf}{55}
\newcommand{\perdriveruf}{13.00\%}
\newcommand{\numdriverut}{3}
\newcommand{\perdriverut}{0.71\%}
\newcommand{\numdriveruti}{11}
\newcommand{\perdriveruti}{2.60\%}
\newcommand{\numdriveruatone}{91}
\newcommand{\perdriveruatone}{21.51\%}

\newcommand{\numhalcompiled}{42}
\newcommand{\perhalcompiled}{73.68\%}
\newcommand{\numhalub}{28}
\newcommand{\perhalub}{66.67\%}
\newcommand{\numhaluf}{20}
\newcommand{\perhaluf}{47.62\%}
\newcommand{\numhalut}{9}
\newcommand{\perhalut}{21.43\%}
\newcommand{\numhaluti}{12}
\newcommand{\perhaluti}{28.57\%}
\newcommand{\numhaluatone}{31}
\newcommand{\perhaluatone}{73.81\%}

\newcommand{\numbspcompiled}{89}
\newcommand{\perbspcompiled}{78.07\%}
\newcommand{\numbspub}{25}
\newcommand{\perbspub}{28.09\%}
\newcommand{\numbspuf}{23}
\newcommand{\perbspuf}{25.84\%}
\newcommand{\numbsput}{0}
\newcommand{\perbsput}{0.00\%}
\newcommand{\numbsputi}{4}
\newcommand{\perbsputi}{4.49\%}
\newcommand{\numbspuatone}{29}
\newcommand{\perbspuatone}{32.58\%}

\newcommand{\numpaccompiled}{560}
\newcommand{\perpaccompiled}{99.12\%}
\newcommand{\numpacub}{508}
\newcommand{\perpacub}{90.71\%}
\newcommand{\numpacuf}{538}
\newcommand{\perpacuf}{96.07\%}
\newcommand{\numpacut}{5}
\newcommand{\perpacut}{0.89\%}
\newcommand{\numpacuti}{528}
\newcommand{\perpacuti}{94.29\%}
\newcommand{\numpacuatone}{547}
\newcommand{\perpacuatone}{97.68\%}

\newcommand{\numarchcompiled}{10}
\newcommand{\perarchcompiled}{66.67\%}
\newcommand{\numarchub}{9}
\newcommand{\perarchub}{90.00\%}
\newcommand{\numarchuf}{10}
\newcommand{\perarchuf}{100.00\%}
\newcommand{\numarchut}{1}
\newcommand{\perarchut}{10.00\%}
\newcommand{\numarchuti}{1}
\newcommand{\perarchuti}{10.00\%}
\newcommand{\numarchuatone}{10}
\newcommand{\perarchuatone}{100.00\%}

\newcommand{\numutilcompiled}{4,473}
\newcommand{\perutilcompiled}{93.89\%}
\newcommand{\numutilub}{1,591}
\newcommand{\perutilub}{35.57\%}
\newcommand{\numutiluf}{1053}
\newcommand{\perutiluf}{23.54\%}
\newcommand{\numutilut}{210}
\newcommand{\perutilut}{4.69\%}
\newcommand{\numutiluti}{554}
\newcommand{\perutiluti}{12.39\%}
\newcommand{\numutiluatone}{1,790}
\newcommand{\perutiluatone}{40.02\%}

\newcommand{\numuncatcompiled}{403}
\newcommand{\peruncatcompiled}{95.72\%}
\newcommand{\numuncatub}{102}
\newcommand{\peruncatub}{25.31\%}
\newcommand{\numuncatuf}{78}
\newcommand{\peruncatuf}{19.35\%}
\newcommand{\numuncatut}{8}
\newcommand{\peruncatut}{1.99\%}
\newcommand{\numuncatuti}{30}
\newcommand{\peruncatuti}{7.44\%}
\newcommand{\numuncatuatone}{133}
\newcommand{\peruncatuatone}{33.00\%}

\newcommand{\qratesnumtotalfailed}{405}
\newcommand{\qratesnumtotalcompiled}{6003}
\newcommand{\qratesnumtotalcompiledperc}{93.68\%}
\newcommand{\qratesnumtotalub}{2336}
\newcommand{\qratesnumtotalubperc}{38.91\%}
\newcommand{\qratesnumtotaluf}{1779}
\newcommand{\qratesnumtotalufperc}{29.64\%}
\newcommand{\qratesnumtotalut}{236}
\newcommand{\qratesnumtotalutperc}{3.93\%}
\newcommand{\qratesnumtotaluti}{1143}
\newcommand{\qratesnumtotalutiperc}{19.04\%}
\newcommand{\numtotalunsafecrates}{2634}
\newcommand{\numtotalunsafecratesperc}{43.88\%}

\newcommand{\lockbudrtoscompiled}{6}
\newcommand{\lockbudrtoscompiledperc}{100.00\%}
\newcommand{\lockbudrtoswarnings}{0}
\newcommand{\lockbudrtoswarningsperc}{0.00\%}

\newcommand{\lockbuddrivercompiled}{466}
\newcommand{\lockbuddrivercompiledperc}{100.00\%}
\newcommand{\lockbuddriverwarnings}{0}
\newcommand{\lockbuddriverwarningsperc}{0.00\%}

\newcommand{\lockbudhalcompiled}{56}
\newcommand{\lockbudhalcompiledperc}{98.25\%}
\newcommand{\lockbudhalwarnings}{0}
\newcommand{\lockbudhalwarningsperc}{0.00\%}

\newcommand{\lockbudbspcompiled}{114}
\newcommand{\lockbudbspcompiledperc}{100.00\%}
\newcommand{\lockbudbspwarnings}{0}
\newcommand{\lockbudbspwarningsperc}{0.00\%}

\newcommand{\lockbudpaccompiled}{565}
\newcommand{\lockbudpaccompiledperc}{100.00\%}
\newcommand{\lockbudpacwarnings}{0}
\newcommand{\lockbudpacwarningsperc}{0.00\%}

\newcommand{\lockbudarchcompiled}{15}
\newcommand{\lockbudarchcompiledperc}{100.00\%}
\newcommand{\lockbudarchwarnings}{0}
\newcommand{\lockbudarchwarningsperc}{0.00\%}

\newcommand{\lockbudutilcompiled}{4,733}
\newcommand{\lockbudutilcompiledperc}{99.35\%}
\newcommand{\lockbudutilwarnings}{5}
\newcommand{\lockbudutilwarningsperc}{0.11\%}

\newcommand{\lockbuduncatcompiled}{417}
\newcommand{\lockbuduncatcompiledperc}{99.05\%}
\newcommand{\lockbuduncatwarnings}{3}
\newcommand{\lockbuduncatwarningsperc}{0.72\%}

\newcommand{\lockbudtotalcompiled}{6,372}
\newcommand{\lockbudtotalcompiledperc}{99.44\%}
\newcommand{\lockbudtotalwarningscompiled}{8}
\newcommand{\lockbudtotalwarningscompiledperc}{0.13\%}

\newcommand{\yugartoscompiled}{5}
\newcommand{\yugartoscompiledperc}{71.43\%}
\newcommand{\yugartoswarnings}{0}
\newcommand{\yugartoswarningsperc}{0.00\%}

\newcommand{\yugadrivercompiled}{421}
\newcommand{\yugadrivercompiledperc}{90.34\%}
\newcommand{\yugadriverwarnings}{1}
\newcommand{\yugadriverwarningsperc}{0.24\%}

\newcommand{\yugahalcompiled}{43}
\newcommand{\yugahalcompiledperc}{75.44\%}
\newcommand{\yugahalwarnings}{0}
\newcommand{\yugahalwarningsperc}{0.00\%}

\newcommand{\yugabspcompiled}{102}
\newcommand{\yugabspcompiledperc}{89.47\%}
\newcommand{\yugabspwarnings}{0}
\newcommand{\yugabspwarningsperc}{0.00\%}

\newcommand{\yugapaccompiled}{560}
\newcommand{\yugapaccompiledperc}{99.12\%}
\newcommand{\yugapacwarnings}{4}
\newcommand{\yugapacwarningsperc}{0.71\%}

\newcommand{\yugaarchcompiled}{10}
\newcommand{\yugaarchcompiledperc}{66.67\%}
\newcommand{\yugaarchwarnings}{0}
\newcommand{\yugaarchwarningsperc}{0.00\%}

\newcommand{\yugautilcompiled}{4,304}
\newcommand{\yugautilcompiledperc}{90.34\%}
\newcommand{\yugautilwarnings}{59}
\newcommand{\yugautilwarningsperc}{1.37\%}

\newcommand{\yugauncatcompiled}{397}
\newcommand{\yugauncatcompiledperc}{94.52\%}
\newcommand{\yugauncatwarnings}{4}
\newcommand{\yugauncatwarningsperc}{1.01\%}

\newcommand{\yugatotalcompiled}{5,842}
\newcommand{\yugatotalcompiledperc}{91.17\%}
\newcommand{\yugatotalwarningscompiled}{68}
\newcommand{\yugatotalwarningscompiledperc}{1.16\%}

\newcommand{\fficheckerrtoscompiled}{2}
\newcommand{\fficheckerrtoscompiledperc}{28.57\%}
\newcommand{\fficheckerrtoswarnings}{0}
\newcommand{\fficheckerrtoswarningsperc}{0.00\%}

\newcommand{\fficheckerdrivercompiled}{246}
\newcommand{\fficheckerdrivercompiledperc}{52.79\%}
\newcommand{\fficheckerdriverwarnings}{0}
\newcommand{\fficheckerdriverwarningsperc}{0.00\%}

\newcommand{\fficheckerhalcompiled}{27}
\newcommand{\fficheckerhalcompiledperc}{47.37\%}
\newcommand{\fficheckerhalwarnings}{0}
\newcommand{\fficheckerhalwarningsperc}{0.00\%}

\newcommand{\fficheckerbspcompiled}{33}
\newcommand{\fficheckerbspcompiledperc}{28.95\%}
\newcommand{\fficheckerbspwarnings}{0}
\newcommand{\fficheckerbspwarningsperc}{0.00\%}

\newcommand{\fficheckerpaccompiled}{394}
\newcommand{\fficheckerpaccompiledperc}{69.73\%}
\newcommand{\fficheckerpacwarnings}{0}
\newcommand{\fficheckerpacwarningsperc}{0.00\%}

\newcommand{\fficheckerarchcompiled}{6}
\newcommand{\fficheckerarchcompiledperc}{40.00\%}
\newcommand{\fficheckerarchwarnings}{0}
\newcommand{\fficheckerarchwarningsperc}{0.00\%}

\newcommand{\fficheckerutilcompiled}{2,787}
\newcommand{\fficheckerutilcompiledperc}{58.50\%}
\newcommand{\fficheckerutilwarnings}{4}
\newcommand{\fficheckerutilwarningsperc}{0.14\%}

\newcommand{\fficheckeruncatcompiled}{271}
\newcommand{\fficheckeruncatcompiledperc}{64.52\%}
\newcommand{\fficheckeruncatwarnings}{1}
\newcommand{\fficheckeruncatwarningsperc}{0.37\%}

\newcommand{\fficheckertotalcompiled}{3,766}
\newcommand{\fficheckertotalcompiledperc}{58.77\%}
\newcommand{\fficheckertotalwarningscompiled}{5}
\newcommand{\fficheckertotalwarningscompiledperc}{0.13\%}

\newcommand{\rcanaryrtoscompiled}{4}
\newcommand{\rcanaryrtoscompiledperc}{66.67\%}
\newcommand{\rcanaryrtoswarnings}{1}
\newcommand{\rcanaryrtoswarningsperc}{25.00\%}

\newcommand{\rcanarydrivercompiled}{438}
\newcommand{\rcanarydrivercompiledperc}{93.99\%}
\newcommand{\rcanarydriverwarnings}{0}
\newcommand{\rcanarydriverwarningsperc}{0.00\%}

\newcommand{\rcanaryhalcompiled}{53}
\newcommand{\rcanaryhalcompiledperc}{92.98\%}
\newcommand{\rcanaryhalwarnings}{0}
\newcommand{\rcanaryhalwarningsperc}{0.00\%}

\newcommand{\rcanarybspcompiled}{110}
\newcommand{\rcanarybspcompiledperc}{96.49\%}
\newcommand{\rcanarybspwarnings}{0}
\newcommand{\rcanarybspwarningsperc}{0.00\%}

\newcommand{\rcanarypaccompiled}{549}
\newcommand{\rcanarypaccompiledperc}{97.17\%}
\newcommand{\rcanarypacwarnings}{0}
\newcommand{\rcanarypacwarningsperc}{0.00\%}

\newcommand{\rcanaryarchcompiled}{10}
\newcommand{\rcanaryarchcompiledperc}{66.67\%}
\newcommand{\rcanaryarchwarnings}{0}
\newcommand{\rcanaryarchwarningsperc}{0.00\%}

\newcommand{\rcanaryutilcompiled}{4,623}
\newcommand{\rcanaryutilcompiledperc}{97.04\%}
\newcommand{\rcanaryutilwarnings}{113}
\newcommand{\rcanaryutilwarningsperc}{2.44\%}

\newcommand{\rcanaryuncatcompiled}{413}
\newcommand{\rcanaryuncatcompiledperc}{98.10\%}
\newcommand{\rcanaryuncatwarnings}{5}
\newcommand{\rcanaryuncatwarningsperc}{1.21\%}

\newcommand{\rcanarytotalcompiled}{6,200}
\newcommand{\rcanarytotalcompiledperc}{96.75\%}
\newcommand{\rcanarytotalwarningscompiled}{119}
\newcommand{\rcanarytotalwarningscompiledperc}{1.92\%}

\newcommand{\rudranumrtoscompiled}{3}
\newcommand{\rudraperrtoscompiled}{50.00\%}
\newcommand{\rudranumrtosud}{0}
\newcommand{\rudraperrtosud}{0.00\%}
\newcommand{\rudranumrtossv}{0}
\newcommand{\rudraperrtossv}{0.00\%}
\newcommand{\rudranumrtosatone}{0}
\newcommand{\rudraperrtosatone}{0.00\%}

\newcommand{\rudranumdrivercompiled}{257}
\newcommand{\rudraperdrivercompiled}{55.15\%}
\newcommand{\rudranumdriverud}{0}
\newcommand{\rudraperdriverud}{0.00\%}
\newcommand{\rudranumdriversv}{5}
\newcommand{\rudraperdriversv}{1.95\%}
\newcommand{\rudranumdriveratone}{5}
\newcommand{\rudraperdriveratone}{1.1.95\%}

\newcommand{\rudranumhalcompiled}{26}
\newcommand{\rudraperhalcompiled}{45.61\%}
\newcommand{\rudranumhalud}{0}
\newcommand{\rudraperhalud}{0.00\%}
\newcommand{\rudranumhalsv}{0}
\newcommand{\rudraperhalsv}{0.00\%}
\newcommand{\rudranumhalatone}{0}
\newcommand{\rudraperhalatone}{0.00\%}

\newcommand{\rudranumbspcompiled}{58}
\newcommand{\rudraperbspcompiled}{50.88\%}
\newcommand{\rudranumbspud}{0}
\newcommand{\rudraperbspud}{0.00\%}
\newcommand{\rudranumbspsv}{2}
\newcommand{\rudraperbspsv}{3.45\%}
\newcommand{\rudranumbspatone}{2}
\newcommand{\rudraperbspatone}{3.45\%}

\newcommand{\rudranumpaccompiled}{385}
\newcommand{\rudraperpaccompiled}{68.14\%}
\newcommand{\rudranumpacud}{0}
\newcommand{\rudraperpacud}{0.00\%}
\newcommand{\rudranumpacsv}{322}
\newcommand{\rudraperpacsv}{83.64\%}
\newcommand{\rudranumpacatone}{322}
\newcommand{\rudraperpacatone}{83.64\%}

\newcommand{\rudranumarchcompiled}{5}
\newcommand{\rudraperarchcompiled}{33.33\%}
\newcommand{\rudranumarchud}{0}
\newcommand{\rudraperarchud}{0.00\%}
\newcommand{\rudranumarchsv}{0}
\newcommand{\rudraperarchsv}{0.00\%}
\newcommand{\rudranumarchatone}{0}
\newcommand{\rudraperarchatone}{0.00\%}

\newcommand{\rudranumutilcompiled}{2830}
\newcommand{\rudraperutilcompiled}{59.40\%}
\newcommand{\rudranumutilud}{69}
\newcommand{\rudraperutilud}{2.44\%}
\newcommand{\rudranumutilsv}{79}
\newcommand{\rudraperutilsv}{2.79\%}
\newcommand{\rudranumutilatone}{140}
\newcommand{\rudraperutilatone}{4.95\%}

\newcommand{\rudranumuncatcompiled}{287}
\newcommand{\rudraperuncatcompiled}{68.66\%}
\newcommand{\rudranumuncatud}{4}
\newcommand{\rudraperuncatud}{1.39\%}
\newcommand{\rudranumuncatsv}{14}
\newcommand{\rudraperuncatsv}{4.88\%}
\newcommand{\rudranumuncatatone}{17}
\newcommand{\rudraperuncatatone}{5.92\%}

\newcommand{\rudratotalsuccess}{3852}
\newcommand{\rudratotalsuccessperc}{60.11\%}
\newcommand{\rudratotalud}{73}
\newcommand{\rudratotaludperc}{1.89\%}
\newcommand{\rudratotalsv}{425}
\newcommand{\rudratotalsvperc}{11.03\%}
\newcommand{\rudratotalatone}{489}
\newcommand{\rudratotalatoneperc}{12.69\%}

\newcommand{\sdrtosnumsuccessful}{4}
\newcommand{\sdrtosnumsuccessfulperc}{66.67\%}
\newcommand{\sdrtosnumuaf}{2}
\newcommand{\sdrtosnumuafperc}{50.00\%}
\newcommand{\sdrtosnumdf}{2}
\newcommand{\sdrtosnumdfperc}{50.00\%}
\newcommand{\sdrtosnumdp}{2}
\newcommand{\sdrtosnumdpperc}{50.00\%}
\newcommand{\sdrtosnumima}{0}
\newcommand{\sdrtosnumimaperc}{0.00\%}
\newcommand{\sdrtosnumatone}{3}
\newcommand{\sdrtosnumatoneperc}{75.00\%}

\newcommand{\sddrivernumsuccessful}{436}
\newcommand{\sddrivernumsuccessfulperc}{93.56\%}
\newcommand{\sddrivernumuaf}{84}
\newcommand{\sddrivernumuafperc}{19.27\%}
\newcommand{\sddrivernumdf}{122}
\newcommand{\sddrivernumdfperc}{27.98\%}
\newcommand{\sddrivernumdp}{97}
\newcommand{\sddrivernumdpperc}{22.25\%}
\newcommand{\sddrivernumima}{0}
\newcommand{\sddrivernumimaperc}{0.00\%}
\newcommand{\sddrivernumatone}{126}
\newcommand{\sddrivernumatoneperc}{28.90\%}

\newcommand{\sdhalnumsuccessful}{52}
\newcommand{\sdhalnumsuccessfulperc}{91.23\%}
\newcommand{\sdhalnumuaf}{24}
\newcommand{\sdhalnumuafperc}{46.15\%}
\newcommand{\sdhalnumdf}{29}
\newcommand{\sdhalnumdfperc}{55.77\%}
\newcommand{\sdhalnumdp}{24}
\newcommand{\sdhalnumdpperc}{46.15\%}
\newcommand{\sdhalnumima}{0}
\newcommand{\sdhalnumimaperc}{0.00\%}
\newcommand{\sdhalnumatone}{29}
\newcommand{\sdhalnumatoneperc}{55.77\%}

\newcommand{\sdbspnumsuccessful}{109}
\newcommand{\sdbspnumsuccessfulperc}{95.61\%}
\newcommand{\sdbspnumuaf}{92}
\newcommand{\sdbspnumuafperc}{84.40\%}
\newcommand{\sdbspnumdf}{96}
\newcommand{\sdbspnumdfperc}{88.07\%}
\newcommand{\sdbspnumdp}{92}
\newcommand{\sdbspnumdpperc}{84.40\%}
\newcommand{\sdbspnumima}{0}
\newcommand{\sdbspnumimaperc}{0.00\%}
\newcommand{\sdbspnumatone}{96}
\newcommand{\sdbspnumatoneperc}{88.07\%}

\newcommand{\sdpacnumsuccessful}{560}
\newcommand{\sdpacnumsuccessfulperc}{99.12\%}
\newcommand{\sdpacnumuaf}{19}
\newcommand{\sdpacnumuafperc}{3.39\%}
\newcommand{\sdpacnumdf}{84}
\newcommand{\sdpacnumdfperc}{15.00\%}
\newcommand{\sdpacnumdp}{20}
\newcommand{\sdpacnumdpperc}{3.57\%}
\newcommand{\sdpacnumima}{0}
\newcommand{\sdpacnumimaperc}{0.00\%}
\newcommand{\sdpacnumatone}{84}
\newcommand{\sdpacnumatoneperc}{15.00\%}

\newcommand{\sdarchnumsuccessful}{10}
\newcommand{\sdarchnumsuccessfulperc}{66.67\%}
\newcommand{\sdarchnumuaf}{2}
\newcommand{\sdarchnumuafperc}{20.00\%}
\newcommand{\sdarchnumdf}{3}
\newcommand{\sdarchnumdfperc}{30.00\%}
\newcommand{\sdarchnumdp}{2}
\newcommand{\sdarchnumdpperc}{20.00\%}
\newcommand{\sdarchnumima}{0}
\newcommand{\sdarchnumimaperc}{0.00\%}
\newcommand{\sdarchnumatone}{3}
\newcommand{\sdarchnumatoneperc}{30.00\%}

\newcommand{\sdutilnumsuccessful}{4,625}
\newcommand{\sdutilnumsuccessfulperc}{97.08\%}
\newcommand{\sdutilnumuaf}{1,109}
\newcommand{\sdutilnumuafperc}{23.98\%}
\newcommand{\sdutilnumdf}{1,406}
\newcommand{\sdutilnumdfperc}{30.40\%}
\newcommand{\sdutilnumdp}{1,317}
\newcommand{\sdutilnumdpperc}{28.49\%}
\newcommand{\sdutilnumima}{0}
\newcommand{\sdutilnumimaperc}{0.00\%}
\newcommand{\sdutilnumatone}{1,679}
\newcommand{\sdutilnumatoneperc}{36.30\%}

\newcommand{\sduncatnumsuccessful}{413}
\newcommand{\sduncatnumsuccessfulperc}{98.10\%}
\newcommand{\sduncatnumuaf}{121}
\newcommand{\sduncatnumuafperc}{29.30\%}
\newcommand{\sduncatnumdf}{141}
\newcommand{\sduncatnumdfperc}{34.14\%}
\newcommand{\sduncatnumdp}{131}
\newcommand{\sduncatnumdpperc}{31.72\%}
\newcommand{\sduncatnumima}{0}
\newcommand{\sduncatnumimaperc}{0.00\%}
\newcommand{\sduncatnumatone}{156}
\newcommand{\sduncatnumatoneperc}{37.77\%}

\newcommand{\sdtotalcompiled}{6.209}
\newcommand{\sdtotalcompiledperc}{96.89\%}
\newcommand{\sdtotaluaf}{1453}
\newcommand{\sdtotaluafperc}{23.40\%}
\newcommand{\sdtotaldf}{1883}
\newcommand{\sdtotaldfperc}{30.33\%}
\newcommand{\sdtotaldp}{1,685}
\newcommand{\sdtotaldpperc}{27.14\%}
\newcommand{\sdtotalima}{0}
\newcommand{\sdtotalimaperc}{0.00\%}
\newcommand{\sdnumatonetotal}{2,176}
\newcommand{\sdnumatonetotalperc}{35.05\%}

\newcommand{\numsafedropfailed}{200}
\newcommand{\numsafedropfailedperc}{3.12\%}

\newcommand{\clartosnumhavstat}{6}
\newcommand{\clartosnumhavstatperc}{100.00\%}
\newcommand{\clartosnumatonexfer}{0}
\newcommand{\clartosnumatonexferperc}{0.00\%}

\newcommand{\cladrivernumhavstat}{462}
\newcommand{\cladrivernumhavstatperc}{99.14\%}
\newcommand{\cladrivernumatonexfer}{0}
\newcommand{\cladrivernumatonexferperc}{0.00\%}

\newcommand{\clahalnumhavstat}{56}
\newcommand{\clahalnumhavstatperc}{98.25\%}
\newcommand{\clahalnumatonexfer}{0}
\newcommand{\clahalnumatonexferperc}{0.00\%}

\newcommand{\clabspnumhavstat}{113}
\newcommand{\clabspnumhavstatperc}{99.12\%}
\newcommand{\clabspnumatonexfer}{0}
\newcommand{\clabspnumatonexferperc}{0.00\%}

\newcommand{\clapacnumhavstat}{562}
\newcommand{\clapacnumhavstatperc}{99.47\%}
\newcommand{\clapacnumatonexfer}{0}
\newcommand{\clapacnumatonexferperc}{0.00\%}

\newcommand{\claarchnumhavstat}{15}
\newcommand{\claarchnumhavstatperc}{100.00\%}
\newcommand{\claarchnumatonexfer}{0}
\newcommand{\claarchnumatonexferperc}{0.00\%}

\newcommand{\clautilnumhavstat}{4727}
\newcommand{\clautilnumhavstatperc}{99.22\%}
\newcommand{\clautilnumatonexfer}{58}
\newcommand{\clautilnumatonexferperc}{1.23\%}

\newcommand{\clauncatnumhavstat}{416}
\newcommand{\clauncatnumhavstatperc}{98.81\%}
\newcommand{\clauncatnumatonexfer}{14}
\newcommand{\clauncatnumatonexferperc}{3.37\%}

\newcommand{\clatotalnumhavstat}{6,357}
\newcommand{\clatotalnumhavstatperc}{99.20\%}
\newcommand{\clatotalnumatonexfer}{73}
\newcommand{\clatotalnumatonexferperc}{1.13\%}

\newcommand{\clametricsatleastonecrates}{198}

\newcommand{\numctorustfailed}{6}

\newcommand{\numctorustfailedperc}{37.5\%}

\newcommand{\numctorustpass}{10}

\newcommand{\numctorustpassperc}{62.5\%}

\newcommand{\numctorustcompilationfailed}{9}

\newcommand{\numsasttools}{6}
\newcommand{\numdev}{225}

\newcommand{\rtosavgffiincompat}{26}

\newcommand{\mcrfscore}{82}

\begin{document}

\title{Rust for Embedded Systems: Current State and Open
Problems
\ifSUPPLEMENT
\\ (Extended Report)
\fi
}


\author{Ayushi Sharma}
\authornote{Both authors contributed equally to this research.}
\affiliation{
  \institution{Purdue University}
  \city{West Lafayette}
  \country{USA}
}
\email{sharm616@purdue.edu}
\orcid{0009-0008-8967-8016}

\author{Shashank Sharma}
\authornotemark[1]
\affiliation{
  \institution{Purdue University}
  \city{West Lafayette}
  \country{USA}
}
\email{sharm611@purdue.edu}
\orcid{0009-0005-4149-884X}

\author{Sai Ritvik Tanksalkar}
\affiliation{
  \institution{Purdue University}
  \city{West Lafayette}
  \country{USA}
}
\email{stanksal@purdue.edu}
\orcid{0009-0008-6428-0511}

\author{Santiago Torres-Arias}
\affiliation{
  \institution{Purdue University}
  \city{West Lafayette}
  \country{USA}
}
\email{torresar@purdue.edu}
\orcid{0000-0002-9283-3557}

\author{Aravind Machiry}
\affiliation{
  \institution{Purdue University}
  \city{West Lafayette}
  \country{USA}
}
\email{amachiry@purdue.edu}
\orcid{0000-0001-5124-6818}



\begin{abstract}
Embedded software is used in safety-critical systems such as medical devices and autonomous vehicles, where software defects, including security vulnerabilities, have severe consequences.
Most embedded codebases are developed in unsafe languages, specifically C/C++, and are riddled with memory safety vulnerabilities.
To prevent such vulnerabilities,~\rust{}, a performant memory-safe systems language, provides an optimal choice for developing embedded software.
\rust{} interoperability enables developing~\rust{} applications on top of existing C codebases.
Despite this, even the most resourceful organizations continue to develop embedded software in C/C++.

This paper performs the first systematic study to holistically understand the current state and challenges of using~\rust{} for embedded systems.
Our study is organized across three research questions.
We collected a dataset of~\totalnumcrates{}~\rust{} embedded software spanning various categories and~\numsasttools{}~\ac{SAST} tools.
We performed a systematic analysis of our dataset and surveys with~\numdev{} developers to investigate our research questions.
We found that existing~\rust{} software support is inadequate,~\ac{SAST} tools cannot handle certain features of~\rust{} embedded software, resulting in failures, and the prevalence of advanced types in existing~\rust{} software makes it challenging to engineer interoperable code.
In addition, we found various challenges faced by developers in using~\rust{} for embedded systems development.
\end{abstract}





\maketitle

\pagestyle{plain}


\section{Introduction}
Our dependence on embedded devices (\eg{} IoT devices), has significantly increased, controlling various aspects of our daily lives, including homes~\cite{Alrawi2019SoK:Deployments}, transportation~\cite{al2020intelligence}, traffic management~\cite{soni2017review}, and the distribution of vital resources like food~\cite{prapti2022internet} and power~\cite{o2013industrial}. The adoption of these devices has seen rapid and extensive growth, with an estimated count of over 50 billion devices~\cite{al-garadi_survey_2020}.
Vulnerabilities in these devices have far-reaching consequences~\cite{antonakakis2017understanding, writer_5_2020} due to the pervasive and interconnected nature of these devices, as exemplified by the infamous Mirai botnet~\cite{margolis2017depth}.

Most embedded software are developed in ``unsafe'' (i.e., not memory-safe) languages, specifically C/C++, because of the low memory footprint, good performance, and the availability of extensive support software.
It is well-known that software developed in unsafe languages is prone to security vulnerabilities, especially memory safety vulnerabilities~\cite{cvetrend, microsoftmemsafe, top25}.
Likewise, embedded systems are riddled with security vulnerabilities~\cite{s21072329, noauthor_ripple20_nodate, noauthor_amnesia33_2020, valja2017study}.
The most recent URGENT/11~\cite{noauthor_urgent11_nodate} vulnerabilities in VxWorks that affected millions of medical~\cite{ganesan2011architecture}, SCADA systems~\cite{yadav2021architecture}, and industrial controllers~\cite{bhamare2020cybersecurity} are all because of memory safety (spatial) violations.
It is important to ensure that embedded systems do not contain memory-safety issues.
Automated memory-safety retrofitting techniques~\cite{duck2016heap, nagarakatte2009softbound,kendall1983bcc,steffen1992adding, necula2005ccured, condit2007dependent} based on compile-time instrumentation add significant overhead (both space and runtime) and are inapplicable to resource-constrained embedded systems.

Our analysis (details in~\apdx{apdx:rtosbugstudy}) of security vulnerabilities in various~\acp{RTOS} (an important class of embedded software) for the past ten years shows that 59 (54.2\%) of them are memory corruption vulnerabilities,~\ie spatial or temporal memory issues.
It is important to use memory-safe languages to prevent such vulnerabilities.
Furthermore, recently, the White House released a report~\cite{whitehousepressrelease} requiring future software to be developed in memory-safe languages.
Traditional memory-safe languages, such as Java, have high overhead and are not suitable for embedded systems.
\rust{}~\cite{noauthor_rust_nodate} is a memory-safe language that is shown to have comparable performance as native code.
Furthermore,~\rust{} can easily interoperate with existing unsafe codebases~\cite{rustffiomnibus}, enabling incremental adoption.
\rust{} team has a special focus on embedded systems~\cite{noauthor_rust_nodate-1}, and several works~\cite{10.1145/3124680.3124717, Levy2017MultiprogrammingA6} demonstrate the feasibility of engineering a complete embedded software stack in~\rust{}.
Furthermore,~\rust{} popularity is rising~\cite{pyplrustnodate}, and it is now adopted in Linux kernel~\cite{noauthor_rust_nodate-2} and Android~\cite{noauthor_android_nodate}.
Unfortunately, embedded systems are continuing to be developed in C. 
Even the most resourceful organizations, such as Microsoft, continue to develop embedded systems in C, as demonstrated by their recent Azure RTOS~\cite{noauthor_azure_nodate}.
Previous works~\cite{Fulton2021BenefitsAD, 9794066} investigated the challenges of adopting~\rust{} for regular software.
However, no work tries to understand factors affecting the use of~\rust{} for embedded systems development.

In this paper, we perform the first systematic study to holistically understand the issues in using~\rust{} for developing embedded systems.
Specifically, we explored the following research questions:
\begin{itemize}[noitemsep,nolistsep,leftmargin=*]
    \item\textbf{RQ1: Software Support.} How effective (quantity and quality) is the existing~\rust{} software support for embedded system development?
    \item\textbf{RQ2: Interoperability of~\rust{}.} Given that most of the existing embedded systems are in C, how well can~\rust{} interoperate with existing C codebases? and what are the challenges specific to embedded codebases?
    \item\textbf{RQ3: Developers Perspective.} What challenges do developers face in using~\rust{} for embedded system development?
\end{itemize}
We collected a dataset of~\totalnumcrates{}~\rust{} embedded software packages (or crates) spanning various categories and~\numsasttools{}~\ac{SAST} tools.
We performed a systematic analysis of our dataset and surveys with~\numdev{} developers to investigate our research questions.
Our study revealed several interesting findings (\numfindings{}), drawbacks of existing tools on embedded crates, and open problems (\numopenproblems{}) to increase the adoption of~\rust{} for embedded systems.
A few interesting findings include the following:
Embedded crates use more ($\sim$2X)~\rustcode{unsafe} blocks than non-embedded crates, significantly increasing the possibility of memory safety issues. However, existing techniques to isolate~\rustcode{unsafe} blocks are not applicable to embedded systems.
Existing developer support tools related to~\rust{}, such as~\ctorust{}, fail on majority of embedded codebases, as these tools fail to handle the diverse build systems and execution semantics of embedded systems.
The state-of-the-art \rust{}~\ac{SAST} tools perform poorly on embedded crates.
The superior type-system of~\rust{} makes it challenging to engineer interoperable embedded systems code.
Our observations are in line with the developer survey, and many developers consider the~\rust{} documentation for embedded systems poorly organized and want the documentation to contain more examples.
In summary, the following are our contributions:
\begin{itemize}[noitemsep,nolistsep,leftmargin=*]
    \item\textbf{Software Study:} We perform a systematic study of the~\rust{} software ecosystem to support the use of~\rust{} for embedded applications and highlight opportunity areas for adoption.
    \item\textbf{Tool Study:} We systematically studied the effectiveness of various (9)~\rust{} related tools,~\ie{}~\ac{SAST} tools, quality checking, and conversion tools, on embedded crates and identified various weaknesses specific to embedded systems.
    \item\textbf{Developer Aspects:} We performed a large-scale developer survey (with~\numdev{} developers) that highlights the challenges for slow adoption of~\rust{} for embedded applications.
\item\textbf{Dataset, Findings, and Open Problems~\footnote{https://zenodo.org/records/12775715}:} We curated a set of~\totalnumcrates{} embedded~\rust{} crates cataloged into various categories along with the necessary infrastructure to run analysis tools.
Our findings shed light on challenges in adopting~\rust{} for embedded systems, insights into open problems, and possible research directions.
\end{itemize}

\section{Background}
\label{sec:background}
This section provides the necessary background information for the rest of our work.
\subsection{Embedded Systems}
\label{subsec:embeddedsystems}
Embedded systems are designed to perform a designated set of tasks in a resource-constrained environment and on battery-powered devices.
There are several ways to categorize embedded systems.
Previous work~\cite{mariusndsspaper} categorizes embedded systems based on underlying~\ac{OS}.




(a) Type-1 systems have feature-rich general-purpose~\acp{OS} retrofitted for embedded systems.

(b) Type-2 systems or constrained devices~\cite{type2constraineddevices} use specialized embedded~\acp{OS}, which are usually designed as~\acf{RTOS},~\eg{} WEMO Light controller~\cite{wemoweb} running FreeRTOS~\cite{freertos}.

(c) Type-3 systems do not use~\ac{OS} abstractions and are rarely used in commercial products.

\par Previous work~\cite{Alrawi2019SoK:Deployments} shows consumer IoT devices, such as door knobs and temperature controllers, are mostly Type-2, which we primarily focus on. Type-2 systems execute on battery-powered and resource-constrained \acfp{MCU}. These systems have a lot of diversity in terms of 
hardware (\ac{MCU} and peripherals) and supported  
software~\cite{mariusndsspaper,white_making_2011}.
For instance, there are 31 different~\acp{RTOS}~\cite{osrtosweb}.
To handle this diversity, Type-2 systems have a layered design~\cite{Shen2023NCMAs} 
\ifSUPPLEMENT
(illustrated in~\fig{fig:background}).
\fi
Application logic is implemented in tasks managed by an~\ac{RTOS}.


\myparagraph{Execution Semantics}
The application and all the layers are compiled into a single monolithic binary and flashed onto the on-chip flash memory.
On reset, the contents of the memory are loaded into RAM, and execution starts from a pre-defined address,~\ie start or reset address. The tasks get scheduled per the scheduling policy, and handlers get triggered on corresponding events.

\subsection{\texorpdfstring{\rust{}}{Rust}}
\label{subsec:rustbackground}
\rust{} is a programming language created by Mozilla to build efficient and safe low-level software~\cite{klabnik_rust_2023, noauthor_rust_nodate, spedofrust, bugden_rust_2022, saligrama_practical_2019}.
\rust{} is targeted to achieve performance comparable to programs written in C while avoiding many safety issues in C, including concurrency and memory safety bugs.
This section provides a brief overview of~\rust{}'s safety features.
We recommend the~\rust{}'s official book~\cite{noauthor_rust_nodate} for a comprehensive understanding of these features.

\myparagraph{Features and Safety Guarantees}
\rust{} has several features, such as scopes, borrowing rules~\cite{noauthor_references_nodate}, single ownership~\cite{noauthorunderstandingrustowner}, and lifetimes~\cite{noauthor_lifetimes_nodate}, which force developers to follow certain practices enabling verification of memory safety properties (mostly) at compile time.
For instance, all read-write variables should be explicitly marked as mutable (\ie~\rustcode{mut}).
\rust{} provides both spatial and temporal memory safety. We provide a discussion of these guarantees in~\apdx{apdx:rustsafetyguarentees}

\myparagraph{Unsafe Rust}
\rust{} features can be too restrictive in a few cases.
For instance,~\rust{} requires all global variables to be read-only,~\ie disallows~\rustcode{mut}.
Similarly, we may need to call a C library function, which is also not allowed.
\rust{} provides~\rustcode{unsafe} blocks~\cite{noauthor_unsafe_nodate} to relax these restrictions and enable interaction with external language (or foreign) functions.
Arbitrary regions of code can be enclosed in an~\rustcode{unsafe} directive, and such code will be permitted certain (otherwise disallowed) actions, such as modifying a mutable global variable, dereferencing a raw pointer, calling an unsafe or external method, etc.

\myparagraph{\acf{FFI} Support}
\rust{} supports easy interaction with functions written in foreign languages through its~\acf{FFI}~\cite{rustffiomnibus}.
Specifically, such functions need to be annotated with special attributes, which enables the~\rust{} compiler to generate appropriate code respecting the ABI of the target language.

\myparagraph{Build System and Package Management}
\rust{} uses an integrated and easy-to-use build system and package manager called \textbf{Cargo}~\cite{noauthor_introduction_nodate}, which downloads library packages, called crates, as needed, during builds. 
Developers specify the build configuration along with all dependencies in a~\texttt{.toml} file~\cite{noauthor_toml_nodate} --- an organized key-value text file.
Rust has an official community package registry called~\url{crates.io}~\cite{noauthor_cratesio_nodate}, which (as of 29 April 2024) has more than 144K crates (\ie libraries) -- a 200\% increase over the last two years.

\myparagraph{\rust{} Compilation Attributes}
\rust{} supports attributes or configurations that enable compilation specialization. These attributes can be at various levels, \eg{} crate level, file level, function level, etc.
\textbf{\rustcode{no_std} attribute~\cite{nostdimpl}} is a crate-level attribute that avoids linking the entire standard module and results in small binaries.
Embedded software in~\rust{} should use this attribute to produce a self-contained binary independent of~\ac{OS} abstractions.
A~\rustcode{no_std} compatible crate should also have all its dependencies to be~\rustcode{no_std} compatible too.




\section{Study Methodology}
Our study aims to perform a holistic analysis to understand various aspects regarding usage of~\rust{} for embedded systems.
We aim to answer the following research questions:
\begin{itemize}[noitemsep,nolistsep,leftmargin=*]
    \item \textbf{RQ1: Software Support (\sect{subsec:adoptabilityofrust})}: How good is the software support for developing~\rust{} based embedded systems?
    \item \textbf{RQ2: Interoperability (\sect{subsec:rq3interop})}: How effective is the interoperability support of~\rust{} to deal with existing C based embedded codebases? 
    \item \textbf{RQ3: Developers Perspective (\sect{sec:rq4developerspers})}: What is developers' perspective on using~\rust{} for embedded systems?
\end{itemize}




\subsection{Embedded Software Dataset}
\label{subsec:embeddedswdataset}
Our goal is to collect~\rust{} crates that are applicable to embedded systems,~\ie{}~\rustcode{no_std} compatible, and can be built using one of the embedded toolchains.
We also want to identify the necessary compilation steps for all the collected crates.
\subsubsection{Crates collection}

As mentioned in~\sect{subsec:rustbackground},~\url{crates.io} is the official repository for all~\rust{} crates (\ie{} libraries).
However, there are other well-known sources, such as~\rust{}-embedded project~\cite{awesome-embedded-rust} and~\url{arewertosyet.com}, that also contain embedded~\rust{} projects.
We used a two-pronged approach to collect our embedded~\rust{} dataset.
\begin{itemize}[noitemsep,nolistsep,leftmargin=*]
\item\textbf{Crawling~\url{crates.io}:}
We crawled~\url{crates.io} (in~\crawldate{}) and got all the crates that are~\code{no_std} compatible.
This is not trivial as crates can declare~\code{no_std} compatibility at various levels.
For instance, ~\code{arduino_nano_connect v0.6.0}~\cite{arduino-nostd} crate declares~\code{no_std} compatibility at the crate level (\ie{} in~\rustcode{lib.rs} file). In contrast,\\
~\code{futures-executor v0.3.30} ~\cite{futures-cfg-nostd} crate uses~\rustcode{cfg} attribute to have only selected code blocks compile for~\code{no_std} environment. 
We perform lightweight static analysis to identify all such crates.

\item\textbf{Well-known Sources}:
We collected additional crates by crawling other well-known sources, specifically~\rust{}-embedded project and~\url{arewertosyet.com}.
\end{itemize}
After deduplication, we collected~\totalnostdunique{} unique crates.

\subsubsection{Identifying Stable Crates}
We tried to build crates using a stable version of~\rust{} and the corresponding compiler.
However, we identified that~\cratesnighgltydepend{} (\cratesnighgltydependperc) embedded crates depend on unstable~\rust{} versions, specifically nightly versions~\cite{ruststablevsnightly}.
These versions contain unstable~\rust{} features and might pose threats to the security guarantees of~\rust{}.
This is also reflected in one of the concerns (in~\sect{subsec:devexperienceusingrust}) raised by developers in using~\rust{} for embedded systems.
We only considered those that build on the stable version of~\rust{}, specifically~\ruststableversion{}.
This resulted in~\totalstablebuildable{} crates.

\subsubsection{Compilation Validation}
The~\rustcode{no_std} compatibility alone is a necessary condition but~\emph{not sufficient} for a crate to be usable on embedded systems.
For instance, ~\code{oc-wasm-futures}~\cite{ocwasm-crate} crate is~\rustcode{no_std} compatible but is for WebAssembly target, which is not an embedded architecture.

In this step, our goal is to validate crates to check for their applicability to embedded targets and identify the corresponding build commands.

\noindent\emph{Identifying Build Command:}
All crates can be built using~\shell{cargo build}, which uses the default configuration specified in the crate's~\shell{cargo.toml} file.
However, not all crates have their default configuration to be~\rustcode{no_std},~\ie{} the default build step (\shell{cargo build}) may not build~\rustcode{no_std} compatible version.
Such crates require special configuration flags to be passed to the build command,~\eg{} we need to use

\shell{cargo build
--no-default-features --features no_std} to build~\rustcode{no_std} variant of \code{async_cell}.
Developers specify such flags through~\rust{}'s conditional compilation attributes~\cite{rustconditionalcompilation} (\rustcode{cfg_attr}) as a propositional logic formula.

\noindent For instance, ~\rustcode{#![cfg_attr(all(feature = "no_std",}

\rustcode{not(feature = "std")), no_std)]} (in ~\code{resize v0.8.4} crate) indicates that we need to pass~\rustcode{no_std} flag and not pass~\rustcode{std} to build for~\rustcode{no_std}.

We use a lightweight static analysis technique to identify the appropriate build command.
First, we identify all~\rustcode{cfg_attr} directly corresponding to~\rustcode{no_std} (\ie{} containing~\rustcode{#![cfg_attr(..., no_std)]})
Second, we analyze the propositional formula to identify the flags that must be enabled or disabled for~\rustcode{no_std}.
Our technique was able to find the build commands for~\totalnostdsuccess (\totalnostdsuccessperc{}) of crates.

The rest~\totalnostdfailureperc{} failed because of the following reasons:
(i)~\emph{Incorrect attributes:} Here, crates have an incorrect~\rustcode{cfg_attr} specification.
For instance, ~\code{zero-crypto v0.1.11} crate has wrong flag name~\rustcode{(featusre = "std")} (Correct: (feature = "std"));
(ii)~\emph{Incorrect dependencies:} As mentioned in~\sect{sec:background}, for a crate to be~\rustcode{no_std} compatible all its dependencies should also be~\rustcode{no_std}.
However, few crates use dependencies that are either not~\rustcode{no_std} compatible or incorrectly configured.
For instance, ~\code{linux-kvm v0.2.0} crate depends on ~\code{linux-io v0.6.0} crate, which is not~\rustcode{no_std} compatible.; 
(iii)~\emph{Complex attributes specification:}
In our analysis, we consider only directly related flags,~\ie{} those specified along with~\rustcode{no-std} in~\rustcode{cfg_attr}.
However, there could be conditional compilation flags that are indirectly related.
For instance, ~\code{ab_glyph v0.2.23} crate requires ~\code{--features="libm"} flag. As we explain in~\apdx{apdx:rustindirectflags}, this flag dependency is specified indirectly and accurately identifying such flags is a combinatorial problem~\cite{acher2019learning}.

\begin{tcolorboxfloat}[htbp]
\noindent\fcolorbox{black}{green!30!white}{
  \parbox{\dimexpr\linewidth-2\fboxsep-2\fboxrule\relax}{\parskip=1pt\parindent=0pt
    \textbf{Open Problem P0.1:} We need techniques to automatically identify embedded system specific (\ie{}\rustcode{no_std} compatible) build configurations for~\rust{} crates --- this also enables identifying mistakes in build configurations (a prevalent problem).
One possible approach is to map the dependencies into a boolean formula for constraint solver and use the solution to derive the appropriate flags.
  }
}
\end{tcolorboxfloat}

\subsubsection{Embedded Targets Filtering}
There are~\totalembeddedtargets embedded targets (\totaltargets{} total targets) supported by the latest stable version of~\rustc (version: 1.77.2).
For all the crates for which we identified the build commands, we further filtered out crates that did not build for any of the embedded targets.
For instance, the~\code{no_std} variant of ~\code{winapi v0.3.9} crate is excluded because it requires an underlying operating system environment, which is not present in embedded targets. This resulted in a total of~\totalnumcrates{} crates after filtering out ~\totalnonembeddednostd{} crates.
Although our study focuses on type-2 systems,~\emph{our crates are not exclusively type-2.}
For instance, type-1 systems exist for \code{aarch64-cpu}, one of our targets.


\subsubsection{Categorization}
Based on the functionality, we categorize each embedded crate into eight categories (\tbl{tab:cratetypes}). We will present details of these categories in~\sect{subsec:rustlibrarysupport}.
We created a Multi-class Random Forest (MCRF) classifier~\cite{improvedmcrf} to categorize a given crate.
We manually categorized 2000 crates into various categories.
Using this as ground truth, we created an MCRF classifier with an F1-score of~\mcrfscore{}\%.
We used our MCRF classifier to categorize the rest of the crates.

\myparagraph{Summary}
We collected a total of~\totalnumcrates{} embedded crates along with appropriate build commands that produce~\rustcode{no_std} compatible binary.

\section{RQ1: Software Support}
\label{subsec:adoptabilityofrust}
In this research question, we want to assess the existing software support for engineering embedded systems in~\rust{}.
We plan to investigate the categories of support software that aid in common software engineering activities. Specifically:
\begin{itemize}[noitemsep,nolistsep,leftmargin=*]
\item\textbf{(For Development) Libraries and Support Software:} As explained in~\sect{sec:background}, applications in embedded systems are developed atop an~\ac{RTOS} and need necessary libraries that enable communicating with peripherals and provide certain common functionality (\eg network protocols).
\item\textbf{(For Testing) \ac{SAST} Tools:} These are an integral part of software development~\cite{roshaidie2020importance}. We need to have effective~\ac{SAST} tools to ensure the quality of newly developed~\rust{} based embedded systems.
\item\textbf{(To Handle Existing Codebases) C to~\rust{} Conversion Tools:} Given that most existing embedded codebases are in C, we should have tools to convert C to~\rust{} effectively.
\end{itemize}

\subsection{Libraries and Support Software}
\label{subsec:rustlibrarysupport}
The available software support,~\ie crates, can be broadly categorized into~\emph{hardware support crates} and~\emph{utility crates}.
\begin{table}[h]
\caption{Categorization of all available embedded crates.}
\label{tab:cratetypes}
\small
\centering
\resizebox{\columnwidth}{!}{   
{
\begin{tabular}{clrr}
\toprule
\multicolumn{1}{c}{\multirow{2}{*}{\textbf{\begin{tabular}[c]{@{}c@{}}Category\\ Abbr\end{tabular}}}} & \multicolumn{1}{c}{\multirow{2}{*}{\textbf{Type}}} & \multicolumn{2}{c}{\textbf{Crates}}                                                                                                  \\ \cline{3-4}
\multicolumn{1}{c}{}                                                                                  & \multicolumn{1}{c}{}                               & \multicolumn{1}{c}{\textbf{Total}} & \multicolumn{1}{c}{\textbf{\begin{tabular}[c]{@{}c@{}}Wrapper\\ Crates (\%Total)\end{tabular}}} \\ \midrule
\abbrrtoscr     &   RTOS Crates                         & \numrtoscrates{}        & \numrtoscrwrappers{} (\perrtoscrwrappers)     \\ \midrule
\abbrdrivercr   &   Driver Crates                       & \numdrivercrates{}      & \numdrivercrwrappers{} (\perdrivercrwrappers)  \\ \midrule
\abbrhalcr      &   HAL Crates                          & \numhalcrates{}         & \numhalcrwrappers{} (\perhalcrwrappers)     \\ \midrule
\abbrbspcr      &   Board Support Package               & \numbspcrates{}         & \numbspcrwrappers{} (\perbspcrwrappers)     \\ \midrule
\abbrpaccr      &   Peripheral Access Crates            & \numpaccrates{}         & \numpaccrwrappers{} (\perpaccrwrappers)     \\ \midrule
\abbrarchcr     &   Architecture Support Crates         & \numarchcrates{}        & \numarchcrwrappers{} (\perarchcrwrappers)    \\ \midrule
\abbrutilcr     &   Utility Crates                      & \numutilcrates{}        & \numutilcrwrappers{} (\perutilcrwrappers)    \\  \midrule
\abbruncatcr     &   Others                      & \numuncategorizedcrates{}        & \numuncatcrwrappers{} (\peruncatcrwrappers)    \\  \midrule
\multicolumn{2}{c}{\textbf{TOTAL}} & \totalnumcrates{} & \numtotalwrappers{} (\numtotalwrappersperc{})\\
\bottomrule
\end{tabular}
}   
}   
\end{table}
\subsubsection{Hardware Support Crates}
These provide software abstraction to interact with the hardware~\ie~\ac{MCU}, Peripherals, etc.
There are at least~\nummcufamilies{} different~\ac{MCU} families with various peripherals~\cite{moharkar2022review, swathi2018performance, abdrakhmanov2016development}.
We use the following categories to further categorize based on the type of interactions the crates provide.
~\tbl{tab:cratetypes} shows the summary of different categories of crates available for embedded systems development.
\begin{itemize}[noitemsep,nolistsep,leftmargin=*]
\item\textbf{Architecture Support:} These help in interacting with the processor and are~\ac{ISA} dependent.
For instance, the~\texttt{aarch64-cpu} crate~\cite{aarch64-cpu_2023} provides the function \rustcode{SPSR_EL2.write} to write to the Saved Program Status Register (SPSR) at EL2 exception level on aarch64 processors.
\emph{For embedded processors (\ie~\ac{RISC}~\acp{ISA}), there are support crates for ARM, MIPS, and RISC-V}.
\item\textbf{Peripheral Access:} These provide necessary functions to access peripherals on different~\acp{MCU}.
\emph{Out of~\nummcufamilies{} different~\ac{MCU} families, peripheral crates are currently available for only 16 (37\%).}
Most of these crates are generated using~\code{svd2rust} utility~\cite{svd2rust}, which automatically converts CMSIS-SVD~\cite{cmsissvd} file (XML description of ARM Cortex-M processors) into~\rust{} structs.
Consequently, most of these crates are for ARM Cortex-M family processors.
However, other~\acp{MCU}, such as AVR-based ATMEGA48PB, do not have SVD files but rather have~\code{.atdf} files.
There exist crates such as~\code{atdf2svd}~\cite{atdf2svd} to convert these into CMSIS-SVD format, but these tools are not robust and have issues.

\begin{tcolorboxfloat}[htbp]
\noindent\fcolorbox{black}{green!30!white}{
  \parbox{\dimexpr\linewidth-2\fboxsep-2\fboxrule\relax}{\parskip=1pt\parindent=0pt
   \textbf{Open Problem P1.1:}
We need effective techniques to automatically generate peripheral access crates for non-ARM architectures.
Recent advancements in LLM-assisted techniques~\cite{yang2024exploring} show promise in solving this problem.
  }
}
\end{tcolorboxfloat}

\item\textbf{HAL Implementation (HAL):}
These are implementations of \code{embedded-hal}~\cite{embeddedhal}, a common Hardware Abstraction Interface for various~\acp{MCU}.
These provide higher-level functions than peripheral crates, which just provide structures encapsulating peripheral registers.
For example,~\rustcode{GPIO::write} is a function provided by~\code{embedded-hal}, which involves multiple interactions through GPIO registers.
\emph{The HAL crates are available for 14 (32\%)~\ac{MCU} families.}
Unlike peripheral access crates, HAL crates are not automatically generated but are manually engineered.
Consequently, a lot of variance exists in~\ac{MCU} families having HAL crates.
For instance, Espressif~\acp{MCU} (with Xtensa~\ac{ISA}) has HAL crates~\cite{esphal} but does not have peripheral crates. 
\item\textbf{Board Support (BSP):}
These crates help in bootstrapping an~\ac{MCU} for an~\ac{RTOS}.
Specifically, these included bootloaders and other code to initialize and recognize other hardware peripherals.
BSPs are built using HAL and peripheral traits and expose higher-level functions to operate the underlying~\ac{MCU} or~\ac{SoC}.
For instance,~\code{hifive1} BSP crate~\cite{riscv-rusthifive1_nodate} (for HiFive1 boards) exposes a function~\rustcode{configure_spi_flash} which uses~\code{e310x_hal}\\~\cite{e310xhal} HAL crate to configure SPI Flash with maximum speed.
\emph{There are BSP crates for 19 (44\%) different boards.}
Unlike peripheral or HAL crates, BSP crates are specific to each board ---- a combination of~\ac{MCU} and peripherals.

\begin{tcolorboxfloat}
    
\noindent\fcolorbox{black}{green!30!white}{
  \parbox{\dimexpr\linewidth-2\fboxsep-2\fboxrule\relax}{\parskip=1pt\parindent=0pt
\textbf{Open Problem P1.2:}
Recent work~\cite{Shen2023NCMAs} exposes layering violations in C embedded systems, \ie{} components breaking the layered abstraction, \eg{}  HAL crate not using peripheral crates.
However, no such work exists for \rust{} crates.
  }
}

\end{tcolorboxfloat}

\item\textbf{Driver Crates:}
These are device drivers and expose functions to access various aspects of a device.
For instance,~\code{eeprom24x} driver crate~\cite{eeprom24x_2023} provides the necessary functions (\eg{}~\rustcode{read_byte}) to access 24x series serial EEPROMs.

\item\textbf{\ac{RTOS} Crates:}
These are complete~\acp{RTOS}, which expose necessary functions for task creation and synchronization, thus enabling easy creation of embedded applications.
\end{itemize}

\begin{tcolorboxfloat}
\noindent\fcolorbox{black}{yellow!30!white}{%
  \parbox{\dimexpr\linewidth-2\fboxsep-2\fboxrule\relax}{\parskip=1pt\parindent=0pt
    \textbf{Finding RQ1.1:}
Existing hardware support crates mainly target ARM Cortex-M family~\acp{MCU} and boards.
Although, there are ongoing efforts~\cite{avrrust} to improve support for other family~\acp{MCU} such as AVR. It is still a work in progress, and more efforts are required.
  }
}
\end{tcolorboxfloat}

\subsubsection{Utility Crates}
These are hardware-independent embedded crates (\ie~\code{no_std}) that provide various capabilities for embedded systems development.
For instance,~\code{tinybmp} embedded crate~\cite{tinybmp_2023} provides functions to parse BMP images.

Despite the existence of a large number of utility crates in the~\rust{} ecosystem, only~\numutilcrates{} can be used in embedded systems because of the requirement to be independent of~\ac{OS} abstractions,~\ie{} should not use~\rust{}'s~\code{std} crate (or be \code{no_std} compatible).
However, it is not easy to convert a crate to be~\code{no_std}~\cite{nostdredditpost} compatible as it requires the ability to perform semantic refactoring of the crate and its dependencies.
Our Extended Report has an example.
\begin{tcolorboxfloat}[htbp]
\noindent\fcolorbox{black}{green!30!white}{
  \parbox{\dimexpr\linewidth-2\fboxsep-2\fboxrule\relax}{\parskip=1pt\parindent=0pt
\textbf{Open Problem P1.3:}
We need techniques to automatically convert~\rust{} crates to be~\code{no_std} compatible to enable existing large quantity of crates to be usable in embedded systems.
Recent work by Sharma~\etal{}~\cite{aunor} demonstrates a possible approach using type-based conditional compilation.
  }
}
\end{tcolorboxfloat}

\begin{table*}[ht!]
\footnotesize
\caption{Summary of~\qrates{} results on embedded crates (uf: Call to Unsafe Function, ptr: Derefercing raw pointer, mstat: Use of Mutable Static, estat: Use of Extern Static, asm: Use of Inline Assembly, union: Access to Union Field)}
\label{tab:qrates}
\centering
\begin{tabular}{crrrrrrl}
\toprule
\multicolumn{1}{c}{\multirow{2}{*}{\textbf{Category}}} & \multicolumn{1}{c}{\multirow{2}{*}{\textbf{\begin{tabular}[c]{@{}c@{}}Num. Crates Successful\\ (\% of Total\\ from~\tbl{tab:cratetypes})\end{tabular}}}} & \multicolumn{5}{c}{\textbf{Number of Crates (\% of Successful) having}} & \multicolumn{1}{c}{\multirow{2}{*}{\textbf{\begin{tabular}[c]{@{}c@{}}Top 3 Reasons for\\ Unsafe Usage \end{tabular}}}} \                  \\ \cmidrule(lr){3-7}
\multicolumn{1}{c}{}                          & \multicolumn{1}{c}{}                                                                                                              & \multicolumn{1}{c}{\textbf{\begin{tabular}[c]{@{}c@{}}Unsafe \\ Blocks\end{tabular}}} & \multicolumn{1}{c}{\textbf{\begin{tabular}[c]{@{}c@{}}Unsafe \\ Functions\end{tabular}}} & \multicolumn{1}{c}{\textbf{\begin{tabular}[c]{@{}c@{}}Unsafe Trait\\ Impl\end{tabular}}} & \multicolumn{1}{c}{\textbf{\begin{tabular}[c]{@{}c@{}}Unsafe\\ Trait\end{tabular}}} & \multicolumn{1}{c}{\textbf{\begin{tabular}[c]{@{}c@{}}At least one\\ unsafe idiom\end{tabular}}} \\ \midrule

\abbrrtoscr     & \numrtoscompiled{} (\perrtoscompiled)    & \numrtosub{} (\perrtosub)    & \numrtosuf{} (\perrtosuf)    & \numrtosuti{} (\perrtosuti)   & \numrtosut{} (\perrtosut)    & \numrtosuatone{} (\perrtosuatone) & ud (82\%), mstat (29\%), estat (6\%)  \\ \midrule
\abbrdrivercr     & \numdrivercompiled{} (\perdrivercompiled)    & \numdriverub{} (\perdriverub)    & \numdriveruf{} (\perdriveruf)    & \numdriveruti{} (\perdriveruti)   & \numdriverut{} (\perdriverut)    & \numdriveruatone{} (\perdriveruatone) & ud (82\%), ptr (19\%), mstat (2\%)  \\ \midrule
\abbrhalcr     & \numhalcompiled{} (\perhalcompiled)    & \numhalub{} (\perhalub)    & \numhaluf{} (\perhaluf)    & \numhaluti{} (\perhaluti)   & \numhalut{} (\perhalut)    & \numhaluatone{} (\perhaluatone)  & uf (72\%), ptr (38\%), mstat (2\%) \\ \midrule
\abbrbspcr     & \numbspcompiled{} (\perbspcompiled)    & \numbspub{} (\perbspub)    & \numbspuf{} (\perbspuf)    & \numbsputi{} (\perbsputi)   & \numbsput{} (\perbsput)    & \numbspuatone{} (\perbspuatone)  & uf (88\%), ptr (77\%), mstat (2\%) \\ \midrule
\abbrpaccr     & \numpaccompiled{} (\perpaccompiled)    & \numpacub{} (\perpacub)    & \numpacuf{} (\perpacuf)    & \numpacuti{} (\perpacuti)   & \numpacut{} (\perpacut)    & \numpacuatone{} (\perpacuatone)  & uf (91\%), ptr (18\%), mstat (1\%) \\ \midrule
\abbrarchcr     & \numarchcompiled{} (\perarchcompiled)    & \numarchub{} (\perarchub)    & \numarchuf{} (\perarchuf)    & \numarchuti{} (\perarchuti)   & \numarchut{} (\perarchut)    & \numarchuatone{} (\perarchuatone)  & uf (52\%), asm (31\%), ptr (22\%) \\ \midrule
\abbrutilcr     & \numutilcompiled{} (\perutilcompiled)    & \numutilub{} (\perutilub)    & \numutiluf{} (\perutiluf)    & \numutiluti{} (\perutiluti)   & \numutilut{} (\perutilut)    & \numutiluatone{} (\perutiluatone)  & uf (89\%), ptr (16\%), union (1\%) \\ \midrule
\abbruncatcr     & \numuncatcompiled{} (\peruncatcompiled)    & \numuncatub{} (\peruncatub)    & \numuncatuf{} (\peruncatuf)    & \numuncatuti{} (\peruncatuti)   & \numuncatut{} (\peruncatut)    & \numuncatuatone{} (\peruncatuatone)  & uf (84\%), ptr (25\%), mstat (2\%) \\ \midrule
\textbf{Total}  & \qratesnumtotalcompiled{} (\qratesnumtotalcompiledperc) & \qratesnumtotalub{} (\qratesnumtotalubperc) & \qratesnumtotaluf{} (\qratesnumtotalufperc) & \qratesnumtotaluti{} (\qratesnumtotalutiperc) & \qratesnumtotalut{} (\qratesnumtotalutperc) & \numtotalunsafecrates{} (\numtotalunsafecratesperc{}) & uf (90\%), ptr (18\%), mstat (1\%) \\
\bottomrule
\end{tabular}
\end{table*}
\begin{table}[htbp]
\footnotesize
\caption{~\clametrics results.}
\centering
\label{tab:cla}
\begin{tabular}{crr}
\toprule
\multicolumn{1}{c}{\textbf{Category}} & \multicolumn{1}{c}{\textbf{\begin{tabular}[c]{@{}c@{}}Num. Crates Successful\\ (\% of Total\\ from ~\tbl{tab:cratetypes})\end{tabular}}} & \multicolumn{1}{c}{\textbf{\begin{tabular}[c]{@{}c@{}}Num. Crates having\\ at least one \\ Transfer Pt.\end{tabular}}} \\ \midrule
\abbrrtoscr     & \clartosnumhavstat{} (\clartosnumhavstatperc)   & \clartosnumatonexfer{} (\clartosnumatonexferperc)   \\ \midrule
\abbrdrivercr   & \cladrivernumhavstat{} (\cladrivernumhavstatperc)   & \cladrivernumatonexfer{} (\cladrivernumatonexferperc)   \\ \midrule
\abbrhalcr      & \clahalnumhavstat{} (\clahalnumhavstatperc)   & \clahalnumatonexfer{} (\clahalnumatonexferperc)   \\ \midrule
\abbrbspcr      & \clabspnumhavstat{} (\clabspnumhavstatperc)   & \clabspnumatonexfer{} (\clabspnumatonexferperc)   \\ \midrule
\abbrpaccr      & \clapacnumhavstat{} (\clapacnumhavstatperc)   & \clapacnumatonexfer{} (\clapacnumatonexferperc)   \\ \midrule
\abbrarchcr     & \claarchnumhavstat{} (\claarchnumhavstatperc)   & \claarchnumatonexfer{} (\claarchnumatonexferperc)   \\ \midrule
\abbrutilcr     & \clautilnumhavstat{} (\clautilnumhavstatperc)   & \clautilnumatonexfer{} (\clautilnumatonexferperc)   \\ \midrule
\abbruncatcr     & \clauncatnumhavstat{} (\clauncatnumhavstatperc)   & \clauncatnumatonexfer{} (\clauncatnumatonexferperc)   \\ \midrule
\textbf{Total}  & \clatotalnumhavstat{} (\clatotalnumhavstatperc)   & \clatotalnumatonexfer{} (\clatotalnumatonexferperc)   \\
\bottomrule
\end{tabular}%
\end{table}

\subsubsection{\texorpdfstring{Quality of Embedded~\rust{}}{Quality of Embedded Rust crates}}
\label{subsubsec:codequality}
At a high level, as shown by the last column of~\tbl{tab:cratetypes}, many (\numtotalwrappers{} (\numtotalwrappersperc{})) of the crates are just wrappers around C libraries (details in~\apdx{apdx:rustwrappercrates}).
We also consider crates that depend on a wrapper crate to be wrapper crates.
These crates are susceptible to the bugs in corresponding C libraries.
In other words, vulnerabilities in the wrapped libraries can be exploited to get complete control of the corresponding~\rust{} program.
This problem has received considerable attention, and several works try to isolate code running as part of libraries (or in general~\rustcode{unsafe} blocks) from the rest of the crate.
These techniques depend on special hardware features~\cite{bang2023trust, gulmez2023friend, 10.1145/3492321.3519582, 10.1145/3485832.3485903, 10.1145/3377811.3380325}, specifically Intel's~\ac{MPK} or~\ac{OS} abstractions, such as~\code{mprotect}~\cite{10.1145/3176258.3176330}, IPC mechanism~\cite{9700352}, sandboxing~\cite{10.1145/3144555.3144562} --- making them inapplicable to type-2 embedded systems,~\ie{}~\ac{RTOS} based embedded systems that run on~\acp{MCU}.

\begin{tcolorboxfloat}
\noindent\fcolorbox{black}{green!30!white}{
  \parbox{\dimexpr\linewidth-2\fboxsep-2\fboxrule\relax}{\parskip=1pt\parindent=0pt
   \textbf{Open Problem P1.4:}
We need techniques (applicable also for embedded systems) to isolate~\rust{} code from~\rustcode{unsafe} code,~\ie{} techniques that do not depend on hardware features,~\ac{OS} abstractions, and have low overhead.
Recent techniques~\cite{10179388,10179285} on C-based embedded software compartmentalization demonstrate possible approaches. However, these should be customized for \rust{}.
  }
}
\end{tcolorboxfloat}

\noindent\textbf{Code Quality:}
We use the following tools to further assess embedded crates' code quality.
\begin{itemize}[noitemsep,nolistsep,leftmargin=*]
\item\textbf{\qrates{}~\cite{10.1145/3428204}:}
This tool finds instances of various~\rustcode{unsafe} idioms,~\ie{} blocks, functions, traits, and trait implementations.
Unfortunately, the tool failed on~\qratesnumtotalfailed{} (6.30\%) crates. We provide a categorization of failures in~\apdx{apdx:qratecompilationfailures}.
Out of the remaining~\qratesnumtotalcompiled{} crates,~\numtotalunsafecrates{} (\numtotalunsafecratesperc{}) contain at least one~\rustcode{unsafe} idiom.
This is much higher than in non-embedded crates where only 23.6\% crates (as reported in~\cite{10.1145/3428204}) contain~\rustcode{unsafe} idioms.
\tbl{tab:qrates} shows the results along with top three reasons for \rustcode{unsafe}ness.
Note that the percentages are not cumulative,~\ie{} there could be multiple reasons for an \rustcode{unsafe} block.
These reasons differ from regular crates, indicating the need for different design decisions when creating analysis tools for embedded crates.
\item\textbf{\clametrics{}~\cite{mergendahl2022cross}:}
Recently, Mergendahl~\etal{}~\cite{mergendahl2022cross} demonstrated the feasibility of Cross-Language-Attacks, wherein interactions of~\rust{} with C/C++ could result in security vulnerabilities.
They released~\clametrics{}, a tool to identify these cross-language interaction points.
\clametrics{} works on binaries and require~\code{ELF} files with specific sections.
As we showed in~\tbl{tab:cratetypes}, there are~\numtotalwrappers{} wrapper crates, which means these contain at least one call from~\rust{} to C/C++,~\ie{} a transfer point.
Interestingly,~\clametrics{} found only~\clametricsatleastonecrates{} crates with interaction points.
These results indicate potential issues with the~\clametrics{} tool and we found that an important class of such transfer points that~\clametrics{} misses is indirect function calls. Indirect function calls are common in embedded systems due to their event driven nature. Recent works ~\cite{BlogLLVMCFIa} show that employing CFI mechanism through LLVM can help detect indirect calls. Although as we see in \ref{subsubsec:ctorustresults}, embedded systems fail to build with clang. We need more work in the area. ~\clametrics{} uses the differences in name mangling used by \rust{} and C++ to determine such transfer points.
This information would not be available for indirect function calls and hence~\clametrics{} misses out these.
\end{itemize}

\noindent\textbf{Security Implications:}
The prevalence of \rustcode{unsafe} idioms indicates that developers should be more cautious in using embedded crates.
Moreover, the robustness issues in analysis tools indicate that security researchers should consider embedded crates as part of their evaluation.


\begin{tcolorboxfloat}[htb]
\noindent\fcolorbox{black}{yellow!30!white}{%
  \parbox{\dimexpr\linewidth-2\fboxsep-2\fboxrule\relax}{\parskip=1pt\parindent=0pt
   \textbf{Finding RQ1.2:} Compared to non-embedded crates, many embedded crates (48.5\% v/s 23.6\%) contain~\rustcode{unsafe}~\rust{} code idioms.

\textbf{Finding RQ1.3:} \clametrics{} fails to identify cross-language interactions through indirect calls~\eg{} calls through function pointers.
  }
}
\end{tcolorboxfloat}

\subsection{\texorpdfstring{\ac{SAST} Tools}{SAST Tools}}
\label{subsec:sasttools}
As we show in~\sect{subsubsec:codequality}, embedded crates contain a large amount of~\rustcode{unsafe} blocks.
The presence of~\rustcode{unsafe} blocks potentially violates~\rust{}'s safety guarantees and results in various memory safety issues~\cite{10.1145/3428204}.
It is important to use~\ac{SAST} tools for embedded systems development in~\rust{}.
We investigate the effectiveness of state-of-the art~\rust{}~\ac{SAST} tools on embedded crates.





\begin{table}[htbp]
\captionsetup[table]{belowskip=-10pt}
\caption{Failure reasons of~\ac{SAST} tools and the number of affected crates.~\apdx{apdx:toolchain_failure} contains detailed and examples of failures.}
\label{tab:toolfailures}
\centering
\footnotesize
\begin{tabular}{c|c|c}
\toprule
\textbf{\begin{tabular}[c]{@{}c@{}}Failure Reason\end{tabular}}      & \textbf{\begin{tabular}[c]{@{}c@{}}Affected\\ Tools and Crates\end{tabular}} & 
\textbf{\begin{tabular}[c]{@{}c@{}}Total\end{tabular}}\\ 
\midrule
\begin{tabular}[c]{@{}c@{}}Toolchain \\ Incompatibility \end{tabular} & \begin{tabular}[c]{@{}c@{}}\ffichecker{} (2559,	39.93\%) \\ \rudra{} (2547, 39.75\%)\\   \yuga{} (539, 8.41\%)\\  \safedrop{} (166,2.59\%) \\ \rcanary{} (156, 2.43\%)\\ \lockbud{} (30, 0.46\%)\end{tabular} & \begin{tabular}[c]{@{}c@{}}2,692\end{tabular}          \\ \hline
\begin{tabular}[c]{@{}c@{}}Tool Crashes \end{tabular} & \begin{tabular}[c]{@{}c@{}}\ffichecker{} (67, 1.046\%) \\ \rcanary{} (9,0.14\%) \\ \safedrop{} (1,	0.015\%) \\\lockbud{} (5, 0.78) \\\yuga{} (1,	0.02\%)\end{tabular} &    \begin{tabular}[c]{@{}c@{}}89\end{tabular}\\ \hline
\begin{tabular}[c]{@{}c@{}}No binary target \\support \end{tabular} & \begin{tabular}[c]{@{}c@{}}\safedrop{} (27, 0.42\%)\\  \rcanary{} (25, 0.39\%)\\ \ffichecker{} (6, 0.094\%)\\  \rudra{} (4, 0.04\%)\end{tabular} &  \begin{tabular}[c]{@{}c@{}}27\end{tabular} \\  \hline
\begin{tabular}[c]{@{}c@{}}Ignoring Project-Specific\\Configurations \end{tabular} & \begin{tabular}[c]{@{}c@{}}\yuga{} (19,	0.30\%)\\ \rudra{} (6,	0.09\%)\\  \ffichecker{} (2, 0.03\%) \\  \safedrop{} (2, 0.03\%) \\ \rcanary{} (1, 0.02\%) \end{tabular}  & \begin{tabular}[c]{@{}c@{}}21\end{tabular}\\ \hline
\begin{tabular}[c]{@{}c@{}}Timeouts (large crates) \end{tabular} & \begin{tabular}[c]{@{}c@{}} \rcanary{} (16, 0.25\%) \end{tabular}  & \begin{tabular}[c]{@{}c@{}}16\end{tabular}\\  \hline
\begin{tabular}[c]{@{}c@{}}Rustc version \\ incompatibility\end{tabular} & \begin{tabular}[c]{@{}c@{}}\ffichecker{} (8,	0.12\%)\\ \end{tabular}  &  \begin{tabular}[c]{@{}c@{}}8\end{tabular} \\  \hline
\begin{tabular}[c]{@{}c@{}}Unknown Reasons \end{tabular} & \begin{tabular}[c]{@{}c@{}}\yuga{} (7, 0.11\%)\end{tabular}        &     \begin{tabular}[c]{@{}c@{}}7\end{tabular} \\ 
\bottomrule
\end{tabular}
\end{table}

\begin{table*}[tb]
    \caption{Summary of~\rust{} SAST tools evaluated as part of the study and the features that are supported (\greencheck) or not (\redcross), type of reports (\colorbox{red!55}{insufficient information},~\colorbox{orange!55}{missing relevant details}, or~\colorbox{green!55}{detailed report}), along with references to results.}
    \label{tab:toolanalysis}
    \centering
    \scriptsize
    \begin{tabular}{c|c|c|cccc|c|c}
        \toprule
        \multirow{2}{*}{\textbf{\begin{tabular}[c]{@{}c@{}}Tool\\ Name\end{tabular}}} 
        & \multirow{2}{*}{\textbf{\begin{tabular}[c]{@{}c@{}}Target\\ Bug Types\end{tabular}}} 
        & \multirow{2}{*}{\textbf{\begin{tabular}[c]{@{}c@{}}Techniques\\ Used\end{tabular}}} 
        & \multicolumn{4}{c|}{\textbf{Flow-Tracking}} 
        & \multirow{2}{*}{\textbf{\begin{tabular}[c]{@{}c@{}}Report \\ Type\end{tabular}}} 
        & \multirow{2}{*}{\textbf{\begin{tabular}[c]{@{}c@{}}Result \\ Reference\end{tabular}}} \\ 
        \cline{4-7}
        & & & \multicolumn{1}{c|}{\textbf{Require}} 
        & \multicolumn{1}{c|}{\textbf{\begin{tabular}[c]{@{}c@{}}Across\\ Unsafe blocks\end{tabular}}} 
        & \multicolumn{1}{c|}{\textbf{\begin{tabular}[c]{@{}c@{}}Across\\ FFI Boundaries\end{tabular}}} 
        & \textbf{\begin{tabular}[c]{@{}c@{}}Handles Async/\\ Indirect Flows\end{tabular}} 
        & \\ 
        \midrule

        \lockbud{}~\cite{10.1145/3385412.3386036} 
        & {\begin{tabular}[c]{@{}c@{}}Concurrency,\\Memory Safety\end{tabular}} 
        & {\begin{tabular}[c]{@{}c@{}}CallGraph Analysis,\\Points-to Analysis,\\Dataflow Analysis\end{tabular}} 
        & \multicolumn{1}{c|}{\greencheck} 
        & \multicolumn{1}{c|}{\redcross} 
        & \multicolumn{1}{c|}{\redcross} 
        & \multicolumn{1}{c|}{\redcross} 
        & {\cellcolor{orange!55}{\begin{tabular}[c]{@{}c@{}}Source Level Traces \\ (\apdx{apdx:lockbudreport})\end{tabular}}} 
        & 
        \ifSUPPLEMENT
            \tbl{tab:lockbud} \\
            \cline{1-9}
        \fi

        \rudra{}~\cite{bae2021rudra} 
        & Temporal \& Spatial safety 
        & {\begin{tabular}[c]{@{}c@{}}Taint Analysis,\\ Dataflow analysis\end{tabular}} 
        & \multicolumn{1}{c|}{\greencheck} 
        & \multicolumn{1}{c|}{\greencheck} 
        & \multicolumn{1}{c|}{\redcross} 
        & \multicolumn{1}{c|}{\redcross} 
        & {\cellcolor{orange!55}{\begin{tabular}[c]{@{}c@{}}Source Level Traces \\ (\apdx{apdx:rudrareport})\end{tabular}}} 
        & 
        \ifSUPPLEMENT
            \tbl{tab:rudra} \\
            \cline{1-9}
        \fi

        \yuga{}~\cite{Nitin2023YugaAD} 
        & Temporal Safety 
        & {\begin{tabular}[c]{@{}c@{}}Taint Analysis,\\ Alias Analysis\end{tabular}} 
        & \multicolumn{1}{c|}{\greencheck} 
        & \multicolumn{1}{c|}{\greencheck} 
        & \multicolumn{1}{c|}{\redcross} 
        & \multicolumn{1}{c|}{\redcross} 
        & {\cellcolor{green!55}{\begin{tabular}[c]{@{}c@{}}Detailed Source\\Level Traces \\ (\apdx{apdx:yugareport})\end{tabular}}} 
        &
        \ifSUPPLEMENT
            \tbl{tab:yuga} \\
            \cline{1-9}
        \fi

        \rcanary{}~\cite{rcanary} 
        & Temporal Safety 
        & {\begin{tabular}[c]{@{}c@{}}Dataflow Analysis,\\ Constraint solving\end{tabular}} 
        & \multicolumn{1}{c|}{\greencheck} 
        & \multicolumn{1}{c|}{\greencheck} 
        & \multicolumn{1}{c|}{\redcross} 
        & \multicolumn{1}{c|}{\redcross} 
        & {\cellcolor{red!55}{\begin{tabular}[c]{@{}c@{}}sat/unsat \\ (\apdx{apdx:rcanaryreport})\end{tabular}}} 
        &
        \ifSUPPLEMENT
            \tbl{tab:rcanary} \\
            \cline{1-9}
        \fi

        \ffichecker{} ~\cite{ffichecker} 
        & Temporal Safety 
        & {\begin{tabular}[c]{@{}c@{}}Taint Analysis,\\ Alias Analysis\end{tabular}} 
        & \multicolumn{1}{c|}{\greencheck} 
        & \multicolumn{1}{c|}{\greencheck} 
        & \multicolumn{1}{c|}{\greencheck} 
        & \multicolumn{1}{c|}{\redcross} 
        & {\cellcolor{red!55}{\begin{tabular}[c|]{@{}c@{}}Generic warning\\ (\apdx{apdx:fficheckerreport})\end{tabular}}} 
        &
        \ifSUPPLEMENT
            \tbl{tab:ffichecker} \\
            \cline{1-9}
        \fi

        \safedrop{}~\cite{Cui2021SafeDropDM} 
        & Temporal Safety
        & {\begin{tabular}[c]{@{}c@{}}Dataflow Analysis,\\ Alias analysis\end{tabular}} 
        & \multicolumn{1}{c|}{\greencheck} 
        & \multicolumn{1}{c|}{\greencheck} 
        & \multicolumn{1}{c|}{\redcross} 
        & \multicolumn{1}{c|}{\redcross} 
        & {\cellcolor{orange!55}{\begin{tabular}[c]{@{}c@{}}Affected Function name\\ (\apdx{apdx:safedropreport})\end{tabular}}} 
        &
        \ifSUPPLEMENT
            \tbl{tab:safedrop} \\
        \fi
        \bottomrule
    \end{tabular}
\end{table*}

\subsubsection{Tool Selection}
\label{subsub:toolselection}
The recent study by Ami~\etal{}~\cite{ami2023false} shows that developers are more likely to use~\ac{SAST} tools that do not require any configuration and can be directly used on a software project.
Following this, we aim to collect state-of-the-art and readily usable~\ac {SAST} tools.
Specifically, these tools should run directly on a given crate and not require any configuration.
We searched~\rust{} forums and the past five years' proceedings of top-tier security and software engineering conferences and collected the set of readily usable~\ac{SAST} tools.
We filtered out tools that did not satisfy our requirements. For instance, we did not select~\mirchecker{}~\cite{10.1145/3460120.3484541} because it requires configuring the abstract domain and specifying analysis entry points.
After filtering such tools, our investigation resulted in six tools as summarized in~\tbl{tab:toolanalysis}. 

Almost all tools except for~\lockbud{} focus on identifying temporal safety issues,~\eg{} incorrect lifetimes, and multiple drops.
All these tools are based on flow-tracking as indicated by~\greencheck{} under the~\emph{Require} column.

\subsubsection{Qualitative Assessment}
As presented in~\sect{subsubsec:codequality}, embedded\\~\rust{} crates have a higher percentage of unsafe blocks, use~\ac{FFI} functions (\ie{} interact with C libraries), and use indirect (or function pointer) calls.
\ac{SAST} tools should be able to handle these idioms to be effective on embedded crates.

\noindent\emph{Supported Features:} We referred to the research papers on the corresponding tools and created simple examples to identify their capabilities to handle idioms common in embedded crates.
The various columns under~\emph{Flow-Tracking} indicate whether each of these idioms is either supported (\greencheck{}) or not (\redcross{}) by the corresponding tools.
All tools, except for~\lockbud{}, handle flows across~\rustcode{unsafe} blocks. 
None of the tools handle data-flows through indirect calls (\ie{} function pointer calls) --- which is one of the common idioms in embedded systems (\sect{subsec:embeddedsystems} and~\cite{Shen2023NCMAs}).
Except for~\ffichecker{}, none of the tools handle flows across~\ac{FFI} boundaries, another common usage in embedded crates.

\noindent\emph{Usability:}
 Despite the existence of standard formats, such as SARIF~\cite{sarifformat}, \rust{} \ac{SAST} tools employ ad-hoc ways to report their warnings.
As shown in the last column of~\tbl{tab:toolanalysis}, these reports do not always contain the necessary information to triage the underlying defect.
The~\tbl{tab:toolanalysis} also contains references to the examples of corresponding warnings.
\emph{All tools, except for~\yuga{}, report their findings in an ad-hoc and hard-to-analyze manner.}
\rcanary{} and~\ffichecker{} just provide a single-line warning without any details about the source location --- which makes these warnings almost impossible to analyze.
\lockbud{},~\rudra{}, and~\safedrop{} provide source level traces.
However, the complex semantics of~\rust{} lifetimes make it hard to triage the reported warnings.
\yuga{} provides a well-formatted HTML report with necessary information about the identified defect.

\subsubsection{Effectiveness}
\label{subsubsec:sasteffectiveness}
There is no existing~\rust{} embedded systems bug dataset.
The situation is the same for C/C++~\cite{shen2023empirical}.
, which also contains references to the complete results
We evaluated the effectiveness of~\ac{SAST} tools on our embedded crates dataset.
The last column of~\tbl{tab:toolanalysis} has references to the complete results for each tool.
\noindent\emph{Robustness Issues:}
\ac{SAST} tools fail to handle the diverse build configurations, code structures, and semantics of embedded~\rust{} crates. Consequently, these tools failed on several crates. The~\tbl{tab:toolfailures} summarizes different classes of failures, affected tools, and crates.
The majority of failures are because of~\emph{``Toolchain Imcompatibilities''},~\ie{} tools fail to identify the backend toolchain required by crates and consequently fail to analyze.

\noindent\emph{Precision:}
Given the large number of warnings, we used a random sampling method to analyze the precision of the tools.
Specifically, we picked 30 crates with more downloads than the median across all the crates.
This is to avoid selecting unimportant or rarely used crates.

We ignored \rcanary{} and \ffichecker{} as their warnings did not contain enough information.
Furthermore, even for other tools (\eg{}~\lockbud{}), the information provided is not always sufficient to triage the corresponding warning.
We categorized each warning into True Positive (TP), False Positive (FP) or Insufficient Information (IsIn).
~\tbl{tab:toolresults} shows the results, the top two reasons for false positives, and the corresponding examples.
First, tools were able to find real defects. 
\ifSUPPLEMENT
The~\lst{lst:lockbudtp} in ~\ shows a real deadlock found by~\lockbud{} in the \texttt{tracing-log} crate.
\fi
However, the true positive rate is very low.
Contrary to tools' claim, all tools suffer from a very high false positive rate (40\%-90\%) on embedded crates.
This is unsurprising as all these tools are evaluated (mostly) on non-embedded crates.
This indicates that the design choices of the current tools fail to consider embedded crates.

\begin{table}[htb]
\caption{Summary of manual analysis of results of various~\rust{} SAST tools with True Positives (TP), False positives (FP), and Insufficient Information (IsIn). We list the top two reasons for FPs here (examples and complete results in \ifSUPPLEMENT ~\apdx{apdx:toolsreporting}\fi). Details in~\sect{subsubsec:sasteffectiveness}.}
\label{tab:toolresults}
\centering
\scriptsize
\begin{tabular}{c|rrr|c}
\toprule
\multirow{2}{*}{\textbf{\begin{tabular}[c]{@{}c@{}}Tool\\ Name\end{tabular}}} 
& \multicolumn{3}{c|}{\textbf{Analysis Results}} 
& \multirow{2}{*}{\textbf{\begin{tabular}[c]{@{}c@{}}Top 2\\ FP Reasons\end{tabular}}} \\ \cline{2-4}
& \multicolumn{1}{c|}{\textbf{TP}} 
& \multicolumn{1}{c|}{\textbf{FP}} 
& \multicolumn{1}{c|}{\textbf{IsIn}} 
& \\ \midrule

\begin{tabular}[c]{@{}c@{}}\lockbud{}\end{tabular} 
& \multicolumn{1}{r}{10 (33\%)} 
& \multicolumn{1}{r}{14 (46\%)} 
& 6 (20\%) 
& \begin{tabular}[c]{@{}c@{}}Lock type ambiguity (42\%)\\ Complex Program Semantics (32\%)\end{tabular} 
\\ \hline

\begin{tabular}[c]{@{}c@{}}\rudra{}\end{tabular} 
& \multicolumn{1}{r}{1 (2.7\%)} 
& \multicolumn{1}{r}{36 (97.2\%)} 
& N/A 
& \begin{tabular}[c]{@{}c@{}}Ignoring Explicit Guards (50\%)\\Ignoring Atomic Types (30\%)\end{tabular} 
\\ \hline

\begin{tabular}[c]{@{}c@{}}\yuga{}\end{tabular} 
& \multicolumn{1}{r}{10 (33\%)} 
& \multicolumn{1}{r}{13 (43\%)} 
& 7 (23.3\%) 
& \begin{tabular}[c]{@{}c@{}}Ignoring Caller Contexts (56\%)\\Complex Program Semantics (30\%)\end{tabular} 
\\ \hline

\begin{tabular}[c]{@{}c@{}}\safedrop{}\end{tabular} 
& \multicolumn{1}{r}{9 (30\%)} 
& \multicolumn{1}{l}{17 (56.6\%)} 
& 4 (13.3\%) 
& \begin{tabular}[c]{@{}c@{}}Infeasible Paths (80\%)\\Analysis Imprecision (10\%)\end{tabular} 
\\ \bottomrule
\end{tabular}
\end{table}

\begin{tcolorboxfloat}
\noindent\fcolorbox{black}{yellow!30!white}{%
  \parbox{\dimexpr\linewidth-2\fboxsep-2\fboxrule\relax}{\parskip=1pt\parindent=0pt
   \textbf{Finding RQ1.4:} Current~\ac{SAST} tools lack the necessary features required to effectively handle embedded crates.

\textbf{Finding RQ1.5:} Current~\ac{SAST} tools do not provide the necessary information to triage the reported defects, making it hard (rather impossible) to verify the reports.

\textbf{Finding RQ1.6:} Current~\ac{SAST} tools fail to effectively handle build idioms and configurations of embedded~\rust{} crates, resulting in robustness issues.

\textbf{Finding RQ1.7:} The design choices of current~\ac{SAST} tools fail to effectively handle the common idioms in embedded crates resulting in a very high false positive rate (40\%-90\%).
  }
}
\end{tcolorboxfloat}

\begin{tcolorboxfloat}
\noindent\fcolorbox{black}{green!30!white}{
  \parbox{\dimexpr\linewidth-2\fboxsep-2\fboxrule\relax}{\parskip=1pt\parindent=0pt
\textbf{Open Problem P1.5:}
There is no dataset of security bugs in~\rust{} embedded crates.
Recent systematic bug dataset creation works~\cite{10.1145/3428334} provide possible approaches to tackle this.
  }
}
\end{tcolorboxfloat}

\noindent\textbf{Security Implications:}
Our results indicate that developers cannot solely rely on existing automated~\ac{SAST} tools to assess their crates and should also perform manual or semi-automated assessments.
\subsection{C to~\rust{} Conversion Tools}
\label{subsec:ctorusttools}
We selected C to~\rust{} conversion tools by following the same approach as for~\ac{SAST} tools (\sect{subsub:toolselection}).
Although several tools satisfy our requirements, we present the results of only the~\ctorust{} tool.
Other recent tools, such as~\laertes{}~\cite{10.1145/3485498} and~\crusts{}~\cite{9793767}, do not work directly on C code but rather improve the~\rust{} code produced by~\ctorust{} through novel post-processing techniques.
As we will show in~\sect{subsubsec:ctorustresults},~\ctorust{} either failed or produced uncompilable~\rust{} code on (almost) the entire dataset.
Consequently, recent tools that depend on~\ctorust{} also failed on the dataset.
\subsubsection{Dataset}
We collected popular C/C++ based~\ac{RTOS} from~\url{osrtos.com}, which maintains the list of all popular~\acp{RTOS} released to date.
We selected well-maintained (\ie{} has build instructions) and compilable~\acp{RTOS}.
This resulted in a total of~\numcrtos{} C/C++~\acp{RTOS} (\nonrustrots).
The compilation of~\acp{RTOS} is specific to an~\ac{MCU} and includes HAL and other peripheral access libraries for the~\ac{MCU}.
Thus, using~\acp{RTOS} enables us to test the effectiveness of~\ctorust{} on codebases across different layers of embedded systems.

\subsubsection{Running~\ctorust{}}
\label{subsubsec:runningctorust}
To convert a project, we first need to capture compilation commands,~\eg{} generating~\code{compile_commands.json} using~\code{scan-build}~\cite{clangtoolsscanbuild}.
Next, we need to run~\ctorust{} on the captured~\code{compile_commands.json}.
\ctorust{} uses~\clang{} to parse C files and uses pattern-based techniques on the resulting~\ac{AST} to produce corresponding~\rust{} code.
Specifically, each compilation command (from~\code{compile_commands.json}) will be executed by replacing the compiler with~\clang{}.
However, just replacing the compiler will not work as embedded systems use non-standard and~\ac{MCU} specific toolchains,~\eg{}~\code{avr-gcc}, whose compiler flags/options may not be supported by \clang{}.
We followed an on-demand approach to convert into a \clang compatible variant and run~\ctorust{}.
Specifically, for each incompatible option leading to an error in conversion/compilation, we refer to \clang{}'s documentation to see if there is an alternative option (case-1), or if it is not supported by \clang{} (case-2). For case-1, we use the corresponding alternative flags, \eg{} we replace \texttt{-march=nehalem} with \texttt{-march=armv8-a}. For case-2, we remove those flags/options (5 flags), \eg \code{-Wformat-overflow}.
The removal of case-2 flags does not affect the conversion (a frontend task), as all of these flags are related to optimization (a middle/backend task).
\subsubsection{Results}
\label{subsubsec:ctorustresults}
\ifSUPPLEMENT
~\tbl{tab:c2rustblinky} (in Appendix) has a summary of the results.
\fi
All~\acp{RTOS}, except for two, required manually fixing

\noindent\code{compile_commands.json} (discussed in~\sect{subsubsec:runningctorust}).
\emph{\ctorust{} failed on~\numctorustfailed{} (\numctorustfailedperc{})~\acp{RTOS}}.
The two main reasons for this are: (i) Embedded system codebases often use (\clang) unsupported C language features, and (ii)~\ctorust{} uses~\rust{}~\code{std} library to generate certain wrapper functions, but as mentioned in~\sect{subsec:rustbackground},~\code{std} library should not be used in an embedded environment.
For instance,~\code{gnucc/oscore.c} file in~\code{stateos/StateOS} uses parameter references in naked functions, which is not supported by~\clang{}~\cite{r217200nodate} and consequently,~\ctorust{} fails.
It executed successfully on~\numctorustpass{} (\numctorustpassperc{})~\acp{RTOS}.
Out of which,~\emph{the generated~\rust{} code was incorrect or syntactically invalid (\eg{} missing semicolon) on~\numctorustcompilationfailed{} (90\%)~\acp{RTOS}.}
The conversion was successful (\ie{}~\ctorust{} produced compilable~\rust{} code) on only one~\ac{RTOS},~\ie{}\\~\code{kmilo17pet/QuarkTS}.

Finally, \ctorust{} uses a syntactic approach and consequently produces \rust{} code with mostly~\rustcode{unsafe} blocks.
Although recent works~\cite{10.1145/3485498} have tried to improve the situation, the progress is rather slow and requires more focused efforts.


\begin{tcolorboxfloat}
\noindent\fcolorbox{black}{yellow!30!white}{%
  \parbox{\dimexpr\linewidth-2\fboxsep-2\fboxrule\relax}{\parskip=1pt\parindent=0pt
    \textbf{Finding RQ1.8:} C to~\rust{} tools fail on most,~\ie{} 93.8\% (15/16), embedded codebases because of the prevalent use of special compiler flags and non-standard C language features.

\textbf{Finding RQ1.9:} C to~\rust{} tools do not consider the~\rustcode{no_std} requirement and consequently will generate~\rust{} code inapplicable for embedded systems.
  }
}
\end{tcolorboxfloat}

\section{RQ2: Interoperability of Rust}
\label{subsec:rq3interop}


Most existing embedded system codebases are written in C~\cite{shen2023empirical}.
Developers should be able to write~\rust{} code that can interoperate with existing C code to avoid reengineering the entire embedded software stack in~\rust{}.
As mentioned in~\sect{subsec:rustbackground},~\rust{} has~\acf{FFI} support enabling interoperability with code written in other languages, especially C.

To answer this research question, we investigate the effort and challenges in developing~\rust{} (or C) code that can interoperate with C (or~\rust{}) code.
We first provide a brief overview of recommended steps to develop interoperable code and quantify the effort and challenges specific to embedded systems.
Second, we will present our experience and challenges in engineering interoperable code in various embedded system development scenarios.
\begin{table*}[htb]
\footnotesize
\caption{Summary of Rust Interoperability Modes. We indicate whether each step is easy (\easyeffort), (\eg{} running a tool on a C file), requires medium effort or~\emph{automation opportunities} (\mediumeffort) (\eg{} configuring linker script), or requires significant effort or~\emph{open-problems} (\hardeffort) (\eg{} rewriting embedded C code in~\rust{}). The challenges affecting embedded systems are \colorbox{red!55}{highlighted}.}
\label{tab:interoptable}
\centering
\begin{tabular}{ccc|c|c|c}
\toprule
\multicolumn{3}{c|}{\textbf{Interoperability Modes}}                                                                                                                                                                                                                                                                                 & \multirow{2}{*}{\textbf{Method}}                                                                                                                                                   & \multirow{2}{*}{\textbf{Effort}}                                                                                                            & \multirow{2}{*}{\textbf{\begin{tabular}[c]{@{}c@{}}Embedded System\\ Specific\\ Challenges\end{tabular}}}           \\ \cline{1-3}
\multicolumn{1}{c|}{\textbf{Mode}}                                                                                                          & \multicolumn{1}{c|}{\textbf{\begin{tabular}[c]{@{}c@{}}Sub\\ Abbr.\end{tabular}}}         & \textbf{Desc.}                                                                             &                                                                                                                                                                                    &                                                                                                                                             &                                                                                                                       \\ \midrule
\multicolumn{1}{c|}{\multirow{2}{*}{\rotatebox[origin=c]{90}{\begin{tabular}[c]{@{}c@{}}R \textless{}-\textgreater C\\ (\sect{subsec:rusttoc})\end{tabular}}}}                    & \multicolumn{1}{c|}{\rotatebox[origin=c]{90}{\begin{tabular}[c]{@{}c@{}}R-\textgreater{}C\\ (\sect{subsubsec:callingcfromrust})\end{tabular}}} & \begin{tabular}[c]{@{}c@{}}Calling C function\\ from Rust\end{tabular}                     & \begin{tabular}[c]{@{}c@{}}1. Use bindgen to get declaration in Rust (\easyeffort)\\ 2. Link with target C object file (\mediumeffort)\end{tabular}                                                                & \begin{tabular}[c]{@{}c@{}}Easy\\ (All C types are\\ FFI Compatible)\end{tabular}                                                           & N/A                                                                                                                   \\ \cline{2-6} 
\multicolumn{1}{c|}{}                                                                                                                       & \multicolumn{1}{c|}{\rotatebox[origin=c]{90}{\begin{tabular}[c]{@{}c@{}}C-\textgreater{}R\\ (\sect{subsubsec:callingrustfromc})\end{tabular}}}                                                    & \begin{tabular}[c]{@{}c@{}}Calling Rust \\ function from C\end{tabular}                    & \begin{tabular}[c]{@{}c@{}}1. Use cbindgen to get declaration in C (\hardeffort).\\ 2. Link with target Rust object file (\mediumeffort).\end{tabular}                                                             & {\begin{tabular}[c]{@{}c@{}}Depends on the use of \\ Rust types not compatible\\ with C types.\\ (i.e., FFI Incompatible types)\end{tabular}} & {\cellcolor{red!55} \begin{tabular}[c]{@{}c@{}}Most functions in \\ Embedded crates use\\ FFI incompatible types.\end{tabular}} \\ 
\midrule
\multicolumn{1}{c|}{\multirow{3}{*}{\rotatebox[origin=c]{90}{\begin{tabular}[c]{@{}c@{}}Interoperability in\\ Embedded System \\ Components\\ (\sect{subsubsec:rustinteroprocrwc})\end{tabular}}}} & \multicolumn{1}{c|}{\rotatebox[origin=c]{90}{\begin{tabular}[c]{@{}c@{}}RoC\\ (\sect{subsubsec:rustappontopofc})\end{tabular}}}                                                                  & \begin{tabular}[c]{@{}c@{}}Developing Rust \\ Application on top of \\ C-RTOS\end{tabular} & \begin{tabular}[c]{@{}c@{}}1. Use bindgen to get C-RTOS \\ functions' declarations in Rust (\easyeffort).\\ 2. Modify the linker script (\mediumeffort).\end{tabular}                                            & \begin{tabular}[c]{@{}c@{}}Easy\\ (All C type are \\ FFI compatible)\end{tabular}                                                           & N/A                                                                                                                   \\ \cline{2-6} 
\multicolumn{1}{c|}{}                                                                                                                       & \multicolumn{1}{c|}{\rotatebox[origin=c]{90}{\begin{tabular}[c]{@{}c@{}}CoR\\ (\sect{subsubsec:capponrustrtos})\end{tabular}}}                                                                  & \begin{tabular}[c]{@{}c@{}}Developing C \\ Application on top of \\ Rust RTOS\end{tabular} & \begin{tabular}[c]{@{}c@{}}1. Use cbindgen to get Rust-RTOS \\ functions' declarations in C (\hardeffort).\\ 2. Modify the linker script (\mediumeffort).\end{tabular}                                           & {\begin{tabular}[c]{@{}c@{}}Depends on the use\\ of FFI incompatible types\\ in Rust RTOSes.\end{tabular}}                                    & {\cellcolor{red!55} \begin{tabular}[c]{@{}c@{}}There is a prevalent\\ use of FFI incompatible\\ types in Rust RTOSes.\end{tabular}}       \\ \cline{2-6} 
\multicolumn{1}{c|}{}                                                                                                                       & \multicolumn{1}{c|}{\rotatebox[origin=c]{90}{\begin{tabular}[c]{@{}c@{}}RwC\\ (\sect{subsubsec:rustcomincrtos})\end{tabular}}}              & \begin{tabular}[c]{@{}c@{}}Converting a component\\ in C-RTOS to Rust\end{tabular}         & \begin{tabular}[c]{@{}c@{}}1. Use bindgen to convert all dependent \\ component C headers to Rust (\easyeffort).\\ 2. Rewrite the target embedded \\ component in Rust (\hardeffort).\\ 3. Modify the Makefile (\mediumeffort).\end{tabular} & {\begin{tabular}[c]{@{}c@{}}Depends on the effort\\ to rewrite C code to Rust.\end{tabular}}                                                    & {\cellcolor{red!55}{\begin{tabular}[c]{@{}c@{}}C to~\rust{} conversion tools \\ fail to handle embedded codebases, \\ forcing manual rewriting.\end{tabular}}}                                                                                                                   \\ \bottomrule
\end{tabular}
\end{table*}

\subsection{\texorpdfstring{\rust{} $\Leftrightarrow$ C}{Rust ↔ C}}
\label{subsec:rusttoc}
The top part of~\tbl{tab:interoptable} summarizes our observations.


\subsubsection{Calling C function from~\rust{} (\rust{} $\rightarrow$ C)}
\label{subsubsec:callingcfromrust}
To invoke a C function from from~\rust{}, first, we need to provide the~\rust{}~\ac{FFI} signature of the function.
This can be done using tools such as~\code{bindgen}~\cite{rustbindgen} to automatically generate~\ac{FFI} signatures from C header files.
Then, they can link the library (\ie object file) containing the C function with the~\rust{} object file to get the final executable.
We illustrate these steps with an example in~\apdx{apdx:callingcfromrust}.
One of the main tasks here is to generate~\ac{FFI} bindings for the C functions.
It is relatively straightforward to create these bindings as the~\rust{}'s type system~\cite{matsakis2014rust} is a superset of C's,~\ie{} every builtin C type has a corresponding type in~\rust{}.
Finally, the target object file created from~\rust{} code should be linked to the source C project.
However, there are no automated tools to achieve this.
In summary, it is relatively straightforward to write~\rust{} code that can invoke C functions, but automation opportunities exist.

\subsubsection{Calling~\rust{} function from C (C $\rightarrow$~\rust{})}
\label{subsubsec:callingrustfromc}
Similar to~\rust{} $\rightarrow$ C (\sect{subsubsec:callingcfromrust}), here we need to generate C declaration for the target~\rust{} function, which can be automated using~\code{cbindgen}~\cite{generatingcbindgen} tool (\apdx{apdx:callingrustfromc} provides details of this process).
The superior~\rust{} type system has several types that are not supported in C. For instance,~\rustcode{Vec}~\cite{rustvectype}, one of the most commonly used~\rust{} types, is not supported in C.
Consequently,~\code{cbindgen} fails for such functions.
Developers need to write type wrappers to handle this manually.
But advanced features of~\rust{} types, such as~\rustcode{trait}~\cite{vanhattum2022verifying}, makes engineering these wrapper functions challenging~\cite{rustffiomnibus}, more details in~\apdx{apdx:incompatibleffitypes}.
We also performed a type compatibility analysis to assess the extent to which external functions in~\rust{} crates use advanced~\rust{} types,~\ie{} library functions for which developers need to engineer corresponding type wrapper functions manually.
\apdx{apdx:nonfficompat} provides details of the same.
This is also the difficulty faced by developers (\textbf{RQ3.4}) as we discuss in~\sect{subsec:devexperienceusingrust}.

\begin{tcolorboxfloat}[htbp]
\noindent\fcolorbox{black}{yellow!30!white}{%
  \parbox{\dimexpr\linewidth-2\fboxsep-2\fboxrule\relax}{\parskip=1pt\parindent=0pt
    \textbf{Finding RQ2.1:}
Although it is relatively straightforward to invoke C functions from~\rust{} code, automation opportunities exist to ease the process.

\textbf{Finding RQ2.2:}
The use of~\ac{FFI} incompatible types makes it hard to invoke~\rust{} functions from C code. The majority ($\sim$70\%) of~\rust{} embedded crates have functions with incompatible types.
  }
}
\end{tcolorboxfloat}

\begin{tcolorboxfloat}[htbp]
\noindent\fcolorbox{black}{green!30!white}{
  \parbox{\dimexpr\linewidth-2\fboxsep-2\fboxrule\relax}{\parskip=1pt\parindent=0pt
\textbf{Open Problem P2.1:} Embedding rust function calls in C application is challenging due to the need for type conversion between C types and FFI-incompatible rust types.
One possible approach is to manually create (once for all) type wrappers for basic complex types (\eg{}~\rustcode{Vec}) and use them to automatically create wrappers for composite types (\eg{}~\rustcode{struct}).
  }
}
\end{tcolorboxfloat}

\subsection{\rust{} Interoperable Challenges in Embedded Systems Development}
\label{subsubsec:rustinteroprocrwc}
We used~\rust{} in various real-world scenarios to investigate this aspect.
Specifically, we explore:~\rust{} application on top of C~\ac{RTOS} (\rustonc{}), C application on top of~\rust{}~\ac{RTOS} (\conrust{}) and converting a component in C~\ac{RTOS} to~\rust{} (\rustwc{}).
The bottom part of~\tbl{tab:interoptable} summarizes our observations.

\subsubsection{Setup}
We chose the blinker application~\cite{nrfpinchangeexample} for our application scenarios (\rustonc{} and~\conrust{}) as it encompasses all the necessary aspects of a typical embedded system,~\ie interacts with~\ac{RTOS}, has event-driven custom interrupt handler, and uses call-backs.
The application periodically (through an interrupt handler) blinks an LED by interacting through GPIO addresses.
We used the~\texttt{nrf52840-dk} \ac{MCU} board~\cite{nRF52840-DK} with~\texttt{ARM Cortex-M4} for our target board, as it is a widely recognized and adopted development platform in the embedded systems community and is well-supported by~\rust{}.
We used~\freertos{}~\cite{freertos} as our C~\acp{RTOS}, because of its widespread popularity in the embedded systems community~\cite{Wire} and extensive documentation~\cite{FreeRTOSBookandReferenceManual}.
As mentioned in~\sect{subsec:embeddedswdataset},~\rust{}~\ac{RTOS} can be either fully developed in~\rust{} (\ie{} native) or wrappers around a C~\ac{RTOS}.
We selected~\lilos{}~\cite{Biffle_2023} and~\freertosrs{}~\cite{FreeRTOS-rust_2023} as our native and wrapper~\acp{RTOS}, respectively.
\lilos{} is a stable and purely~\rust{} based and completly asynchronous~\ac{RTOS}. This is a representative~\rust{} based~\ac{RTOS} using the strongly suggested~\rustcode{async} design pattern~\cite{Media}.

\subsubsection{\rust{} application on top of C~\ac{RTOS} (\rustonc{})}
\label{subsubsec:rustappontopofc}
Our goal is to create a~\rust{} blinky application on top of C~\freertos{}.
We followed similar steps as described in~\sect{subsubsec:callingcfromrust}.
First, we generated embedded system compatible (\ie{}~\rustcode{no_std})~\rust{}~\ac{FFI} bindings from~\freertos{} header files using~\code{bindgen}.
Second, we developed blinky application using these~\ac{FFI} bindings.
\ifSUPPLEMENT
\lst{lst:rustapponcrots} (in Appendix) shows the snippet of creating a task using~\freertos{} through its~\ac{FFI} bindings.
\fi
Specifically, we converted~\rust{} types into appropriate~\ac{FFI} types and invoked the target function.
We followed a similar procedure for all other steps,~\ie{} registering interrupts, etc.
Finally, we created a static library of C~\freertos{} and linked it with our~\rust{} application to get the final executable.
We tested the final executable and ensured that it worked as expected.
\emph{The entire process was straightforward.}
The only issue was creating a linker script suitable for the target board.
As mentioned before in~\sect{subsubsec:callingrustfromc}, the availability of automated tools will make this process easier.





\begin{lstlisting}[
  language=Rust, 
  caption={FFI incompatible function and FFI-friendly wrapper function to create tasks in Lilos scheduler}, 
  label={lst:Lilos_wrapper_scheduler},
  xleftmargin=0.5cm,               % Set left margin
  escapeinside={||},               % Escape inside || like in minted
  basicstyle=\scriptsize\ttfamily,  % Font size and style
  numbers=left,                    % Line numbers on the left
  breaklines=true,                 % Break long lines
]
// FFI incompatible function
pub fn run_tasks(
futures: &mut [Pin<&mut dyn Future<Output = Infallible>>],
initial_mask: usize,
) -> !

#[no_mangle]
pub extern "C" fn lilos_run_two_tasks(fn1: *mut fn(), fn2: *mut fn(), initial_mask: usize) -> ! {
  unsafe {
     let fut1 = *fn1;
     let future1 = pin!(async move {
         loop { fut1() } });
     let fut2 = *fn2;
     let future2 = pin!(async move {
         loop { fut2() } });
     run_tasks(&mut [future1, future2], initial_mask); 
  } 
}
\end{lstlisting}

\subsubsection{C application on top of~\rust{}~\ac{RTOS} (\conrust{})}
\label{subsubsec:capponrustrtos}
This interoperable modality is crucial for developers who seek to build secure systems by leveraging existing components.
Furthermore, as shown in~\fig{fig:rq4interop}, 36\% of developers claim to have developed C code calling \rust{} functions.
Here, our goal is to create a C blinky application on top of~\rust{}~\acp{RTOS}, specifically on~\freertosrs{} (\rust{} wrapper of C~\freertos{}) and~\lilos{} (a pure~\rust{}~\ac{RTOS}).
We followed similar steps as described in~\sect{subsubsec:callingrustfromc}.
\begin{itemize}[noitemsep,nolistsep,leftmargin=*]
\item\emph{On~\freertosrs{}:} Being a wrapper, all external functions used C compatible types, and~\code{cbindgen} was able to create C declarations for all the required functions.
This made it easy to create the main task of the C blinky application.
However, accessing GPIO pins required us to use~\code{nrf52840_pac}~\cite{nrf52840-pac-rust}~\rust{} create, which uses a C incompatible type,~\ie{}~\rustcode{RegisterBlock}.
Consequently,~\code{cbingen} failed to create corresponding C declarations.
We manually created an~\ac{FFI} compatible~\rust{} function (\rustcode{togglePin}) to access GPIO pins and used it in our application.
\ifSUPPLEMENT
Refer to ~\lst{lst:little_rust_w_c} (of Appendix) for details.
\fi


\item\emph{On~\lilos{}:} This presented an extreme case wherein none of the external functions are~\ac{FFI} compatible, and consequently,~\code{cbindgen} failed to create C declarations.
We had to manually create~\ac{FFI} compatible wrapper functions (\eg{}~\rustcode{lilos_run_two_tasks} for~\rustcode{run_tasks} in~\lst{lst:Lilos_wrapper_scheduler}).
For accessing GPIO pins, we followed the same approach as described before in~\emph{On~\freertosrs{}}.
\end{itemize}

The main challenge in both cases was dealing with incompatible~\rust{} types.
We found (from our analysis in~\sect{subsubsec:callingrustfromc}) that on-average of~\rtosavgffiincompat{} interface functions in~\rust{}~\acp{RTOS} use incompatible~\rust{} types.

\begin{tcolorboxfloat}
\noindent\fcolorbox{black}{yellow!30!white}{%
  \parbox{\dimexpr\linewidth-2\fboxsep-2\fboxrule\relax}{\parskip=1pt\parindent=0pt
\textbf{Finding RQ2.3:}
Significant development effort is required to engineer a C-embedded application on top of~\rust{}~\acp{RTOS} because of the prevalent use of incompatible~\rust{} types.  
  }
}
\end{tcolorboxfloat}

\subsubsection{\rust{} component in C~\ac{RTOS} (\rustwc{})}
\label{subsubsec:rustcomincrtos}
Here, we aim to convert a component in C~\ac{RTOS} into~\rust{} to mimic an incremental porting scenario.
We selected~\code{list} component in C~\freertos{}, as it is self-contained (\ie{} no calls to other components).
We followed a similar procedure as described in~\sect{subsubsec:callingcfromrust}.
First, we used~\code{bindgen} on~\code{list.h} to create the required~\rust{} types.
\ifSUPPLEMENT
The~\rustcode{xLIST} in~\lst{lst:RwC_list} (in Appendix) shows the type generated by~\code{bindgen}.
\fi
Second, we reimplemented the list functions (in~\code{list.rs}) using the types generated by~\code{bindgen}.
Unfortunately, as mentioned in~\sect{subsec:ctorusttools}, the recommended way to convert C to~\rust{} code does not work on embedded codebases.
We manually translated the corresponding C implementation line-by-line into~\rust{}, which required considerable effort.
Our Extended Report shows a snippet of~\rustcode{vListInitialise} function in~\rust{}.
%
%
Finally, we modified the~\code{Makefile} to build~\code{list.rs} into a static library and linked it with the final~\freertos{} object file.

\begin{tcolorboxfloat}
\noindent\fcolorbox{black}{yellow!30!white}{%
  \parbox{\dimexpr\linewidth-2\fboxsep-2\fboxrule\relax}{\parskip=1pt\parindent=0pt
\textbf{Finding RQ2.4:}The lack of embedded codebase support in C-to-\rust{} conversion tools (described in~\sect{subsec:ctorusttools}) poses a considerable challenge in adopting the (recommended) incremental porting approach~\cite{xiaempirical, 10.1145/3485498} to convert embedded codebases to~\rust{}. 
  }
}
\end{tcolorboxfloat}

\section{RQ3: Developers Perspective}
\label{sec:rq4developerspers}
\begin{table*}[tb]
\footnotesize
\caption{Summary of Survey Questions. Exact questions in~\apdx{apdx:surveyquestions}}
\label{tab:quessum}
\centering
\begin{tabular}{c|>{\centering\arraybackslash}p{10cm}|c}
\toprule
\textbf{Category} & \textbf{Description} & \textbf{No. of questions} \\
\midrule
Familiarity and Experience & Examines participants’ familiarity, experience, and preferred languages for embedded systems development. & 9 \\
\hline
Acquaintance with Rust & Explores familiarity with Rust and its specific features of participants. & 15 \\
\hline
Reasons to Use Rust & Gathers opinions of participants on reasons to use Rust, its advantages, and perceived challenges. & 13 \\
\hline
Hardware Support, Integration, and Performance & Enquires about issues related to hardware support, integration, and performance when using Rust for embedded development. & 11 \\
\hline
Memory Safety and Debugging & Focuses on the importance of memory safety, ease of debugging with Rust, and related practices in embedded systems development. & 9 \\
\hline
Documentation and Community Support & Evaluates the quality of Rust documentation and the level of community support available for embedded systems. & 6 \\
\hline
Development Time and Code Quality & Investigates views on potential gains in development time and code quality when using Rust. & 3 \\
\bottomrule
\end{tabular}
\end{table*}

We aim to shed light on developers' perspectives on using~\rust for embedded systems development.
Specifically, (i) Reasons for not using~\rust.; (ii) Challenges faced by developers in using~\rust.; and (iii) Developer's perspective on~\rust{}'s performance, safety and interoperability.

\subsection{Study Methodology}
We used an anonymous online survey with questions spanning various categories as shown in \tbl{tab:quessum}.
We recruited participants by sending the link to our survey to various embedded systems communities and~\rust{} embedded developers' mailing lists (Details in~\apdx{apdx:solicitationmailinglists}). Also, we used our industry collaborations to circulate our survey to multiple organizations.
Our Institutional Review Board (IRB) reviewed and approved our study protocol.

\myparagraph{Survey Respondents}
We got 268 responses, out of which we filtered out 43 responses from inattentive participants (through attention-checking questions), resulting in 225 valid responses.
There is considerable diversity in the embedded systems experience of participants, indicating a representative developers group.
Our Extended Report shows the distribution of embedded systems experience of participants.

\subsection{Not Using~\rust{} for Embedded Systems: Expectations v/s Reality}
\begin{figure}[ht]
\captionsetup[figure]{belowskip=-5pt}
     \includegraphics[width=\linewidth]{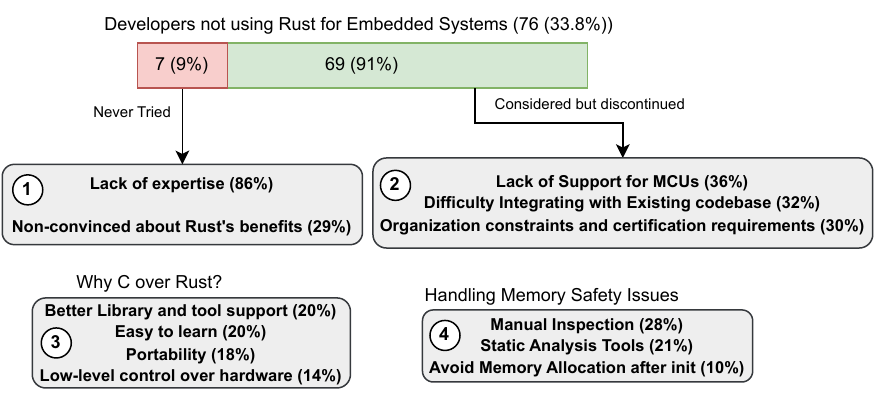}
    \caption{Response summary of Developers not using~\rust.}
    \label{fig:rq4norust}
    \Description[Response summary of Developers not using~\rust.]{Response summary of Developers not using~\rust.}
\end{figure}

The~\fig{fig:rq4norust} shows the summary of ~\totalnorust{} (\totalnorustperc{}) participants who currently do not use~\rust{}.
Only,~\didnotconsiderrust{} (\didnotconsiderrustperc{}) participants never tried to use~\rust{}, mainly because of the lack expertise (\encircle{1}).
Furthermore, 29\% of developers are not convinced about~\rust security benefits as~\emph{embedded systems rarely use dynamic memory allocation and do not need~\rust{}'s ownership features --- an important safety feature of~\rust{}}.

However, the other~\considerrust{} (\considerrustperc{}) participants considered~\rust{}, but discontinued because of three main reasons (\encircle{2}): (i) Lack of support for~\acp{MCU}, this is inline with our analysis in~\sect{subsec:adoptabilityofrust}. (ii) Integrating with existing codebases. (iii) Organizational and certification constraints. Source code used as part of critical infrastructure, such as airplanes, undergo rigorous certification~\cite{dodd2012safety, kornecki2008airborne, kornecki2009certification}.
This is expensive and time-consuming. Switching to~\rust{} requires re-certification, which may not be desirable for organizations.

All developers in~\fig{fig:rq4norust} use C, and the~\encircle{3} box shows the reasons for choosing C.
The first two reasons are expected, as C is an old language with many libraries and toolchain support.
The third reason,~\ie Portability, is interesting. In C, there are no language-specific considerations for embedded systems. Consequently, it is relatively easy to port (or repurpose) an existing library for the embedded use case by linking it with embedded versions of standard libraries.
However, in~\rust{},~\emph{embedded libraries (\ie crates) should be developed with~\code{no_std} environment --- which restricts the uses of certain language-level features.
Consequently, porting existing libraries to be~\code{no_std} compatible and to use in embedded systems is challenging~\cite{nostdimpl, nostdcompat}}.

Interestingly, as shown in~\encircle{4} of~\fig{fig:rq4norust}, many (28\%) embedded systems developers (using C) do not use any automated security tools and rely on manual inspection.
Only 21\% of the developers use static analysis tools.
This confirms observations made by a recent study~\cite{shen2023empirical}.
Finally, none of the developers use any dynamic analysis tools.

\begin{tcolorboxfloat}
\noindent\fcolorbox{black}{yellow!30!white}{%
  \parbox{\dimexpr\linewidth-2\fboxsep-2\fboxrule\relax}{\parskip=1pt\parindent=0pt
\textbf{Finding RQ3.1}: To improve adoption of~\rust{} for embedded systems:
\begin{itemize}[noitemsep,nolistsep,leftmargin=*]
\item Support needs to be added for more MCUs.
\item Techniques and methods should be developed to ease the certification of~\rust{} code ported from already certified C code.
\item Automated techniques should be developed to convert~\rust{} crates to~\code{no_std} compatible.
\end{itemize}
  }
}
\end{tcolorboxfloat}

\subsection{Experiences in Using~\rust for Embedded Systems}
\label{subsec:devexperienceusingrust}
There were~\totalyesrust{} (\totalyesrustperc{}) participants who currently use~\rust{} for embedded systems development.
These participants have varied development experience with~\rust{}, specifically, 19\% with $<$ 6 months,  28\%  with 6 months - 1 year and 18\% with 1-2 years, and 35\% with more than 2 years.

\begin{figure}[ht]
\captionsetup{skip=3pt, belowskip=-8pt}
\centering
\begin{minipage}[b]{1\linewidth}
    \centering
    \includegraphics[width=\linewidth]{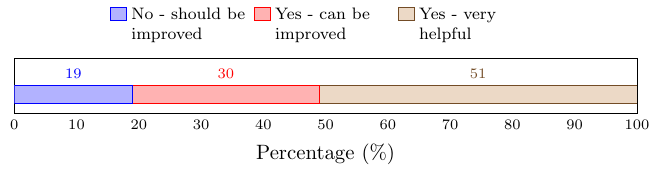} 
    \caption{Is Rust Documentation and Community Helpful?}
    \label{fig:rustsuppportembedded}
\end{minipage}
\end{figure}

\noindent\textbf{Adopting~\rust and Motivation:}
The two main motivations to learn~\rust{} for embedded systems are safety and reliability (94\%) and familiarity with the language (57\%).
Although there exists good support for~\rust{} in the embedded systems community (\fig{fig:rustsuppportembedded}), the majority percentage (85\%) of developers claim that it still requires considerable effort (\ie Moderate (40\%) + Hard (45\%)) to adopt~\rust{}.

\begin{figure}[ht]
    \hspace*{-1cm}
    \includegraphics[scale=0.7]{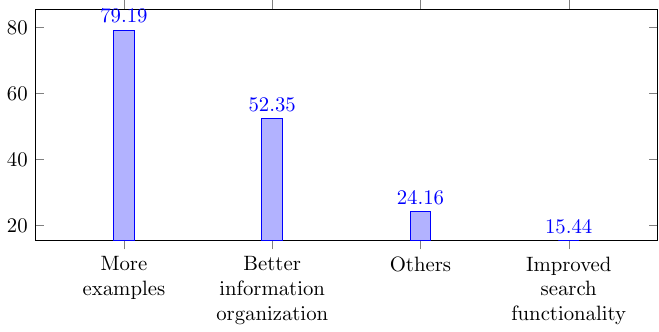}
   \caption{Required Improvements to Rust Documentation.}
    \label{fig:rustdocspecificimprovements}
    \Description[Required Improvements to Rust Documentation.]{Required Improvements to Rust Documentation.}
\end{figure}
\begin{figure}[ht]
\captionsetup{skip=7pt, belowskip=-10pt}
\centering
\begin{minipage}[b]{1\linewidth}
    \centering
    \includegraphics[width=\linewidth]{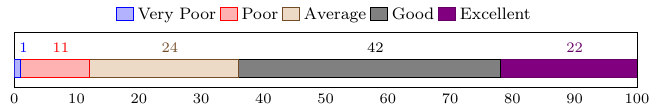} 
    \caption{Support for Rust in Embedded Systems Community}
    \label{fig:rustdocimprovements}
\end{minipage}
\end{figure}

\noindent\textbf{\rust{} Documentation and Community Support (\fig{fig:rustdocimprovements} and~\ref{fig:rustdocspecificimprovements}):}
The majority,~\ie{} 81\% (51 + 30) of developers, agree that the available documentation and community are helpful.
However, 49\% (30 + 19) of developers mention that documentation should be improved.
The~\fig{fig:rustdocspecificimprovements} shows specific suggestions to improve the documentation.
Specifically,~\emph{\rust{} documentation should contain more examples and be organized better}.

\noindent\textbf{Developer Tools and Crates:}
68\% of the developers think the currently available crates provide sufficient support (\ie very satisfied -- somewhat satisfied), whereas the rest,~\emph{32\%, observe that it is not adequate (\ie dissatisfied -- very dissatisfied)}.
This is also in line with our analysis (\sect{subsec:adoptabilityofrust}), where we noticed that necessary support (\ie HAL and other necessary crates) is unavailable for certain~\acp{MCU}.

\begin{figure}[ht]
\captionsetup[figure]{belowskip=-10pt}
    \includegraphics[width=\linewidth]{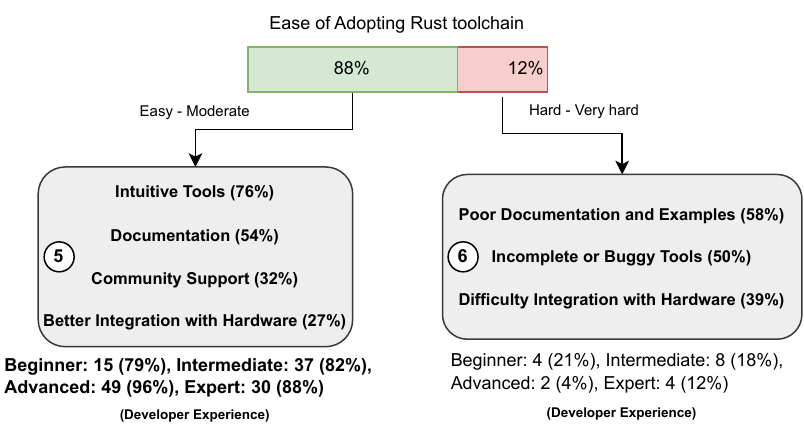}
    \caption{Response summary of Developers on ease of adopting~\rust toolchain.}
    \label{fig:rq4easeofdevtools}
    \Description[Response summary of Developers on ease of adopting~\rust toolchain.]{Response summary of Developers on ease of adopting~\rust toolchain.}
\end{figure}

~\fig{fig:rq4easeofdevtools} shows the opinion of users w.r.t the~\rust{} toolchain support.
Most developers (across all experience levels) mention that it is easy to adopt~\rust{} toolchain for embedded systems development --- reasons are intuitive tools and their documentation (\encircle{5}).
This is in line with our analysis in~\sect{subsec:rq3interop}.
Nonetheless, 12\% (18) developers expressed concerns,~\ie poor documentation (missing examples) and buggy tools (\encircle{6}).
These could be because of using tools from non-stable branches.
This also further confirms observations in~\fig{fig:rustdocspecificimprovements}, where developers require more examples to be included in the documentation.

\noindent\textbf{Performance of~\rust{}:}
It is interesting to see that only 54\% of developers mentioned that they performed a systematic comparative evaluation of their~\rust{} implementation with existing C implementation.
Wherein 28.5\% noticed similar performance, 22\% noticed that~\rust{} was faster, and the remaining 3.5\% noticed that~\rust{} implementation was slower.


The slowdown observations contradict the common belief that given the asynchronous nature of embedded systems, the performance of~\rust{}'s implementation can be significantly improved by carefully using its built-in features, such as closures~\cite{heyman2020comparison}.
These observations also highlight~\emph{the need for a systematic performance evaluation of using~\rust{} for embedded systems}.

\begin{tcolorboxfloat}
\noindent\fcolorbox{black}{yellow!30!white}{%
  \parbox{\dimexpr\linewidth-2\fboxsep-2\fboxrule\relax}{\parskip=1pt\parindent=0pt
  \textbf{Finding RQ3.2}:~\rust{} documentation should be improved with more embedded system-specific examples.
  }
}
\end{tcolorboxfloat}

\begin{tcolorboxfloat}
\noindent\fcolorbox{black}{green!30!white}{
  \parbox{\dimexpr\linewidth-2\fboxsep-2\fboxrule\relax}{\parskip=1pt\parindent=0pt
\textbf{Open Problem P3.1:}
Developers have contradictory views on~\rust{}'s performance on embedded systems.
Existing performance studies~\cite{rustvsc,spedofrust,junperformance} could be extended to include \rust{} embedded systems.
  }
}
\end{tcolorboxfloat}

\begin{figure}[ht]
\captionsetup[figure]{belowskip=-17pt}
    \includegraphics[width=\linewidth]{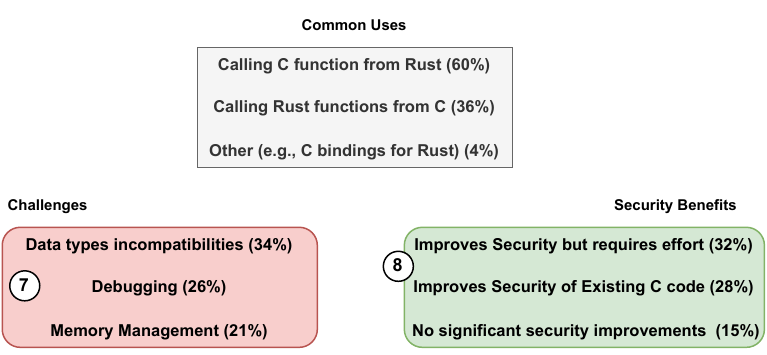}
    \caption{Response summary of Developers perspectives on~\rust{}'s Interoperability.}
    \label{fig:rq4interop}
    \Description[Response summary of Developers perspectives]{Response summary of Developers perspectives on~\rust{}'s Interoperability.}
\end{figure}

\noindent\textbf{Interoperability with Existing Codebase:}
All developers agree that interoperability is needed, and most developers (98\%) were aware of~\rust{}'s interoperability support.
However, 56\% of developers mentioned that they face challenges in using interoperability support of~\rust{}.
%
The~\encircle{7} box in~\fig{fig:rq4interop} shows developers' common challenges in using interoperability support.

\begin{tcolorboxfloat}
\noindent\fcolorbox{black}{yellow!30!white}{%
  \parbox{\dimexpr\linewidth-2\fboxsep-2\fboxrule\relax}{\parskip=1pt\parindent=0pt
   \textbf{Finding RQ3.3}: The majority (\ie 34\%) of developers face issues handling type incompatibilities between~\rust{} and C code. This is in line with our analysis (\sect{subsubsec:callingrustfromc}), where we show that handling data types is one of the challenges in using C on top of~\rust{} (\conrust).
  }
}
\end{tcolorboxfloat}

\emph{The second major (26\%) issue is debugging}, which is expected because, as explained in~\sect{sec:background}, embedded systems follow an asynchronous and event-driven design. This results in frequent cross-language domain interactions and makes debugging hard.

As shown by~\encircle{8} in~\fig{fig:rq4interop}, 60\% (32 + 28) of developers agree that using interoperable~\rust{} improves security. However, 32\% mention that secure usage (\ie through~\code{unsafe} blocks) requires significant effort --- which is in line with existing works~\cite{10.1145/3428204} that show that engineering interoperable code in~\code{unsafe} blocks is challenging and prone to security issues.

\begin{figure}[ht]
\captionsetup{skip=7pt, belowskip=0pt}
\centering
\begin{minipage}[b]{1\linewidth}
    \centering
    \includegraphics[width=\linewidth]{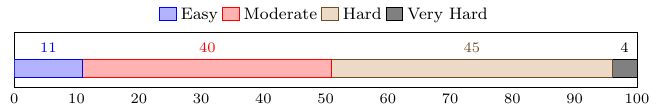} 
    \caption{Ease of Adopting Rust for Embedded Systems Development}
    \label{fig:rustdocumentationhelpful}
\end{minipage}
\end{figure}


\textbf{\rust{} v/c C:}
92\% of developers mentioned that they also used C for embedded systems development.
Out of which,~\emph{64\% of developers claim that development time significantly decreased and also the code quality improved after switching to~\rust{}}.
This is in line with recent findings at Google~\cite{claburnrustnodate}.
%
For embedded systems development, 30\% of developers recommend~\rust unconditionally, whereas~\emph{61\% recommend~\rust only if the developer is well-versed in it}, and the 9\% recommend~\rust only if safety is of high importance.


\section{Limitations and Threats to Validity}
We acknowledge the following limitations and threats to validity:
\begin{itemize}[noitemsep,nolistsep,leftmargin=*]
\item Our findings are based on the analysis of the collected dataset and~\ac{SAST} tools.
The dataset and the tools may not be representative enough. We tried to avoid this by collecting crates and tools from diverse sources.
\item We did not analyze all the alerts raised by various~\ac{SAST} tools (\sect{subsec:sasttools}).
Consequently, using these alerts to assess the quality of crates could be exaggerated because of potential false positives.
\item Our RQ2 (\sect{subsubsec:rustinteroprocrwc}) observations are based on limited scenarios and may not be generalizable.
However, the developer survey (in~\sect{subsec:devexperienceusingrust}) confirmed our findings, reducing the risk.
\end{itemize}

\section{Related Work}


%






\noindent \textbf{Rust Studies:}
Several works~\cite{Fulton2021BenefitsAD, 9794066, 10.1145/3428204, 10.1145/3385412.3386036, DBLP:journals/corr/abs-1806-04929} evaluate various aspects of~\rust{} from the usability perspective.
Fulton~\etal{}~\cite{Fulton2021BenefitsAD} surveyed and interviewed~\rust{} developers to understand challenges and barriers to adoption.
Similarly, Zeng ~\etal{}~\cite{9794066} performed a mixed-methods study of~\rust{} related forums to identify common challenges and corresponding solutions.
However, other works assess specific aspects of~\rust{}.
Astrauskas~\etal{}~\cite{10.1145/3428204} focused on identifying common uses of~\rustcode{unsafe} blocks.
Whereas, Qin~\etal{}~\cite{10.1145/3385412.3386036} focused on identifying challenges in using concurrency constructs and identified common causes of concurrency issues in~\rust{} code.
Similarly, Mindermann~\etal{}~\cite{DBLP:journals/corr/abs-1806-04929} exclusively studied the usability of~\rust{}'s cryptography APIs, providing crucial recommendations for developing these APIs to enhance usability and reduce misuse.

Pinho ~\etal{}~\cite{8990314} and Ashmore ~\etal{}~\cite{9985123} evaluated the feasibility of using~\rust{} for safety critical systems (a class of embedded systems).
Specifically, using evaluation criteria for programming languages, aligning with the standards set by RTCA DO-178C, they demonstrated that~\rust{} meets all the criteria.
Levy~\etal{}~\cite{10.1145/3124680.3124717} focused on using~\rust{} for kernel development and shared their firsthand experience in creating a kernel for low-power~\acp{MCU}.
They also demonstrated~\cite{10.1145/2818302.2818306, Levy2017MultiprogrammingA6} the feasibility of using~\rust{} to engineer common kernel building blocks with only a few~\rustcode{unsafe} blocks.
This paper assesses the applicability and challenges of using~\rust{} for embedded system software, such as~\acp{RTOS}, by performing a systematic analysis and developer study.

\noindent \textbf{Embedded Systems Vulnerabilities:}
Several works~\cite{7097358, 9878283, article} try to understand vulnerabilities in embedded systems and analyze challenges and possible solutions for effective vulnerability detection.
Several embedded systems vulnerability detection techniques use various approaches ranging from static analysis~\cite{Redini2020KaronteDI}, symbolic execution~\cite{10.5555/3277203.3277227}, and rehosting-based dynamic analysis~\cite{Scharnowski2022FuzzwareUP} or fuzzing~\cite{Chen2018IoTFuzzerDM}.
In this work, we do not propose any new techniques but rather use state-of-the-art tools (\sect{subsec:sasttools}) to assess various aspects of~\rust{} embedded software.

\section{Conclusion}
We performed a systematic analysis and a comprehensive (with~\numdev{} developers) survey to understand the current state and challenges in using~\rust{} for embedded systems development.
Our findings provide insights into the current state and expose open problems and potential improvements that can facilitate easy adoption of~\rust{} for embedded system development.

\section{Acknowledgements}
This research was supported in part by the National Science Foundation (NSF) under Grant CNS-2340548, Rolls-Royce Grant on ``Dynamic Analysis of Embedded Systems'', and Defense Advanced Research Projects Agency (DARPA) under
contract numbers N660012\newline0C4031 and N660012224037. The U.S.Government is authorized to reproduce and distribute reprints for Governmental purposes notwithstanding any copyright notation thereon. Any opinions, findings, conclusions, or recommendations expressed in this material are those of the author(s) and do not necessarily reflect the views of the NSF or the United States Government.


{
\begin{spacing}{0.9}
\footnotesize
\bibliographystyle{plain}
\bibliography{bibs/RustPaper}
\end{spacing}
}

\ifSUPPLEMENT
\appendix
\section{Indirect flags}
\label{apdx:rustindirectflags}
The~\lst{lst:incorrectcfgattribute} shows an example of indirect dependency.
Here, we need to use ~\shell{--no-default-features --features=libm} flag to build the crate for~\rustcode{no_std}, as the crate requires \texttt{libm} feature to compile for ~\rustcode{no_std}.
However, this dependency is hard to infer from the provided~\rustcode{cfg_attr}s.

\section{Tools Reporting}
\label{apdx:toolsreporting}
\begin{lstlisting}[
  language=Rust, 
  escapeinside=@@,captionpos=b,
  caption={Example of indirect dependency. The features for \lstinline{no_std} cannot be inferred from \lstinline{cfg_attr} alone.}, 
  label={lst:incorrectcfgattribute},
  xleftmargin=0.5cm,               % Set left margin
  basicstyle=\scriptsize\ttfamily,  % Font size and style
  numbers=left,                    % Line numbers on the left
  breaklines=true                  % Break long lines
]
#![cfg_attr(not(feature = "std"), no_std)]
...
#[cfg(all(feature = "libm", not(feature = "std")))]
mod nostd_float;
\end{lstlisting}

\begin{lstlisting}[
  language=Bash, 
  captionpos=b,
  caption={Example of warning produced by \texttt{\lockbud}.}, 
  label={lst:lockbudwarning},
  xleftmargin=0.5cm,               % Set left margin
  basicstyle=\scriptsize\ttfamily,  % Font size and style
  numbers=left,                    % Line numbers on the left
  breaklines=true                  % Break long lines
]
{
  "DoubleLock": {
    "bug_kind": "DoubleLock",
    "possibility": "Possibly",
    "diagnosis": {
      "first_lock_type": 
          "ParkingLotMutex(gatt::server::ClientCtx)",
      "first_lock_span": 
          "src/gatt/server.rs:852:13: 852:19 (#0)",
      "second_lock_type": 
          "ParkingLotMutex(gatt::server::ClientCtx)",
      "second_lock_span": 
          "src/gatt/server.rs:510:13: 510:19 (#0)",
      "callchains": [
        [
          [
            "src/gatt/server.rs:866:13: 866:30 (#0)"
          ],
          [
            "src/gatt/server.rs:496:66: 496:83 (#0)"
          ]
        ]
      ]
    },
    "explanation": "The first lock is not released 
                    when acquiring the second lock"
  }
}
\end{lstlisting}

\begin{lstlisting}[
  language=Rust, 
  escapeinside=@@,captionpos=b,
  caption={A false positive example (\textcolor{orange}{\faWarning}) detected by lockbud where it does not take into account the nature of locks in crate \textit{vls-persist-0.10.0} from dependency \textit{lightning-0.0.115}}, 
  label={lst:lockbud_locks_fp},
  xleftmargin=0.5cm,               % Set left margin
  basicstyle=\scriptsize\ttfamily,  % Font size and style
  numbers=left,                    % Line numbers on the left
  breaklines=true                  % Break long lines
]
// This code uses FairRwLock which has a RWLock member and AtomicUsize
// writers. This guarantees that at a given point, multiple readers can
// hold the lock but only 1 writer can write.
pub struct FairRwLock<T> {
    lock: RwLock<T>,
    waiting_writers: AtomicUsize,
}

// The struct ChannelManager is a fairly large structure with a lot of
// members, for the sake of brevity, we show redacted functionality
pub struct ChannelManager {
    per_peer_state: FairRwLock<HashMap<PublicKey,
    Mutex<PeerState<<SP::Target as SignerProvider>::Signer>>>>,
    // Other member declarations follow
}

// lightning-0.0.115/src/ln/channelmanager.rs:2032
fn close_channel_internal(&self, channel_id: &[u8; 32], 
counterparty_node_id:
&PublicKey, target_feerate_sats_per_1000_weight: Option<u32>)
-> Result<(), APIError> {
    // Some initializations follow
    let mut failed_htlcs: Vec<(HTLCSource, PaymentHash)>;
    let result: Result<(), _> = loop {
        // First read lock held here
        let per_peer_state =
                self.per_peer_state.read().unwrap();
        // Rest of the function follows
        ..
    }
    // assume handle_error is called
    let _ = handle_error!(self, result, *counterparty_node_id);
    Ok(())
}

// lightning-0.0.115/src/ln/channelmanager.rs:1466
macro_rules! handle_error {
($self: ident, $internal: expr, $counterparty_node_id: expr) => { {
    // Some assert checks
    match $internal {
        Ok(msg) => Ok(msg),
        Err(MsgHandleErrInternal 
                { err, chan_id, shutdown_finish })
            => {
            if !msg_events.is_empty() {
                // Read lock is acquired again, which is ok
                // since we're using RWLock but Lockbud flags it as a 
                // double lock
                // FALSE POSITIVE @\textcolor{orange}{\faWarning}@
                let per_peer_state = 
                        $self.per_peer_state.read().unwrap();
            }

            // Return error in case higher-API need one
            Err(err)
        },
    }
}
\end{lstlisting}

\begin{lstlisting}[
  language=Rust, 
  escapeinside=@@,captionpos=b,
  caption={A false positive example (\textcolor{orange}{\faWarning}) detected by lockbud where it does not take into account the application semantics and the order in which the application tries to acquire the locks, due to a notification from a conditional variable in crate \textit{smoldot-light-0.15.0} from dependency \textit{blocking-1.5.1}.}, 
  label={lst:lockbud_application_semantics_fp},
  xleftmargin=0.5cm,               % Set left margin
  basicstyle=\scriptsize\ttfamily,  % Font size and style
  numbers=left,                    % Line numbers on the left
  breaklines=true                  % Break long lines
]

fn main_loop(&'static self) {
    // First lock acquired here before the loop
    // Any thread wishing to enter the loop must acquire this lock first
    let mut inner = self.inner.lock().unwrap();
    loop {
        // Run tasks in the queue.
        while let Some(runnable) = inner.queue.pop_front() {
            // We have found a task - grow the pool if needed.
        }

        // This thread is now becoming idle.
        inner.idle_count += 1;

        // Put the thread to sleep until another task is scheduled.
        let timeout = Duration::from_millis(500);
        
        // Double lock detected by lockbud is a false positive because 
        // after waking up from notification from cvar, the thread needs to
        // re-acquire the lock.
        let (lock, res) = self.cvar.wait_timeout(inner, timeout).unwrap();
        inner = lock; @\textcolor{orange}{\faWarning}@
    }
}

/// Spawns more blocking threads if the pool is overloaded with work.
fn grow_pool(&'static self, mut inner: MutexGuard<'static, Inner>) {
    while inner.queue.len() > inner.idle_count * 5
        && inner.thread_count < inner.thread_limit.get()
    {
        // Notify all existing idle threads because we need to hurry up.
        self.cvar.notify_all();
        // Rest of the function follows which spawns new threads
    }
}
\end{lstlisting}

\begin{lstlisting}[
  language=bash,
  captionpos=b,
  caption={Example of warning produced by \rudra{}.},
  label={lst:rudrawarning},
  xleftmargin=0.5cm, % Set left margin
  basicstyle=\scriptsize\ttfamily, % Font size and style
  numbers=left, % Line numbers on the left
  breaklines=true % Break long lines
]
Warning 
  (SendSyncVariance:/ApiSyncforSync/NaiveSyncForSync/RelaxSync): \
Suspicious impl of `Sync` found
-> src/lib.rs:63:1: 63:50
unsafe impl<T> Sync for Mutex<T> where T: Send {}
\end{lstlisting}

\begin{lstlisting}[
  language=Rust, 
  escapeinside=@@,captionpos=b,
  caption={A True Positive example (\textcolor{red}{\faBug}) detected by Rudra in crate \textit{imbl-sized-chunks-0.1.2}.}, 
  label={lst:rudra_tp},
  xleftmargin=0.5cm, % Set left margin
  basicstyle=\scriptsize\ttfamily, % Font size and style
  numbers=left, % Line numbers on the left
  breaklines=true % Break long lines
]
// src/inline_array/mod.rs:450
fn clone(&self) -> Self {
    let mut copy = Self::new();
    for i in 0..self.len() {
        unsafe {
            // If clone panics on an unchecked index, it can leave the
            // variables involved in an inconsistent state.
            copy.write_at(i, self.get_unchecked(i).clone()); @\textcolor{red}{\faBug}@
        }
    }
    unsafe {
        *copy.len_mut() = self.len();
    }
    copy
}
\end{lstlisting}

\begin{lstlisting}[
  language=Rust, 
  escapeinside=@@,captionpos=b,
  caption={A false positive example (\textcolor{orange}{\faWarning}) detected by Rudra shows that it is unaware of the presence of special entities such as \textit{panic\_guards}, which ensure that the surrounding code does not result in panic. This was found in crate \textit{array-init-2.1.0}.}, 
  label={lst:rudra_panic_guard},
  xleftmargin=0.5cm, % Set left margin
  basicstyle=\scriptsize\ttfamily, % Font size and style
  numbers=left, % Line numbers on the left
  breaklines=true % Break long lines
]

unsafe {
    let mut array: MaybeUninit<[T; N]> = MaybeUninit::uninit();
    // panic guard is initialized
    let mut panic_guard = UnsafeDropSliceGuard {
        base_ptr: ptr_i,
        initialized_count: 0,
    };

    for i in 0..N {
        // Invariant: `i` elements have already been initialized
        panic_guard.initialized_count = i;
        // If this panics or fails, `panic_guard` is dropped, thus
        // dropping the elements in `base_ptr[.. i]` for D > 0 or
        // `base_ptr[N - i..]` for D < 0.
        let value_i = initializer(i)?;
        // this cannot panic
        // the previously uninit value is overwritten without being read
        // or dropped
        if D < 0 {
            ptr_i = ptr_i.sub(1);
            panic_guard.base_ptr = ptr_i;
        }
        // Rudra detects unsafe flow to write and assumes that write can
        // panic. But the context ensures that the code does not
        // panic by the use of panic_guard.
        ptr_i.write(value_i); @\textcolor{orange}{\faWarning}@
    }
    // From now on, the code can no longer `panic!`, let's take the
    // symbolic ownership back
    mem::forget(panic_guard);
    Ok(array.assume_init())
}
\end{lstlisting}

\begin{lstlisting}[
  language=Rust,
  escapeinside=@@,captionpos=b,
  caption={A false positive example (\textcolor{orange}{\faWarning}) detected by Rudra shows that it is unable to handle Atomic types and determine if they are being used to secure the underlying unsafe entity. This was found in crate \textit{exclusive\_cell-0.1.0}.}, 
  label={lst:rudra_atomic_fp},
  xleftmargin=0.5cm, % Set left margin
  basicstyle=\scriptsize\ttfamily, % Font size and style
  numbers=left, % Line numbers on the left
  breaklines=true % Break long lines
]
// exclusive_cell-0.1.0/src/lib.rs:167:1
// Rudra flags impl of Send/Sync on ExclusiveCell as Suspicious
// CallOnce is an AtomicBool variable and there are no guarantees on data
pub struct ExclusiveCell<T: ?Sized> {
    taken: CallOnce,
    data: UnsafeCell<T>,
}

unsafe impl<T: ?Sized + Send> Send for ExclusiveCell<T> {}
unsafe impl<T: ?Sized + Send> Sync for ExclusiveCell<T> {}

// This is how one of functions use the ExclusiveCell as self
pub fn take(&self) -> Option<&mut T> {
    // The exclusivity to data is in fact provided by taken where only when
    // call_once returns ok will the reference to data be returned
    self.taken
        .call_once()
        .ok()
        .map(... unsafe { &mut *self.data.get() }) @\textcolor{orange}{\faWarning}@
}
\end{lstlisting}

\begin{lstlisting}[
  language=Markdown,
  captionpos=b,
  caption={Example of warning produced by \textit{Yuga}.}, 
  label={lst:yugawarning},
  xleftmargin=0.5cm, % Set left margin
  basicstyle=\scriptsize\ttfamily, % Font size and style
  numbers=left, % Line numbers on the left
  breaklines=true % Break long lines
]
## Potential use-after-free!
src/stream.rs:83:5: 83:36
```rust
pub fn from(src: &[u8]) -> Self
```
`*(src)` is of type `u8` and outlives the lifetime corresponding to `'_`, 
It is (probably) returned as `*(ret.data.elements)` which is of type `u8`,
and lives for the entire duration that it is owned. Here, `ret` denotes the
value returned by the function.
The latter can be longer than the former, which could lead to use-after-free!
**Detailed report:**

`src` is of type `&[u8]`

`ret` is of type `Self`
```rust
pub struct MemoryStreamReader {
    data    : Vec<u8>,
    cursor  : usize,
}
```
`Self` has a custom `Drop` implementation.
```rust
fn drop(&mut self) {}
```
`ret.data` is of type `Vec<u8>`
```rust
pub struct Vec<T> {
    elements    : *mut T,
    count       : usize,
    capacity    : usize,
}
```
`Self` has a custom `Drop` implementation.
```rust
fn drop(&mut self) {
        self.dropElements();
        unsafe { free(self.elements) }
    }
```
`ret.data.elements` is of type `*mut T`


Here is the full body of the function:

```rust
pub fn from(src: &[u8]) -> Self{
        let mut v = Vec::new();
        v.append(src);
        Self { data : v, cursor: 0 }
    }
```
\end{lstlisting}

\begin{lstlisting}[
  language=Rust,
  captionpos=b,
  escapeinside=@@,
  caption={A True Positive example (\textcolor{red}{\faBug}) detected by Rudra in crate \textit{concurrent-list-0.0.2}. Since \textbf{value} is passed directly (without passing reference), its ownership is also transferred to push. When \textbf{push} finishes, \textbf{value} goes out of scope, and the entry in the node with the value is no longer valid (since value is dropped), this can cause use-after-free (uaf).}, 
  label={lst:yuga_tp},
  xleftmargin=0.5cm, % Set left margin
  basicstyle=\scriptsize\ttfamily, % Font size and style
  numbers=left, % Line numbers on the left
  breaklines=true % Break long lines
]
/// Appends `value` to the list.
pub fn push(&mut self, value: T) {
    // Some size checks here
    ...  
    let index = self.inner.size.load(Ordering::Relaxed);
    let node_index = index_in_node(index);

    if index == 0 {
       // size is not big enough for any Reader to be looking
       // in this yet, do the necessary
       ... 
    } else if node_index == 0 {
      // self.tail points to a valid Node<T> that contains the
      // absolute index `index - 1` , do the necessary
      ...
    }

    let node = Node::from_raw_with_size(self.tail, size_from_index(index));
    // SAFETY: self.tail points to a valid Node<T> that contains the
    // absolute index `index`
    let node: &Node<T> = unsafe { &*node };

    let raw_cell = node.data[node_index].as_ptr();
    // We don't know lifetimes of neither self nor value, and since 
    // value is passed
    // directly (without passing reference), its ownership is also 
    // transferred to push. 
    unsafe { UnsafeCell::raw_get(raw_cell).write(value) }; @\textcolor{red}{\faBug}@

    self.inner.size.fetch_add(1, Ordering::Release);
}
\end{lstlisting}

\begin{lstlisting}[
  language=Rust,
  escapeinside=@@,captionpos=b,
  caption={A false positive example (\textcolor{orange}{\faWarning}) detected by Yuga shows that it is unaware of callee context. This was found in crate \textit{alt-std-0.2.9}.}, 
  label={lst:yuga_caller_context_fp},
  xleftmargin=0.5cm, % Set left margin
  basicstyle=\scriptsize\ttfamily, % Font size and style
  numbers=left, % Line numbers on the left
  breaklines=true % Break long lines
]

// alt-std-0.2.9/src/stream.rs:92
impl MemoryStreamReader {
    // Yuga claims that *(src) is of type u8 and outlives the lifetime
    // corresponding to '_. However, it has no information about
    // the callee context
    pub fn from(src: &[u8]) -> Self {
        let mut v = Vec::new();
        v.append(src);
        Self { data : v, cursor: 0 }
    }
}

fn testMemoryReadStream() {
    // @\textcolor{orange}{\faWarning}@ Called with constant string
    // argument, which has a static lifetime
    let mut msr = MemoryStreamReader::from("hello world".as_bytes());
    // Rest of the function
}
\end{lstlisting}

\begin{lstlisting}[
  language=Rust,
  escapeinside=@@,captionpos=b,
  caption={A false positive example (\textcolor{orange}{\faWarning}) detected by Yuga where it does not take into account the application semantics and the fact that due to functionalities such as swap in this context, lifetimes of both the arguments and the return value become independent. This was found in crate \textit{atom\_box-0.2.0}.}, 
  label={lst:yuga_application_semantics_fp},
  xleftmargin=0.5cm, % Set left margin
  basicstyle=\scriptsize\ttfamily, % Font size and style
  numbers=left, % Line numbers on the left
  breaklines=true % Break long lines
]

// self is of type AtomBox<'domain, T, DOMAIN_ID> with struct def -
pub struct AtomBox<'domain, T, const DOMAIN_ID: usize> {
    ptr: AtomicPtr<T>,
    domain: &'domain Domain<DOMAIN_ID>,
}

// Definition of StoreGuard struct is as follows
pub struct StoreGuard<'domain, T, const DOMAIN_ID: usize> {
    ptr: *const T,
    domain: &'domain Domain<DOMAIN_ID>,
}

// atom_box-0.2.0/src/lib.rs:232:5 (Functionality redacted for brevity)
// The function takes self as a borrowed reference, new_value of type
// StoreGuard and returns an object of type StoreGuard . 
pub fn swap_from_guard(
    &self,
    new_value: StoreGuard<'domain, T, DOMAIN_ID>,
) -> StoreGuard<'domain, T, DOMAIN_ID>{
    // get the ptr from new_value
    let new_ptr = new_value.ptr;

    // mem::forget to ensure new_value does not get deallocated once it goes
    // out of scope
    core::mem::forget(new_value);

    // Calls swap on self.ptr and swaps it atomically with the new_ptr and
    // finally returns the StoreGuard object by saving the old_ptr
    // as a result of swap.
    let old_ptr = self.ptr.swap(new_ptr as *mut T, Ordering::AcqRel);

    // Yuga detects that there is a possibility of ptr outliving the 'domain
    // lifetime after the function returns and reference to self is dropped 
    // (but storeguard continues to live).

    // It does not take into account the fact that due to the swap, it is 
    // guaranteed that the old_ptr will only be accessible from and by
    // storeguard and the new_value.ptr is only accessible by self,
    // both of which are independent of each other's lifetimes.
    
    StoreGuard {
        ptr: old_ptr, @\textcolor{orange}{\faWarning}@
        domain: self.domain,
    }
}
\end{lstlisting}

\begin{lstlisting}[
  language=Bash,
  caption={Example of warning produced by \safedrop{}.}, 
  label={lst:safedropwarning},
  xleftmargin=0.5cm, % Set left margin
  basicstyle=\scriptsize\ttfamily, % Font size and style
  numbers=left, % Line numbers on the left
  breaklines=true, % Break long lines
  captionpos=b,
]
Function:DefId(0:167 ~ spin[372e]::rw_lock::{impl#6}::downgrade);
   DefIndex(167)
Dangling Pointer Bug Exist 
  /path/to/crate/src/rw_lock.rs:438:5: 448:6 (#0)
\end{lstlisting}

\begin{figure}[tb]
    \centering
    \includegraphics[width=0.2\textwidth]{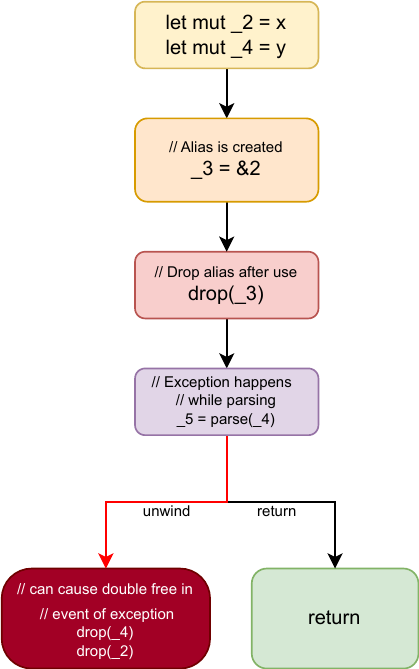}
    \caption{A true positive example detected by safedrop in crate \textit{nuttx-embedded-hal-1.0.10}. Safedrop detects the possibility of double free where it is indeed probable that the alias reference is dropped again in the event of an exception during parsing.}
    \label{fig:safedrop_tp}
    \Description[Safedrop detects the possibility of double free where it is indeed probable that the alias reference is dropped again in the event of an exception during parsing.]{Safedrop detects the possibility of double free where it is indeed probable that the alias reference is dropped again in the event of an exception during parsing.}
\end{figure}


\begin{figure}[tb]
    \centering
    \includegraphics[width=0.3\textwidth]{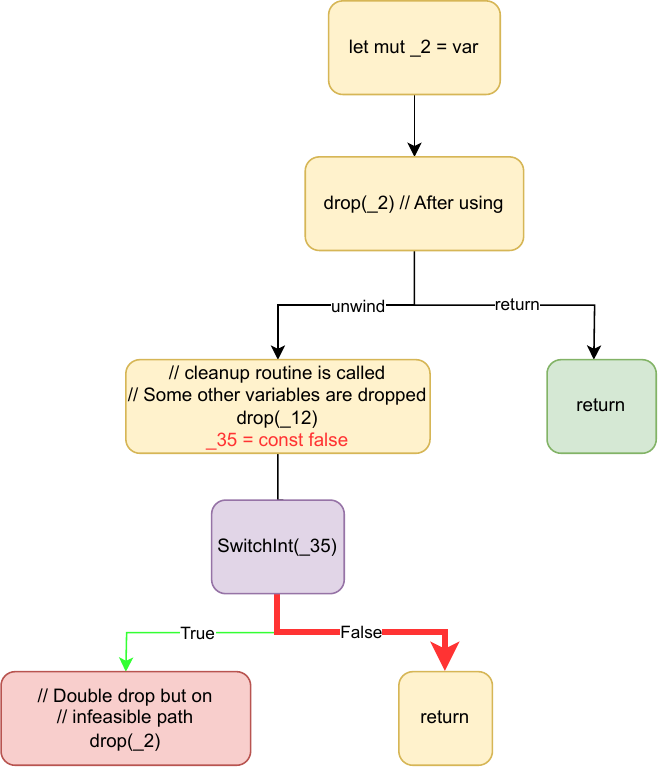}
    \caption{Safedrop fails to detect the presence of infeasible paths. This is from crate ~\textit{arduino\_mkr1000-0.5.0} in one of it's dependency \textit{seq-macro-0.2.2} in function \textit{require\_value}.}
    \label{fig:safedrop_infeasible_paths_fp}
    \Description[Safedrop fails to detect the presence of infeasible paths. This is from crate ~\textit{arduino\_mkr1000-0.5.0} in one of it's dependency \textit{seq-macro-0.2.2} in function \textit{require\_value}.]{Safedrop fails to detect the presence of infeasible paths. This is from crate ~\textit{arduino\_mkr1000-0.5.0} in one of it's dependency \textit{seq-macro-0.2.2} in function \textit{require_value}.}
\end{figure}

\begin{lstlisting}[
  language=Rust,
  escapeinside=@@,captionpos=b,
  caption={A false positive example (\textcolor{orange}{\faWarning}) detected by Safedrop shows that it incorrectly detects the presence of double free in the above function from crate \textit{drogue-es-wifi-0.1.2}.}, 
  label={lst:safedrop_incorrect_analysis_fp},
  xleftmargin=0.5cm, % Set left margin
  basicstyle=\scriptsize\ttfamily, % Font size and style
  numbers=left, % Line numbers on the left
  breaklines=true % Break long lines
]

fn version_and_date_from_rustc_verbose_version(s: &str) -> 
(Option<String>, Option<String>) {
    // Initialize two mutable variables version and date
    // let (mut version, mut date) = (None, None);
    // Iterate over each line in the input string s
    for line in s.lines() {
        // Define a closure split that splits a string s into two parts 
        let split = |s: &str| s.splitn(2, ":").nth(1).map(|s| 
        s.trim().to_string());
        // Match the first word in the line
        match line.trim().split(" ").nth(0) {
            // If the first word is "rustc", get the version and date
            ...
            // If the first word is "release:", extract the version
            Some("release:") => version = split(line),
            // If the first word is "commit-date:", extract the date
            Some("commit-date:") => date = split(line),
            // Ignore all other lines
            _ => continue
        }
    }

    // Return the version and date options. No double drops are possible
    // in the context of this snippet as reported by Safedrop. This is
    // further confirmed in our analysis on the IR level.
    (version, date) @\textcolor{orange}{\faWarning}@
}
\end{lstlisting}

\begin{lstlisting}[
  language=Bash, % Set the language to Bash
  caption={Example of warning produced by \textit{RCanary}.}, 
  label={lst:rcanarywarning},
  basicstyle=\scriptsize\ttfamily, % Font size and typewriter style
  breaklines=true, % Break long lines
  captionpos=b,
]
Unsat DefId(0:89 ~ arcmut[2d40]::{impl#18}::new) src/lib.rs:263:3: 273:4 (#0)
\end{lstlisting}

\begin{lstlisting}[
  language=Bash, % Set the language to Bash
  caption={Example of warning produced by \textit{FFIChecker}.}, 
  label={lst:fficheckerwarning},
  xleftmargin=0.5cm, % Set left margin
  basicstyle=\scriptsize\ttfamily, % Font size and typewriter style
  numbers=left, % Line numbers on the left
  breaklines=true, % Break long lines
  captionpos=b,
]
("rustsecp256k1_v0_9_2_callback_call", Bug info: LLVM IR of C code is
unknown. Possible bugs: Use After Free, Call by function 
pointer, argument is `Tainted`., seriousness: Low)
\end{lstlisting}

\begin{lstlisting}[
  language=Rust,
  caption={A false positive example (\textcolor{orange}{\faWarning}) detected by FFIChecker where it incorrectly judges that argument to a foreign function is borrowed where in fact the function takes no argument.}, 
  label={lst:ffichecker_fp},
  xleftmargin=0.5cm, % Set left margin
  basicstyle=\scriptsize\ttfamily, % Font size and style
  numbers=left, % Line numbers on the left
  breaklines=true, % Break long lines
  escapeinside=||,
  captionpos=b,
]

// Warning by FFIChecker : 
// "smart_contract::payload::Parameters::load", 
// Bug info: LLVM IR of C code is unknown. Possible bugs: 
// Use After Free, FFI _payload is unknown, argument is in state
// `Borrowed`

// smart-contract-0.2.2/src/payload.rs:
pub fn load() -> Parameters {
    // Call to _payload_len() which is a foreign function in this context
    let payload_len = unsafe { crate::sys::_payload_len() };
    let mut payload_bytes = Vec::with_capacity(payload_len);
    // Other operations on payload_bytes follow
    parameters
}

// Declaration of _payload_len() in src/sys.rs:2 shows that the 
// function takes no argument
// FFIChecker claims that the argument is in `Borrowed` state which is 
// incorrect.
pub fn _payload_len() -> usize; |\textcolor{orange}{\faWarning}|
\end{lstlisting}

\begin{lstlisting}[
  language=Rust,
  caption={\ac{FFI} friendly Function to toggle GPIO pin in \rust{}}, 
  label={lst:little_rust_w_c},
  xleftmargin=0.5cm, % Set left margin
  basicstyle=\scriptsize\ttfamily, % Font size and style
  numbers=left, % Line numbers on the left
  breaklines=true, % Break long lines
  captionpos=b,
]
#[no_mangle]
pub unsafe extern "C" fn togglePin(pin_num: u32) {
    // Cast the address to a mutable reference of NRF_GPIO_Type
    let gpio: &mut NRF_GPIO_Type = unsafe { &mut *(NRF_P0_BASE as *mut NRF_GPIO_Type) };

    // Toggle the specific pin in the OUT field
    let pins_state = gpio.OUT;
    ...
}
\end{lstlisting}

\begin{lstlisting}[
  language=Rust,
  caption={A deadlock bug (\textcolor{red}{\faLock}) detected by \lockbud{} in crate \textit{asset-hub-polkadot-runtime-0.13.0} in one of its dependencies \textit{tracing-log v0.1.4}.}, 
  label={lst:lockbudtp},
  xleftmargin=0.5cm, % Set left margin
  basicstyle=\scriptsize\ttfamily, % Font size and style
  numbers=left, % Line numbers on the left
  breaklines=true, % Break long lines
  escapeinside=||,captionpos=b,
]

// tracing-log-0.1.4/src/trace_logger.rs:356
fn event(&self, event: &Event<'_>) {
    // First lock is acquired here
    let spans = self.spans.lock().unwrap();
    // Call to current_id happens
    let current = self.current_id()...
}
// This function in turn calls clone_span on self
fn current_id(&self) -> Option<Id> {...self.clone_span(span)))...}

fn clone_span(&self, id: &Id) -> Id {
    // Tries to re-acquire the same lock on self.spans
    let mut spans = self.spans.lock().unwrap(); |\textcolor{red}{\faLock}|
    ...
}
\end{lstlisting}

\subsection{\lockbud{} Report}
\label{apdx:lockbudreport}

~\lst{lst:lockbudwarning} gives an example of the type of warnings ~\lockbud{} emits. The warnings have abundant information about the callchain flow leading to double lock, easing the triage process. ~\lst{lst:lockbudtp} presents a true positive example where ~\lockbud{} detects double lock. The \rustcode{event} function acquires lock on \rustcode{self.spans} and calls \rustcode{current_id} on \code{self}. This function inturn calls \rustcode{clone_span} on \rustcode{self} which tries to hold lock on the same variable resulting in a double lock. However, ~\lockbud{} does not account for the nature of locks as evident from~\lst{lst:lockbud_locks_fp} where it flags double lock occuring on a special type of lock \rustcode{RWLock} which can allow concurrent reads and exclusive writes. ~\lockbud{} fails to detect such locks and flags concurrent reads as double lock. ~\lst{lst:lockbud_application_semantics_fp} shows that ~\lockbud{} also fails to analyze convoluted application specific semantics such as threads re-acquiring locks after being woken by conditional variables.

\subsection{\rudra{} Report}
\label{apdx:rudrareport}

~\lst{lst:rudrawarning} gives an example of a warning that Rudra emits. It reasons about how unsafe data flows across the context of the function and in this case, mentions about unsafe usage of send, sync variances which make it relatively easier to triage.  ~\lst{lst:rudra_tp} presents a true positive example where Rudra detects unsafe dataflow accurately. If an unchecked index goes out of bounds or accesses invalid memory, \rustcode{clone} function can indeed panic and leave the variables involved in an inconsistent state. However, Rudra fails to account for the presence of special variables which guarantee that a particular function call will not panic.~\lst{lst:rudra_panic_guard} makes use of \rustcode{panic_guard} variable to ensure that the subsequent functions dont panic, or even if they panic, state consistency is properly handled. Rudra also fails to detect guarantees provided by applications which make use of unsafe send, sync variances.~\lst{lst:rudra_atomic_fp} shows that at the instance of usage of the \rustcode{ExclusiveCell}, the \rustcode{AtomicBool} variable \rustcode{taken} ensures exclusive access to the \rustcode{UnsafeCell} data.

\subsection{\yuga{} Report}
\label{apdx:yugareport}

~\yuga{} gives an elaborate markdown report about the reasoning it chooses to detect temporal safety issues which is shown in ~\lst{lst:yugawarning}. This makes it very convenient to follow and triage.~\lst{lst:yuga_tp} shows that ~\yuga{} can accurately analyze lifetime differences between arguments and the return value. In this case, the argument \rustcode{value} is written to the ~\rustcode{node.data[node_index]} and later it's reference is dropped after the function ends. The value that is dropped will remain as a stale entry in the node, which presents a possibility of UAF later on if this value is dereferenced. However, as evident from~\lst{lst:yuga_caller_context_fp}, ~\yuga{} seems to be unaware of caller context. In this case, a constant string is passed as argument to the ~\rustcode{from} function, where ~\yuga{} assumes possibility of Use after Free since lifetimes have not been highlighted. It does not take into account the factors such as constant strings always have static lifetimes. From~\lst{lst:yuga_application_semantics_fp}, ~\yuga{} also seems to be imprecise in judging the application semantics in the effected function. In this case, after the swap function is called, an atomic swap of ~\rustcode{self.ptr} with ~\rustcode{new_ptr} happens and returns ~\rustcode{old_ptr} which guarantees that the ~\rustcode{old_ptr} will only be accessible by ~\rustcode{StoreGuard} and the ~\rustcode{new_value.ptr} is only accessible by self, both of which are independent of each other's lifetime.
 
\subsection{\safedrop{} Report}
\label{apdx:safedropreport}


~\lst{lst:safedropwarning} shows the kinds of warnings ~\safedrop{} emits. It emits just the source location where the temporal safety bug is present. In our triage on effected crates, we observed that most of these source locations are at the end of the function call which make it really difficult to analyze the exact reason for the temporal safety issue. Hence, we chose to follow the approach of analyzing on the MIR level just as ~\safedrop{} does it and attempted to reason our findings. From~\fig{fig:safedrop_tp}, we get a glimpse of the kind of temporal safety bugs ~\safedrop{} accurately finds. A reference to an aliased variable is dropped before calling ~\rustcode{parse} function on another variable. If ~\code{parse} somehow panics, the reference to the original variable \code{_2} is also dropped which can indeed cause a double free. On the flip side,~\fig{fig:safedrop_infeasible_paths_fp} shows that ~\safedrop{} can result in false positives resulting from double drops via infeasible paths. In this case, The \rustcode{SwitchInt} happens on a variable which is by default initialized by a \code{false} value meaning the \code{true} branch is never taken. In many cases of our analysis, we also observed that ~\safedrop{} detects presence of double free in a function where no double drops are happening in the function as well as any of the functions that the parent function calls. We did this analysis both on source as well as IR level to confirm our findings.~\lst{lst:safedrop_incorrect_analysis_fp} shows one such case where seemingly no unintended drops happen. When version and date are returned from the function, their lifetime is extended to match the lifetime of the returned tuple that contains them, hence there is no possibility of double drops in this context.


\subsection{\rcanary{} Report}
\label{apdx:rcanaryreport}
\rcanary{} provides a single line about boolean satisfiability of presence of memory leaks as evident from~\lst{lst:rcanarywarning}. It becomes almost impossible to reason about the validity of the warning without further information about how the tool came to the conclusion. Hence, we did not analyze warnings from \rcanary{}.

\subsection{\ffichecker{} Report}
\label{apdx:fficheckerreport}
~\lst{lst:fficheckerwarning} shows an example of the type of warnings emitted by ~\ffichecker{}. Here as well, we are greeted with one line information about the presence of tainted dataflow into an FFI call. Warnings from ~\ffichecker{} included only the function name which was either the foreign function call or the parent function that called a foreign function. In many cases from our analysis, we found no calls to any foreign functions at all. Hence, we did not detect any true positives in our triage of it's warnings.~\lst{lst:ffichecker_fp} shows an example of a false positive analysis done by ~\ffichecker{} where it assumes argument to the FFI call to be tainted, where infact the FFI call did not expect any argument.

\section{Developer Survey}

\subsection{Solicitation Methodology}
\label{apdx:solicitationmailinglists}
In order to conduct the developer survey for this study, a systematic solicitation methodology was employed to gather responses from a diverse range of participants in the Rust and embedded systems communities. The solicitation process involved posting the survey on various forums, Discord servers, and mailing lists associated with Rust and relevant~\acp{RTOS} and platforms.

To ensure a broad reach and participation, the survey was prominently shared on platforms such as Rust Reddit, which proved to be a valuable source of responses. Additionally, the survey was disseminated through public mailing lists of prominent~\acp{RTOS} and platforms, including Zephyr, Raspberry Pi, and ChibiOS. These mailing lists provided an opportunity to engage with developers who have expertise and experience in embedded systems development.

\subsection{Survey Questions}
\label{apdx:surveyquestions}

\begin{enumerate}
\item How familiar are you with embedded systems programming?
\begin{itemize}
    \item Not familiar
    \item Beginner
    \item Intermediate
    \item Advanced
    \item Expert
\end{itemize}
\item Which languages you often use to develop embedded applications? (select all that apply)
\begin{itemize}
    \item C/ Embedded C/ C++
    \item Rust
    \item Python
    \item Java
    \item Assembly
    \item Other
\end{itemize}
\item How many years of experience do you have in embedded programming?
\begin{itemize}
    \item Less than a year
    \item 1-3 years
    \item 3-5 years
    \item 5-10 years
    \item More than 10 years
\end{itemize}
\item Have you worked with any Real-Time Operating Systems (RTOSes)?
\begin{itemize}
    \item Yes, have used them in multiple projects
    \item Yes, but have not used them extensively
    \item No
\end{itemize}
\item Which RTOSes have you worked with in the past? (select all that apply)
\begin{itemize}
    \item FreeRTOS
    \item Zephyr
    \item VxWorks
    \item C/OS
    \item Other (Text)
\end{itemize}
\item Approximately how many embedded software projects have you developed/contributed to?
\begin{itemize}
    \item Less than 5
    \item 5-10
    \item 10-50
    \item 50-100
    \item More than 100
\end{itemize}
\item Have you worked with Rust before?
\begin{itemize}
    \item Yes
    \item No
\end{itemize}
\item How many years of experience do you have with programming in Rust?
\begin{itemize}
    \item Less than a year
    \item 1-3 years
    \item 3-5 years
    \item 5-10 years
    \item More than 10 years
\end{itemize}
\item Which of the following is not primitive type in Rust?
\begin{itemize}
    \item i32
    \item String
    \item ()
    \item char
\end{itemize}
\item Given the code snippet:

~\rustcode{let mut a: i32 = 1;}

Which of the following is correct
\begin{itemize}
    \item'a' can be reassigned to a new value of any type later in program  
    \item'a' can be reassigned to a new value of same type later in program 
    \item'a' cannot be shadow copied  
    \item Reassigning a value of same type to 'a' later in the program, will throw an error
    \item Code will not compile properly
\end{itemize}
\item Which of the following statements is true about Rust ownership and borrowing?
\begin{itemize}
    \item At any given time, you can have either one mutable reference OR any number of immutable references.
    \item At any given time, you can have one mutable reference AND any number of immutable references
    \item At any given time, we can have any number of mutable AND immutable references
    \item At any given time we can have one immutable reference AND any number of mutable references
    \item At any given time, we can either have one immutable reference OR any number of mutable references.
\end{itemize}
\item How many applications have you developed using Rust programming language (Approximately)?
\begin{itemize}
    \item 1-5
    \item 5-20
    \item 20-50
    \item More than 50
\end{itemize}
\item How many lines of Rust code have you written (Approximately)
\begin{itemize}
    \item Less than 100
    \item 100-1K
    \item 1K-5K
    \item 5K-10K
    \item More than 10K
\end{itemize}
\item Have you ever used rust for embedded development?
\begin{itemize}
    \item Yes
    \item No
\end{itemize}
\item Do you currently use Rust for embedded development?
\begin{itemize}
    \item Yes
    \item No
\end{itemize}
\item Which programming language are you currently using for embedded development
\begin{itemize}
    \item C/ C++
    \item Python
    \item Java
    \item Other (Text)
\end{itemize}

If not using rust for embedded
\item Have you ever considered using Rust for embedded systems development?
\begin{itemize}
    \item Yes
    \item No
\end{itemize}

\item What are your primary reasons for not considering Rust for embedded systems development? (select all that apply)
\begin{itemize}
    \item Lack of experience with Rust
    \item Organization constraints (E.g., My company doesn't allow as the existing frameworks are all in C)
    \item Not convinced of Rust's benefits for embedded systems
    \item Lack of community support for Rust in embedded systems
    \item Other (Please specify)
\end{itemize}
\item What specific concerns do you have about Rust's benefits for embedded systems development? (select all that apply)
\begin{itemize}
    \item Performance concerns
    \item Lack of support for specific hardware platforms
    \item More tedious development process
    \item Difficulty integrating with existing codebases
    \item Other (Please specify)
\end{itemize}
\item Why don’t you use Rust for embedded systems? (select all that apply)
\begin{itemize}
    \item Organization constraints (E.g., My company doesn’t allow as the existing frameworks are all in C)
    \item Steep Learning Curve
    \item Lack of community support for Rust in embedded systems
    \item Other (Please specify)
\end{itemize}
\item Why are you not using Rust? (select all that apply)
\begin{itemize}
    \item Organization constraints (E.g., My company doesn’t allow as the existing frameworks are all in C)
    \item Steep Learning Curve
    \item Lack of community support for Rust in embedded systems
    \item Other (Please specify)
\end{itemize}
\item Are you interested in learning Rust for embedded systems?
\begin{itemize}
    \item Yes
    \item No
\end{itemize}
\item What benefits do you perceive from your current programming language for embedded systems development compared to Rust?
\begin{itemize}
    \item Better library and tool support
    \item Familiarity with language/ Easier to learn and use
    \item Portability
    \item Better performance
    \item Low-level control over hardware
    \item Other (Please specify)
\end{itemize}
\item How important is memory safety in embedded systems development to you?
\begin{itemize}
    \item Very important
    \item Somewhat important
    \item Not very important
    \item Not at all important
\end{itemize}
\item How do you ensure memory safety in your embedded programs with the language you use?
\begin{itemize}
    \item Manual memory management
    \item Use of static analysis tools
    \item Use of run-time checks or garbage collection
    \item Other (Please specify)
\end{itemize}
\item How long have you been using Rust for embedded systems?
\begin{itemize}
    \item Less than 6 months
    \item 6 months to 1 year
    \item 1 year to 2 years
    \item More than 2 years
\end{itemize}
\item Lines of code in embedded programming using Rust(approx)?
\begin{itemize}
    \item Less than 100
    \item 100-1K
    \item 1K-5K
    \item 5K-10K
    \item More than 10K
\end{itemize}
\item How was their learning/coding experience for developing embedded systems with Rust?
\begin{itemize}
    \item Very easy
    \item Somewhat easy
    \item Moderate difficulty
    \item Very difficult
\end{itemize}
\item What motivates you to learn Rust for Embedded systems? (select all that apply)
\begin{itemize}
    \item Better performance
    \item Safety and reliability
    \item Familiarity with Rust
    \item Other (Please specify)
\end{itemize}
\item Is rust documentation helpful for embedded development?
\begin{itemize}
    \item It is helpful
    \item It is somewhat helpful
    \item Technicalities are missing
    \item Less examples
    \item Other (Please specify)
\end{itemize}
\item What can make them more helpful for developers (select all of that apply)?
\begin{itemize}
    \item Better organization of information
    \item Improved search functionality
    \item Other (Please specify)
\end{itemize}
\item Do you often have to refer to external documentation to find solutions for issues?
\begin{itemize}
    \item Yes, frequently
    \item Yes, occasionally
    \item Rarely
\end{itemize}
\item How familiar are you with embedded systems programming?
        \begin{itemize}
            \item Not familiar
            \item Beginner
            \item Intermediate
            \item Advanced
            \item Expert
        \end{itemize}
        \item Which languages you often use to develop embedded applications? (select all that apply)
        \begin{itemize}
            \item C/ Embedded C/ C++
            \item Rust
            \item Python
    \item Never
    \item Not sure
\end{itemize}
\item Which sources have helped you in your rust embedded journey? (select all that apply)
\begin{itemize}
    \item Official Rust documentation
    \item Third-party blogs and tutorials
    \item Online forums and communities
    \item Other (Please specify)
\end{itemize}
\item How do you find the support for Rust in the embedded systems community?
\begin{itemize}
    \item Excellent
    \item Good
    \item Average
    \item Poor
    \item Very poor
\end{itemize}
\item Have you written user level embedded applications in rust?
\begin{itemize}
    \item Yes
    \item No
\end{itemize}
\item Do you prefer writing user level embedded applications in rust?
\begin{itemize}
    \item Yes, always
    \item Yes, sometimes
    \item No, prefer other languages
\end{itemize}
\item Mention the other programming language for embedded user applications and the possible reasons? (select all that apply)
\begin{itemize}
    \item C - widely used, mature, efficient
    \item C++ - object-oriented, high-performance
    \item Python - easy to learn, rapid development
    \item Java - platform-independent, easy to use
    \item Other (Please specify)
\end{itemize}
\item Why not Rust for embedded system's application development? (select all that apply)
\begin{itemize}
    \item Lack of resources (eg library support for the hardware)
    \item Performance issues
    \item Other programming languages better suited
    \item Other (please specify)
\end{itemize}

\item Have you written kernel level code using rust for embedded systems?
\begin{itemize}
    \item Yes
    \item No
\end{itemize}
\item Do you prefer writing kernel level code for embedded systems in rust?
\begin{itemize}
    \item Yes, always
    \item Yes, sometimes
    \item No, prefer other languages
\end{itemize}
\item What programming languages do you use for embedded kernel development? (select all that apply)
\begin{itemize}
    \item C
    \item Embedded C
    \item C++
    \item Assembly
    \item Other (Please specify)
\end{itemize}
\item Why not Rust for embedded system's kernel development? (select all that apply)
\begin{itemize}
    \item Technical difficulties (difficulty in translating the embedded kernel concepts)
    \item Steep learning curve
    \item Limited library and tool support
    \item Other (Please specify)
\end{itemize}
\item How satisfied are you with the availability of resources and crates provided by Rust for embedded systems, such as hardware interaction and drivers?
\begin{itemize}
    \item Very satisfied
    \item Somewhat satisfied
    \item Neutral
    \item Somewhat dissatisfied
    \item Very dissatisfied
\end{itemize}
\item Is rust toolchain for embedded development easy to use?
\begin{itemize}
    \item Yes, very easy
    \item Somewhat easy
    \item Moderately difficult
    \item Very difficult
\end{itemize}
\item What makes it easier for you to use Rust toolchain for embedded development compared to other programming languages? (select all that apply)
\begin{itemize}
    \item Better documentation
    \item More intuitive tools
    \item Wider community support
    \item Better integration with hardware
    \item Other (Please specify)
\end{itemize}
\item Could you help us know the possible reasons of difficulty in use?
\begin{itemize}
    \item Poor documentation
    \item Incomplete or buggy tools
    \item Lack of community support
    \item Difficulty integrating with hardware
    \item Other (Please specify)
\end{itemize}
\item Do you feel that there are enough resources or crates in Rust to deal with embedded systems, such as hardware interaction, drivers, etc.?
\begin{itemize}
    \item Yes, there are plenty of resources available
    \item No, there are not enough resources available
    \item I am not sure
\end{itemize}
\item How often do you encounter compatibility issues with libraries when using Rust for embedded systems?
\begin{itemize}
    \item Very frequently
    \item Occasionally
    \item Rarely
    \item Never
\end{itemize}
\item Are you familiar with Rust's interoperability with C?
\begin{itemize}
    \item Yes
    \item Somewhat
    \item No
\end{itemize} 
\item Do you feel interoperability would help you code better in Rust
\begin{itemize}
    \item Yes
    \item No
\end{itemize}
\item Have you ever used Rust and C together in a project?
\begin{itemize}
    \item Yes
    \item No
\end{itemize}
\item In what scenarios have you used Rust and C together? (check all that apply)
\begin{itemize}
    \item Calling C code from Rust
    \item Calling Rust code from C
    \item Other (Please specify)
\end{itemize}
\item What challenges have you faced while using Rust and C together? (check all that apply)
\begin{itemize}
    \item Managing memory allocation and deallocation between Rust and C  
    \item Handling different data types between Rust and C
    \item Debugging issues with interoperation between Rust and C
    \item None
    \item Other (Please specify)
\end{itemize}
\item How do you perceive the impact of Rust's interoperability with C on improving the safety of existing C codebases?
\begin{itemize}
    \item Yes, it significantly helps to make the existing C codebases safer with ease.
    \item Somewhat, it helps in building safer codebases, but requires significant effort.
    \item No, it doesn't improve the safety of existing C codebases
\end{itemize}
\item Which IDE do you prefer for Rust programming language?
\begin{itemize}
    \item Visual Studio Code/Rust analyzer
    \item IntelliJ Rust
    \item Others (Please specify)
\end{itemize}
\item How useful are rust compiler warning/error messages?
\begin{itemize}
    \item Not useful
    \item Slightly useful
    \item Moderately useful
    \item Very useful
    \item Extremely useful
\end{itemize}
\item Do you feel the chosen programming language support testing of functionalities in a developer-friendly manner (Is it easy to debug the programs)?
\begin{itemize}
    \item Yes, it is very easy
    \item It is somewhat easy
    \item It is moderately difficult
    \item It is very difficult
    \item Not sure
\end{itemize}

\item How much time do you spend debugging your Rust embedded applications?
\begin{itemize}
    \item Less than 25\% of development time
    \item 25\%-50\% of development time
    \item 50\%-75\% of development time
    \item More than 75\% of development time
\end{itemize}
\item Have you encountered any performance issues when using Rust for embedded systems?
\begin{itemize}
    \item Yes
    \item No
    \item I am not sure
\end{itemize}
\item Have you ever evaluated the performance of Rust compared to the other programming languages common in the field of embedded development?
\begin{itemize}
    \item Yes, and Rust outperformed other languages.
    \item Yes, and Rust performed similarly to other languages.
    \item Yes, but other languages outperformed Rust.
    \item No, I have not evaluated the performance of Rust compared to other languages in embedded development.
\end{itemize}
\item How likely are you to recommend rust to other developers for use in embedded systems?
\begin {itemize}
    \item Very likely
    \item Somewhat likely
    \item Neutral
    \item Somewhat unlikely
    \item Very unlikely
\end{itemize}

    \item Have you ever used any other programming language before Rust in embedded system development?
\begin{itemize}
    \item Yes
    \item No
\end{itemize}
\item Have you noticed any significant improvements in development time or code quality since switching to Rust for embedded systems?
\begin{itemize}
    \item Yes, significant improvements
    \item Some improvements
    \item No noticable improvements
    \item It has actually slowed down development
\end{itemize}
\item Have you encountered any performance issues when using Rust for embedded systems ?
\begin{itemize}
    \item Yes
    \item No
    \item I am not sure
\end{itemize}
\item Compared to the other programming languages you have used for embedded programming, how well does Rust integrate with hardware?
\begin{itemize}
    \item Much better integration than other languages
    \item Somewhat better integration than other languages
    \item About the same level of integration as other languages
    \item Worse integration than other languages
\end{itemize}

\item Is there anything else you would like to share or any other thoughts you have regarding Rust for embedded programming?
(Open-ended)

\end{enumerate}

\section{Additional Background}

\begin{figure}[ht]
    \hspace*{-1cm}
    \includegraphics[scale=0.9]{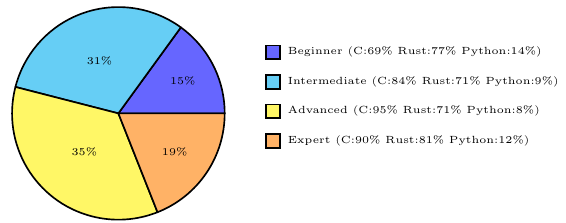}
    \caption{Survey participants (Total: 225) categorized based on their self-declared expertise in embedded systems development and languages frequently used for it.}
\label{fig:surveyparticipants}
\Description[Survey Participants]{Survey participants (Total: 225) categorized based on their self-declared expertise in embedded systems development and languages frequently used for it.}
\end{figure}

\subsection{Wrapper Crates}
\label{apdx:rustwrappercrates}
Wrapper crates provide a layer of safe rust idiomatic abstraction, over the libraries or codebases written in foreign languages such as C/C++. This enables rust developers to reuse existing C/C++ codebases effortlessly and safely, 
eliminating the need to directly handle the foreign function calls. The safe API offered insulates Rust applications from the risks associated with the direct interaction with unsafe external functions. Rust's Foreign Function Interface (FFI) capabilities, detailed in Section ~\ref{subsec:rustbackground}, enable Rust to interact effectively with non-Rust code. Wrapper crates utilise this feature internally, by creating bindings to call unsafe foreign code, which can be automated using bindgen. 


\begin{table}[htb]
\caption{Tools used as part of our study. The column \rust{} Ver. indicates version dependencies of the corresponding tool.}
\label{tab:tooltable}
\scriptsize
\centering
\begin{tabular}{ccc}
\toprule
\multicolumn{1}{c|}{\textbf{Tool Name}} & \multicolumn{1}{c|}{\textbf{Description}} & \multicolumn{1}{c}{\textbf{\rust{} Ver.}}   \\ \midrule 
\multicolumn{3}{c}{\textbf{\rust{} Vulnerability Detection Tools}}                                                                                                                 \\ \midrule
\multicolumn{1}{c|}{\lockbud{}~\cite{10.1145/3385412.3386036}}                                    & \multicolumn{1}{c|}{\begin{tabular}[c]{@{}c@{}}Static analyzer to detect\\ use-after-free and \\ concurrency bugs. \end{tabular}}                                      & \multicolumn{1}{c}{1.66.0-nightly}              \\ \hline
\multicolumn{1}{c|}{\rudra{}~\cite{bae2021rudra}}                                    & \multicolumn{1}{c|}{\begin{tabular}[c]{@{}c@{}}Static analyzer to detect \\ memory safety bugs.\end{tabular}}                                      & \multicolumn{1}{c}{1.56.0-nightly}              \\ \hline
\multicolumn{1}{c|}{\safedrop{}~\cite{Cui2021SafeDropDM}}                           & \multicolumn{1}{c|}{\begin{tabular}[c]{@{}c@{}}\rust compiler pass to\\ detect memory deallocation bugs.\end{tabular}}                             & \multicolumn{1}{c}{based on Rust-1.63}    \\ \midrule
\multicolumn{3}{c}{\textbf{\rust{} Code Quality Tools}}                                                                                                                            \\ \midrule
\multicolumn{1}{c|}{\qrates{}~\cite{10.1145/3428204}}                           & \multicolumn{1}{c|}{\begin{tabular}[c]{@{}c@{}} Static analyzer to detect \\ prevalence of unsafe~\rust{}\end{tabular}}                             & \multicolumn{1}{c}{1.71.0-nightly}     \\ \hline
\multicolumn{1}{c|}{\clametrics{}~\cite{mergendahl2022cross}}                           & \multicolumn{1}{c|}{\begin{tabular}[c]{@{}c@{}} Binary analysis tool to detect \\cross language attack gadgets.\end{tabular}}                             & \multicolumn{1}{c}{N/A} \\ \midrule
\multicolumn{3}{c}{\textbf{Conversion Tools from C to~\rust{}}}                                                                                                                    \\ \midrule
\multicolumn{1}{c|}{\ctorust{}~\cite{c2rustnodate}}                           & \multicolumn{1}{c|}{\begin{tabular}[c]{@{}c@{}}Transforming C programs\\ to \rust{}.\end{tabular}}                             & \multicolumn{1}{c}{N/A}     \\ \bottomrule
\end{tabular}
\end{table}

\begin{lstlisting}[
  language=JSON,
  caption={Target specification for processors using ARMv7.}, 
  label={lst:rusttarget},
  xleftmargin=0.5cm, % Set left margin
  basicstyle=\scriptsize\ttfamily, % Font size and style
  numbers=left, % Line numbers on the left
  breaklines=true % Break long lines
]
{
  "abi": "eabi",
  "arch": "arm",
  "c-enum-min-bits": 8,
  "data-layout": "E-m:e-p:32:32-Fi8-i64:
                  64-v128:64:128-a:0:32-n32-S64",
  "emit-debug-gdb-scripts": false,
  "is-builtin": true,
  "linker": "rust-lld",
  "linker-flavor": "ld.lld",
  "llvm-target": "armebv7r-unknown-none-eabi",
  "max-atomic-width": 64,
  "panic-strategy": "abort",
  "relocation-model": "static",
  "target-endian": "big",
  "target-pointer-width": "32"
}
\end{lstlisting}

\subsection{Bug Study}
\label{apdx:rtosbugstudy}
To understand common security issues in embedded systems, we collected publicly reported vulnerabilities from the CVE database~\cite{noauthor_cve_nodate} across various~\acp{RTOS} --- the important component of type-2 systems.

We focused on vulnerabilities with the necessary technical details to understand the root cause; this resulted in a total of 109 vulnerabilities.
We manually categorized each of them according to their root cause,~\tbl{tab:bugsinrtos} shows the categorization.
There are 25 (22.9\%) application-specific vulnerabilities that any programming language cannot prevent.
However, there are 59 (54.2\%) memory corruption vulnerabilities,~\ie spatial or temporal memory issues.

\begin{table*}[]
\caption{Categorization of publicly reported vulnerabilities (CVEs) in various embedded~\acp{RTOS}.~\emph{The bugs in~\textcolor{green}{green} (Total: 59 (54.2\%)) are prevented in~\rust{}} and the bugs in~\textcolor{red}{red} (Total: 25 (22.9\%)) are application specific and cannot be prevented by any programming language.}
\label{tab:bugsinrtos}

\tiny
\begin{tabular}{crrrrrrrrrrrrrr}
\toprule
\multicolumn{5}{c|}{\textbf{Spatial Safety Issues}}                                                                                                                                                                                                                                                                                                                                                                                & \multicolumn{2}{c|}{\textbf{Temporal Memory Issues}}                                                                                                                                                           & \multicolumn{1}{c|}{\multirow{3}{*}{\textbf{\begin{tabular}[c]{@{}c@{}}Integer\\ Errors\end{tabular}}}} & \multicolumn{1}{c|}{\multirow{3}{*}{\textbf{\begin{tabular}[c]{@{}c@{}}Privilege \\ Escalation\end{tabular}}}} & \multicolumn{1}{c|}{\multirow{3}{*}{\textbf{\begin{tabular}[c]{@{}c@{}}SQL \\ Injection\end{tabular}}}} & \multicolumn{1}{c|}{\multirow{3}{*}{\textbf{\begin{tabular}[c]{@{}c@{}}Side\\ Channels\end{tabular}}}} & \multicolumn{1}{c|}{\multirow{3}{*}{\textbf{\begin{tabular}[c]{@{}c@{}}Open\\ Ports\end{tabular}}}} & \multicolumn{1}{c|}{\multirow{3}{*}{\textbf{\begin{tabular}[c]{@{}c@{}}Improper\\ Permissions\end{tabular}}}} & \multicolumn{1}{c|}{\multirow{3}{*}{\textbf{\begin{tabular}[c]{@{}c@{}}Logical\\ Bug\end{tabular}}}} & \multicolumn{1}{c}{\multirow{3}{*}{\textbf{\begin{tabular}[c]{@{}c@{}}Infinite\\ Loop\end{tabular}}}} \\ \cline{1-7}
\multicolumn{2}{c|}{\textbf{Out of Bounds Access}}                                  & \multicolumn{1}{c|}{\multirow{2}{*}{\textbf{\begin{tabular}[c]{@{}c@{}}Null -ptr\\ Deref\end{tabular}}}} & \multicolumn{1}{c|}{\multirow{2}{*}{\textbf{\begin{tabular}[c]{@{}c@{}}Type\\ Confusion\end{tabular}}}} & \multicolumn{1}{c|}{\multirow{2}{*}{\textbf{\begin{tabular}[c]{@{}c@{}}Uninitialized\\ Pointer Access\end{tabular}}}} & \multicolumn{1}{c|}{\multirow{2}{*}{\textbf{\begin{tabular}[c]{@{}c@{}}Use After\\ Free\end{tabular}}}} & \multicolumn{1}{c|}{\multirow{2}{*}{\textbf{\begin{tabular}[c]{@{}c@{}}Memory\\ Leak\end{tabular}}}} & \multicolumn{1}{c|}{}                                                                                   & \multicolumn{1}{c|}{}                                                                                          & \multicolumn{1}{c|}{}                                                                                   & \multicolumn{1}{c|}{}                                                                                  & \multicolumn{1}{c|}{}                                                                               & \multicolumn{1}{c|}{}                                                                                         & \multicolumn{1}{c|}{}                                                                                & \multicolumn{1}{c}{}                                                                                  \\ \cline{1-2}
\multicolumn{1}{c|}{\textbf{Read}}        & \multicolumn{1}{c|}{\textbf{Write}}     & \multicolumn{1}{c|}{}                                                                                    & \multicolumn{1}{c|}{}                                                                                   & \multicolumn{1}{c|}{}                                                                                                   & \multicolumn{1}{c|}{}                                                                                   & \multicolumn{1}{c|}{}                                                                                & \multicolumn{1}{c|}{}                                                                                   & \multicolumn{1}{c|}{}                                                                                          & \multicolumn{1}{c|}{}                                                                                   & \multicolumn{1}{c|}{}                                                                                  & \multicolumn{1}{c|}{}                                                                               & \multicolumn{1}{c|}{}                                                                                         & \multicolumn{1}{c|}{}                                                                                & \multicolumn{1}{c}{}                                                                                  \\ 
\midrule
\rowcolor{black!15}  \multicolumn{15}{c}{\textbf{FreeRTOS}~\cite{freertos}}                                                                                                                                                                                                                                                                                                                                                                                                                                                                                                                                                                                                                                                                                                                                                                                                                                                                                                                                                                                                                                                                                                                                                                                                                                                                                                                                                                                                                                                                                                 \\ \midrule
\multicolumn{1}{r|}{8}                    & \multicolumn{1}{r|}{2}                  & \multicolumn{1}{r|}{0}                                                                                   & \multicolumn{1}{r|}{0}                                                                                  & \multicolumn{1}{r|}{1}                                                                                                  & \multicolumn{1}{r|}{1}                                                                                  & \multicolumn{1}{r|}{0}                                                                               & \multicolumn{1}{r|}{3}                                                                                  & \multicolumn{1}{r|}{1}                                                                                         & \multicolumn{1}{r|}{0}                                                                                  & \multicolumn{1}{r|}{0}                                                                                 & \multicolumn{1}{r|}{0}                                                                              & \multicolumn{1}{r|}{0}                                                                                        & \multicolumn{1}{r|}{1}                                                                               & 0                                                                                                      \\ \midrule
\rowcolor{black!15} \multicolumn{15}{c}{\textbf{RIOT}~\cite{riot}}                                                                                                                                                                                                                                                                                                                                                                                                                                                                                                                                                                                                                                                                                                                                                                                                                                                                                                                                                                                                                                                                                                                                                                                                                                                                                                                                                                                                                                                                                                     \\ \midrule
\multicolumn{1}{r|}{0}                    & \multicolumn{1}{r|}{12}                 & \multicolumn{1}{r|}{1}                                                                                   & \multicolumn{1}{r|}{1}                                                                                  & \multicolumn{1}{r|}{0}                                                                                                  & \multicolumn{1}{r|}{0}                                                                                  & \multicolumn{1}{r|}{1}                                                                               & \multicolumn{1}{r|}{}                                                                                   & \multicolumn{1}{r|}{2}                                                                                         & \multicolumn{1}{r|}{2}                                                                                  & \multicolumn{1}{r|}{1}                                                                                 & \multicolumn{1}{r|}{0}                                                                              & \multicolumn{1}{r|}{0}                                                                                        & \multicolumn{1}{r|}{5}                                                                               & 0                                                                                                      \\ \midrule
\rowcolor{black!15}\multicolumn{15}{c}{\textbf{VXWorks}~\cite{vxworks}}                                                                                                                                                                                                                                                                                                                                                                                                                                                                                                                                                                                                                                                                                                                                                                                                                                                                                                                                                                                                                                                                                                                                                                                                                                                                                                                                                                                                                                                                                                  \\ \midrule
\multicolumn{1}{r|}{0}                    & \multicolumn{1}{r|}{4}                  & \multicolumn{1}{r|}{0}                                                                                   & \multicolumn{1}{r|}{0}                                                                                  & \multicolumn{1}{r|}{0}                                                                                                  & \multicolumn{1}{r|}{0}                                                                                  & \multicolumn{1}{r|}{0}                                                                               & \multicolumn{1}{r|}{1}                                                                                  & \multicolumn{1}{r|}{5}                                                                                         & \multicolumn{1}{r|}{0}                                                                                  & \multicolumn{1}{r|}{0}                                                                                 & \multicolumn{1}{r|}{3}                                                                              & \multicolumn{1}{r|}{2}                                                                                        & \multicolumn{1}{r|}{6}                                                                               & 0                                                                                                      \\ \midrule
\rowcolor{black!15}\multicolumn{15}{c}{\textbf{Zephyr}~\cite{zephyr}}                                                                                                                                                                                                                                                                                                                                                                                                                                                                                                                                                                                                                                                                                                                                                                                                                                                                                                                                                                                                                                                                                                                                                                                                                                                                                                                                                                                                                                                                                                   \\ \midrule
\multicolumn{1}{r|}{3}                    & \multicolumn{1}{r|}{19}                 & \multicolumn{1}{r|}{4}                                                                                   & \multicolumn{1}{r|}{0}                                                                                  & \multicolumn{1}{r|}{0}                                                                                                  & \multicolumn{1}{r|}{2}                                                                                  & \multicolumn{1}{r|}{0}                                                                               & \multicolumn{1}{r|}{4}                                                                                  & \multicolumn{1}{r|}{3}                                                                                         & \multicolumn{1}{r|}{2}                                                                                  & \multicolumn{1}{r|}{0}                                                                                 & \multicolumn{1}{r|}{0}                                                                              & \multicolumn{1}{r|}{0}                                                                                        & \multicolumn{1}{r|}{8}                                                                               & 1                                                                                                      \\ \midrule
\rowcolor{blue!30}\multicolumn{15}{c}{\textbf{Total}}                                                                                                                                                                                                                                                                                                                                                                                                                                                                                                                                                                                                                                                                                                                                                                                                                                                                                                                                                                                                                                                                                                                                                                                                                                                                                                                                                                                                                                                                                                    \\ \midrule
\multicolumn{1}{r|}{\textcolor{green}{\textbf{11 (10.1\%)}}} & \multicolumn{1}{r|}{\textcolor{green}{\textbf{37 (34\%)}}} & \multicolumn{1}{r|}{\textcolor{green}{\textbf{5 (4.6\%)}}}                                                                  & \multicolumn{1}{r|}{\textcolor{green}{\textbf{1 (0.9\%)}}}                                                                 & \multicolumn{1}{r|}{\textcolor{green}{\textbf{1 (0.9\%)}}}                                                                                 & \multicolumn{1}{r|}{\textcolor{green}{\textbf{3 (2.8\%)}}}                                                                 & \multicolumn{1}{r|}{\textcolor{green}{\textbf{1 (0.9\%)}}}                                                              & \multicolumn{1}{r|}{8 (7.3\%)}                                                                          & \multicolumn{1}{r|}{11 (10.1\%)}                                                                               & \multicolumn{1}{r|}{4 (3.7\%)}                                                                          & \multicolumn{1}{r|}{1 (0.9\%)}                                                                         & \multicolumn{1}{r|}{\textcolor{red}{\textbf{3 (2.8\%)}}}                                                             & \multicolumn{1}{r|}{\textcolor{red}{\textbf{2 (1.8\%)}}}                                                                       & \multicolumn{1}{r|}{\textcolor{red}{\textbf{20 (18.3\%)}}}                                                            & 1 (0.9\%)                                                                                              \\ 
\bottomrule
\end{tabular}
\end{table*}

\begin{figure}[tb]
    \centering
    \includegraphics[width=0.48\textwidth]{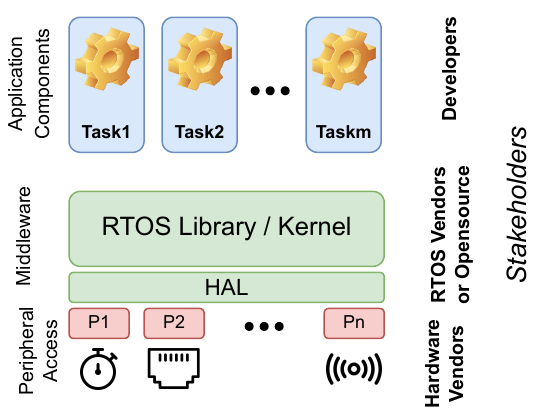}
    \caption{Software design of Type-2 Embedded Systems. 
    Application tasks execute on a \ac{RTOS}, which uses HAL to access peripherals.}
    \label{fig:background}
    \Description[Software design of Type-2 Embedded Systems.]{Software design of Type-2 Embedded Systems.}
\end{figure}

\subsection{\texorpdfstring{\rust{} Safety Guarantees}{Rust Safety Guarantees}}
\label{apdx:rustsafetyguarentees}
\noindent\textbf{Spatial Safety.}~\lst{lst:rustspatialsafety} shows an example where spatial safety issues (\eg out-of-bounds access) are detected in~\rust{} both at compile time and run time.
At Line 12, the variable~\code{x} can be greater than 2, potentially accessing out-of-bounds of the array~\code{a}.
Each array has an associated length, and~\rust{} inserts checks to verify that all indexes into an array are valid.
\rust{} also prevents accessing uninitialized variables (Line 18 in~\lst{lst:rustspatialsafety}) by powerful compile-time analysis to check that a variable is always initialized before access.
\rust{} uses its type system and runtime checks to detect null-ptr dereferences.

\noindent\textbf{Temporal Safety.}~\lst{lst:rusttemporarlsafety} shows a snippet demonstrating~\rust{} ownership features and how it prevents temporal issues, specifically use-after-free.
\rust{}'s ownership feature requires that every object has only one owner.
An object will be released (or dropped) when the owner goes out of scope.
Ownership of an object will be transferred to another variable on assignments. from the owner.
For instance,~\rustcode{x = y}, the ownership of the object pointed by~\rustcode{y} will be transferred to~\code{x}.
Also,~\rust{} allows borrowing, wherein a variable can have a read-only reference to an object (without owning it),~\eg Line 4 and 8 in~\lst{lst:rusttemporarlsafety}.
The borrow checker ensures that the lifetime of all references is more than that of the objects they point to, and thus prevents use-after-free.
In~\lst{lst:rusttemporarlsafety},~\rustcode{z} can be a reference (Line 8) to object~\rustcode{O2} (allocated at Line 6), whose owner is~\rustcode{x}.
The object \rustcode{O2} will freed after the~\rustcode{if} at Line 10 because the owner (\rustcode{x}) goes out-of-scope.
But,~\rustcode{z} can hold a reference to~\rustcode{O2}, which we try to access at Line 13 (a potential use-after-free).
The borrow checker detects and throws a compiler error as the lifetime of~\rustcode{O2} (referenced by~\rustcode{z}) is shorter,~\ie ends at Line 10.

\begin{lstlisting}[
  language=Rust,
  caption={Example demonstrating how \rust{} prevents spatial safety issues (\textcolor{orange}{\faWarning}) at compile time (\textcolor{green}{\faShield}) and run time (\textcolor{red}{\faShield}).},
  escapeinside=||,captionpos=b,
  label={lst:rustspatialsafety},
  xleftmargin=0.5cm, % Set left margin
  basicstyle=\footnotesize\ttfamily, % Font size and style
  numbers=left, % Line numbers on the left
  breaklines=true % Break long lines
]
use std::io;
fn main() {
  // read-only
  let a = [1, 2, 3];
  // read-write (mutable)
  let mut il = String::new();
  io::stdin()
  .read_line(&mut il)
  .expect("Failed to read line");
  let x: usize = il.trim().parse().expect("Invalid");
  // Potential out-of-bounds
  println!("{:?}", a[x]);  |\textcolor{orange}{\faWarning}||\textcolor{red}{\faShield}|
  
  let b;
  if x > 1 {
    b = 4;  
  }
  // Uninitialized variable access.
  println!("Value of b: {}", b); |\textcolor{orange}{\faWarning}||\textcolor{green}{\faShield}|
  }
\end{lstlisting}

\begin{lstlisting}[
  language=Rust,
  caption={Snippet demonstrating how \rust{} prevents temporal safety issues (i.e., use-after-free).}, 
  escapeinside=&&,captionpos=b,
  label={lst:rusttemporarlsafety},
  xleftmargin=0.5cm, % Set left margin
  basicstyle=\footnotesize\ttfamily, % Font size and style
  numbers=left, % Line numbers on the left
  breaklines=true, % Break long lines
  escapeinside=||,captionpos=b,
]
use std::io;
fn main() {
  let y = "HELLO".to_string(); // O1
  let mut z = &y;
  if (...) {
     let x = "Hello".to_string(); // O2
     // z is a reference to O2 owned by x
     z = &x;
     // x goes out of scope, and O2 gets dropped.
  }
  // z might be referencing O2 (a freed object)
  // Potential use-after-free
  println!("{}", z); |\textcolor{orange}{\faWarning}||\textcolor{green}{\faShield}|
}
\end{lstlisting}

\section{\qrates{} Compilation Failures}
\label{apdx:qratecompilationfailures}

Refer~\tbl{tab:qratesfailed} for categorization of~\qrates{} tool failures.
\begin{table}[]
\caption{Reasons for~\qrates{} tool failures and the corresponding number of affected crates.}
\label{tab:qratesfailed}
\small
\centering
\begin{tabular}{|c|r|}
\hline
\textbf{Reason for Compilation Failure} & \textbf{Number of Crates} \\ \hline
Toolchain Incompatibility               & 322                        \\ \hline
Unknown failure                         & 13                        \\ \hline
Failed custom build command             & 10                       \\ \hline
Tool Failure                            & 60                        \\ \hline
\textbf{Total}                          & \textbf{405}              \\ \hline
\end{tabular}
\end{table}

\section{SAST Tools}
\begin{table}[t!]
\footnotesize
\caption{Summary of~\ffichecker{} on embedded crates.}
\label{tab:ffichecker}
    \centering
    \begin{tabular}{cccc}
    \toprule
\multicolumn{1}{c}{\textbf{Category}} & 
\multicolumn{1}{c}{\textbf{\begin{tabular}[c]{@{}c@{}}Num.Crates Successful\\ (\% of Total \\ from~\tbl{tab:cratetypes})\end{tabular}}} &
\multicolumn{1}{c}{\textbf{\begin{tabular}[c]{@{}c@{}}Num. Crates\\ (\% of Successful)\\ having warnings\end{tabular}}} & \multicolumn{1}{c}{\textbf{}} \\ \midrule
        \abbrrtoscr     & \fficheckerrtoscompiled{} (\fficheckerrtoscompiledperc) & \fficheckerrtoswarnings{} (\fficheckerrtoswarningsperc) \\ \midrule
        \abbrdrivercr     & \fficheckerdrivercompiled{} (\fficheckerdrivercompiledperc) & \fficheckerdriverwarnings{} (\fficheckerdriverwarningsperc) \\ \midrule
        \abbrhalcr     & \fficheckerhalcompiled{} (\fficheckerhalcompiledperc) & \fficheckerhalwarnings{} (\fficheckerhalwarningsperc) \\ \midrule
        \abbrbspcr     & \fficheckerbspcompiled{} (\fficheckerbspcompiledperc) & \fficheckerbspwarnings{} (\fficheckerbspwarningsperc) \\ \midrule
        \abbrpaccr     & \fficheckerpaccompiled{} (\fficheckerpaccompiledperc) & \fficheckerpacwarnings{} (\fficheckerpacwarningsperc) \\ \midrule
        \abbrarchcr     & \fficheckerarchcompiled{} (\fficheckerarchcompiledperc) & \fficheckerarchwarnings{} (\fficheckerarchwarningsperc) \\ \midrule
        \abbrutilcr     & \fficheckerutilcompiled{} (\fficheckerutilcompiledperc) & \fficheckerutilwarnings{} (\fficheckerutilwarningsperc) \\ \midrule
        \abbruncatcr     & \fficheckeruncatcompiled{} (\fficheckeruncatcompiledperc) & \fficheckeruncatwarnings{} (\fficheckeruncatwarningsperc) \\ \midrule
        \textbf{Total}  & \fficheckertotalcompiled{} (\fficheckertotalcompiledperc) & \fficheckertotalwarningscompiled{} (\fficheckertotalwarningscompiledperc) \\
         \bottomrule
    \end{tabular}
\end{table}
\begin{table}[t!]
\footnotesize
\caption{Summary of~\yuga{} on embedded crates.}
\label{tab:yuga}
    \centering
    \begin{tabular}{cccc}
    \toprule
\multicolumn{1}{c}{\textbf{Category}} & 
\multicolumn{1}{c}{\textbf{\begin{tabular}[c]{@{}c@{}}Num.Crates Successful\\ (\% of Total \\ from~\tbl{tab:cratetypes})\end{tabular}}} &
\multicolumn{1}{c}{\textbf{\begin{tabular}[c]{@{}c@{}}Num. Crates\\ (\% of Successful)\\ having warnings\end{tabular}}} & \multicolumn{1}{c}{\textbf{}} \\ \midrule
        \abbrrtoscr     & \yugartoscompiled{} (\yugartoscompiledperc) & \yugartoswarnings{} (\yugartoswarningsperc) \\ \midrule
        \abbrdrivercr     & \yugadrivercompiled{} (\yugadrivercompiledperc) & \yugadriverwarnings{} (\yugadriverwarningsperc) \\ \midrule
        \abbrhalcr     & \yugahalcompiled{} (\yugahalcompiledperc) & \yugahalwarnings{} (\yugahalwarningsperc) \\ \midrule
        \abbrbspcr     & \yugabspcompiled{} (\yugabspcompiledperc) & \yugabspwarnings{} (\yugabspwarningsperc) \\ \midrule
        \abbrpaccr     & \yugapaccompiled{} (\yugapaccompiledperc) & \yugapacwarnings{} (\yugapacwarningsperc) \\ \midrule
        \abbrarchcr     & \yugaarchcompiled{} (\yugaarchcompiledperc) & \yugaarchwarnings{} (\yugaarchwarningsperc) \\ \midrule
        \abbrutilcr     & \yugautilcompiled{} (\yugautilcompiledperc) & \yugautilwarnings{} (\yugautilwarningsperc) \\ \midrule
        \abbruncatcr     & \yugauncatcompiled{} (\yugauncatcompiledperc) & \yugauncatwarnings{} (\yugauncatwarningsperc) \\ \midrule
        \textbf{Total}  & \yugatotalcompiled{} (\yugatotalcompiledperc) & \yugatotalwarningscompiled{} (\yugatotalwarningscompiledperc) \\
         \bottomrule
    \end{tabular}
\end{table}
\begin{table}[t!]
\footnotesize
\caption{Summary of~\rcanary{} on embedded crates.}
\label{tab:rcanary}
    \centering
    \begin{tabular}{cccc}
    \toprule
\multicolumn{1}{c}{\textbf{Category}} & 
\multicolumn{1}{c}{\textbf{\begin{tabular}[c]{@{}c@{}}Num.Crates Successful\\ (\% of Total \\ from~\tbl{tab:cratetypes})\end{tabular}}} &
\multicolumn{1}{c}{\textbf{\begin{tabular}[c]{@{}c@{}}Num. Crates\\ (\% of Successful)\\ having warnings\end{tabular}}} & \multicolumn{1}{c}{\textbf{}} \\ \midrule
        \abbrrtoscr     & \rcanaryrtoscompiled{} (\rcanaryrtoscompiledperc) & \rcanaryrtoswarnings{} (\rcanaryrtoswarningsperc) \\ \midrule
        \abbrdrivercr     & \rcanarydrivercompiled{} (\rcanarydrivercompiledperc) & \rcanarydriverwarnings{} (\rcanarydriverwarningsperc) \\ \midrule
        \abbrhalcr     & \rcanaryhalcompiled{} (\rcanaryhalcompiledperc) & \rcanaryhalwarnings{} (\rcanaryhalwarningsperc) \\ \midrule
        \abbrbspcr     & \rcanarybspcompiled{} (\rcanarybspcompiledperc) & \rcanarybspwarnings{} (\rcanarybspwarningsperc) \\ \midrule
        \abbrpaccr     & \rcanarypaccompiled{} (\rcanarypaccompiledperc) & \rcanarypacwarnings{} (\rcanarypacwarningsperc) \\ \midrule
        \abbrarchcr     & \rcanaryarchcompiled{} (\rcanaryarchcompiledperc) & \rcanaryarchwarnings{} (\rcanaryarchwarningsperc) \\ \midrule
        \abbrutilcr     & \rcanaryutilcompiled{} (\rcanaryutilcompiledperc) & \rcanaryutilwarnings{} (\rcanaryutilwarningsperc) \\ \midrule
        \abbruncatcr     & \rcanaryuncatcompiled{} (\rcanaryuncatcompiledperc) & \rcanaryuncatwarnings{} (\rcanaryuncatwarningsperc) \\ \midrule
        \textbf{Total}  & \rcanarytotalcompiled{} (\rcanarytotalcompiledperc) & \rcanarytotalwarningscompiled{} (\rcanarytotalwarningscompiledperc) \\
         \bottomrule
    \end{tabular}
\end{table}
\begin{table}[t!]
\footnotesize
\caption{Summary of~\lockbud{} on embedded crates.}
\label{tab:lockbud}
    \centering
    \begin{tabular}{cccc}
    \toprule
\multicolumn{1}{c}{\textbf{Category}} & 
\multicolumn{1}{c}{\textbf{\begin{tabular}[c]{@{}c@{}}Num.Crates Successful\\ (\% of Total \\ from~\tbl{tab:cratetypes})\end{tabular}}} &
\multicolumn{1}{c}{\textbf{\begin{tabular}[c]{@{}c@{}}Num. Crates\\ (\% of Successful)\\ having warnings\end{tabular}}} & \multicolumn{1}{c}{\textbf{}} \\ \midrule
        \abbrrtoscr     & \lockbudrtoscompiled{} (\lockbudrtoscompiledperc) & \lockbudrtoswarnings{} (\lockbudrtoswarningsperc) \\ \midrule
        \abbrdrivercr     & \lockbuddrivercompiled{} (\lockbuddrivercompiledperc) & \lockbuddriverwarnings{} (\lockbuddriverwarningsperc) \\ \midrule
        \abbrhalcr     & \lockbudhalcompiled{} (\lockbudhalcompiledperc) & \lockbudhalwarnings{} (\lockbudhalwarningsperc) \\ \midrule
        \abbrbspcr     & \lockbudbspcompiled{} (\lockbudbspcompiledperc) & \lockbudbspwarnings{} (\lockbudbspwarningsperc) \\ \midrule
        \abbrpaccr     & \lockbudpaccompiled{} (\lockbudpaccompiledperc) & \lockbudpacwarnings{} (\lockbudpacwarningsperc) \\ \midrule
        \abbrarchcr     & \lockbudarchcompiled{} (\lockbudarchcompiledperc) & \lockbudarchwarnings{} (\lockbudarchwarningsperc) \\ \midrule
        \abbrutilcr     & \lockbudutilcompiled{} (\lockbudutilcompiledperc) & \lockbudutilwarnings{} (\lockbudutilwarningsperc) \\ \midrule
        \abbruncatcr     & \lockbuduncatcompiled{} (\lockbuduncatcompiledperc) & \lockbuduncatwarnings{} (\lockbuduncatwarningsperc) \\ \midrule
        \textbf{Total}  & \lockbudtotalcompiled{} (\lockbudtotalcompiledperc) & \lockbudtotalwarningscompiled{} (\lockbudtotalwarningscompiledperc) \\
         \bottomrule
    \end{tabular}
\end{table}
\begin{table}[ht]
\scriptsize
\centering
\caption{\rudra{} Results (UD: Unsafe Dataflow, SV: Send/Sync Variance).}
\label{tab:rudra}
\begin{tabular} {crrrr}
\toprule
\multirow{2}{*}{\textbf{Category}} & \multirow{2}{*}{\textbf{\begin{tabular}[c]{@{}c@{}}Crates Successful\\ (\% of Total\\ from~\tbl{tab:cratetypes})\end{tabular}}} & \multicolumn{3}{c}{\textbf{Number of Crates (\% of Successful) with}}                                                                                                                                     \\ \cmidrule(lr){3-5}
                                   &                                                                                                                   & \textbf{\begin{tabular}[c]{@{}c@{}}Unsafe\\ Dataflow\end{tabular}} & \multicolumn{1}{c}{\textbf{\begin{tabular}[c]{@{}c@{}}Send/Sync\\ Variance\end{tabular}}} & \multicolumn{1}{c}{\textbf{At least one}} \\ \midrule
\abbrrtoscr  & \rudranumrtoscompiled{} (\rudraperrtoscompiled) & \rudranumrtosud{} (\rudraperrtosud) & \rudranumrtossv{} (\rudraperrtossv) & \rudranumrtosatone{} (\rudraperrtosatone) \\ \midrule
\abbrdrivercr  & \rudranumdrivercompiled{} (\rudraperdrivercompiled) & \rudranumdriverud{} (\rudraperdriverud) & \rudranumdriversv{} (\rudraperdriversv) & \rudranumdriveratone{} (\rudraperdriveratone) \\ \midrule
\abbrhalcr  & \rudranumhalcompiled{} (\rudraperhalcompiled) & \rudranumhalud{} (\rudraperhalud) & \rudranumhalsv{} (\rudraperhalsv) & \rudranumhalatone{} (\rudraperhalatone) \\ \midrule
\abbrbspcr  & \rudranumbspcompiled{} (\rudraperbspcompiled) & \rudranumbspud{} (\rudraperbspud) & \rudranumbspsv{} (\rudraperbspsv) & \rudranumbspatone{} (\rudraperbspatone) \\ \midrule
\abbrpaccr  & \rudranumpaccompiled{} (\rudraperpaccompiled) & \rudranumpacud{} (\rudraperpacud) & \rudranumpacsv{} (\rudraperpacsv) & \rudranumpacatone{} (\rudraperpacatone) \\ \midrule
\abbrarchcr  & \rudranumarchcompiled{} (\rudraperarchcompiled) & \rudranumarchud{} (\rudraperarchud) & \rudranumarchsv{} (\rudraperarchsv) & \rudranumarchatone{} (\rudraperarchatone) \\ \midrule
\abbrutilcr  & \rudranumutilcompiled{} (\rudraperutilcompiled) & \rudranumutilud{} (\rudraperutilud) & \rudranumutilsv{} (\rudraperutilsv) & \rudranumutilatone{} (\rudraperutilatone) \\ \midrule
\abbruncatcr  & \rudranumuncatcompiled{} (\rudraperuncatcompiled) & \rudranumuncatud{} (\rudraperuncatud) & \rudranumuncatsv{} (\rudraperuncatsv) & \rudranumuncatatone{} (\rudraperuncatatone) \\ \midrule
\textbf{Total}  & \textbf{\rudratotalsuccess{} (\rudratotalsuccessperc)} & \textbf{\rudratotalud{} (\rudratotaludperc)} & \textbf{\rudratotalsv{} (\rudratotalsvperc)} & \textbf{\rudratotalatone{} (\rudratotalatoneperc)} \\
\bottomrule
\end{tabular}
\end{table}

\begin{table*}[]
\scriptsize
\caption{Summary of~\safedrop{} results on embedded crates.}
\label{tab:safedrop}
\centering
\begin{tabular}{crrrrrr}
\toprule
\multirow{2}{*}{\textbf{Category}} & \multirow{2}{*}{\textbf{\begin{tabular}[c]{@{}c@{}}Num. Crates Successful\\ (\% of Total\\ from~\tbl{tab:cratetypes})\end{tabular}}} & \multicolumn{5}{c}{\textbf{Number of Crates (\% of Successful) having}}                                                                                                                                                                                                                                                                                                                                         \\ \cmidrule(lr){3-7}
                                   &                                                                                                                           & \textbf{Use-After-Free} & \multicolumn{1}{c}{\textbf{\begin{tabular}[c]{@{}c@{}}Double\\ Free\end{tabular}}} & \multicolumn{1}{c}{\textbf{\begin{tabular}[c]{@{}c@{}}Dangling\\ Pointer\end{tabular}}} & \multicolumn{1}{c}{\textbf{\begin{tabular}[c]{@{}c@{}}Invalid\\ Memory Access\end{tabular}}} & \multicolumn{1}{c}{\textbf{\begin{tabular}[c]{@{}c@{}}At least one of\\ UAF/DF/DP/IMA\end{tabular}}} \\
\midrule
\abbrrtoscr  & \sdrtosnumsuccessful{} (\sdrtosnumsuccessfulperc) & \sdrtosnumuaf{} (\sdrtosnumuafperc) & \sdrtosnumdf{} (\sdrtosnumdfperc) & \sdrtosnumdp{} (\sdrtosnumdpperc) & \sdrtosnumima{} (\sdrtosnumimaperc) & \sdrtosnumatone{} (\sdrtosnumatoneperc) \\ \midrule
\abbrdrivercr  & \sddrivernumsuccessful{} (\sddrivernumsuccessfulperc) & \sddrivernumuaf{} (\sddrivernumuafperc) & \sddrivernumdf{} (\sddrivernumdfperc) & \sddrivernumdp{} (\sddrivernumdpperc) & \sddrivernumima{} (\sddrivernumimaperc) & \sddrivernumatone{} (\sddrivernumatoneperc) \\ \midrule
\abbrhalcr  & \sdhalnumsuccessful{} (\sdhalnumsuccessfulperc) & \sdhalnumuaf{} (\sdhalnumuafperc) & \sdhalnumdf{} (\sdhalnumdfperc) & \sdhalnumdp{} (\sdhalnumdpperc) & \sdhalnumima{} (\sdhalnumimaperc) & \sdhalnumatone{} (\sdhalnumatoneperc) \\ \midrule
\abbrbspcr  & \sdbspnumsuccessful{} (\sdbspnumsuccessfulperc) & \sdbspnumuaf{} (\sdbspnumuafperc) & \sdbspnumdf{} (\sdbspnumdfperc) & \sdbspnumdp{} (\sdbspnumdpperc) & \sdbspnumima{} (\sdbspnumimaperc) & \sdbspnumatone{} (\sdbspnumatoneperc) \\ \midrule
\abbrpaccr  & \sdpacnumsuccessful{} (\sdpacnumsuccessfulperc) & \sdpacnumuaf{} (\sdpacnumuafperc) & \sdpacnumdf{} (\sdpacnumdfperc) & \sdpacnumdp{} (\sdpacnumdpperc) & \sdpacnumima{} (\sdpacnumimaperc) & \sdpacnumatone{} (\sdpacnumatoneperc) \\ \midrule
\abbrarchcr  & \sdarchnumsuccessful{} (\sdarchnumsuccessfulperc) & \sdarchnumuaf{} (\sdarchnumuafperc) & \sdarchnumdf{} (\sdarchnumdfperc) & \sdarchnumdp{} (\sdarchnumdpperc) & \sdarchnumima{} (\sdarchnumimaperc) & \sdarchnumatone{} (\sdarchnumatoneperc) \\ \midrule
\abbrutilcr  & \sdutilnumsuccessful{} (\sdutilnumsuccessfulperc) & \sdutilnumuaf{} (\sdutilnumuafperc) & \sdutilnumdf{} (\sdutilnumdfperc) & \sdutilnumdp{} (\sdutilnumdpperc) & \sdutilnumima{} (\sdutilnumimaperc) & \sdutilnumatone{} (\sdutilnumatoneperc) \\ \midrule
\abbruncatcr  & \sduncatnumsuccessful{} (\sduncatnumsuccessfulperc) & \sduncatnumuaf{} (\sduncatnumuafperc) & \sduncatnumdf{} (\sduncatnumdfperc) & \sduncatnumdp{} (\sduncatnumdpperc) & \sduncatnumima{} (\sduncatnumimaperc) & \sduncatnumatone{} (\sduncatnumatoneperc) \\ \midrule
\textbf{Total}  & \sdtotalcompiled{} (\sdtotalcompiledperc) & \sdtotaluaf{} (\sdtotaluafperc) & \sdtotaldf{} (\sdtotaldfperc) & \sdtotaldp{} (\sdtotaldpperc) & \sdtotalima{} (\sdtotalimaperc) & \sdnumatonetotal{} (\sdnumatonetotalperc) \\
\bottomrule
\end{tabular}
\end{table*}


\subsection{Failure Analysis}

\subsubsection{Toolchain Incompatibility}
~\label{apdx:toolchain_failure}
The ~\ac{SAST} tools analyze the High Intermediate Representation (HIR), Mid-level Intermediate Representation (MIR), or LLVM Intermediate Representation (LLVM IR) of the target crate. To do this they associate themselves with a particular version of rustc compiler. However, issues might arise if the target crate is not compatible with tool's rustc version and this can lead to failure in analyzing the target crate. For instance, ~\ffichecker{} which is dependent on rustc vesion ~\texttt{1.59.0-nightly} fails to analyse the crate ~\code{smartdeploy-macros v0.1.6} because it requires rustc verion ~\texttt{>=1.69.0} 

\subsubsection{Ignoring Project-Specific Configurations}
~\label{apdx:config_failure}
Some crates, use\\ ~\code{.cargo} folder to configure rustc to build crate for a specific target. The configuration file within this directory specifies the default flags to be passed to the \shell{rustc} command. However, we noticed that many of the ~\ac{SAST} tools overlook this optimization and hence they fail to fetch the necessary info from crate to do their analysis. For example, analysis of ~\code{embedded-test v0.3.0} by ~\yuga, ends up in a failure because one of its dependencies ~\code{semihosting v0.1.9} fails to build.

\subsubsection{No binary target support}
~\label{apdx:nobin_failure}
By default, ~\ac{SAST} tools try to analyse all the targets mentioned in a crate's toml file, which might include binary targets. In such cases, if the binary depends on specific feature flags to be passed in, such as ~\code{std}, the tool ends up returning a non zero exit code. Ideally, tools should try to handle these cases similar to how cargo handles them. An example of this issue was seen in ~\safedrop{} where it returned a non-zero exit code while analysing crate ~\code{ais v0.11.0}, because binary target required ~\code{std} feature.

\subsubsection{Tool Crashes}
~\label{apdx:toolcrash_failure}
Failures were also observed due to the tool's codebase being unable to handle corner cases, for example ~\ffichecker{} was unable to analyse the crate ~\code{bachue-auto_impl v0.5.1} because the tool was hitting an erroneous codepath.

\subsubsection{Timeouts}
~\label{apdx:timeout_failure}
Timeout failures were specifically observed in the case of ~\rcanary{}. The tool employs z3 constraint solver to identify the unsatisfiable functions with improper handling of manual memory allocations. This issue is specifically present in larger crates, where the size and complexity can increase the probability of timeouts. One of such crates was: ~\code{k64 v0.1.0}.

\subsubsection{Rustc version incompatibility}
~\label{apdx:version_failure}
~\ffichecker reported the issue during the version checking process of rustc compiler.  ~\ffichecker{} failed to run the build script in case of ~\code{binread v2.2.0} because of the mismatch in requirements.

\section{Bugs Examples}
\label{apdx:lockbudbugexamples}

\begin{lstlisting}[
  language=Rust,
  caption={Configuring bindgen using build.rs script to get Rust bindings for \texttt{no\_std} environment.}, 
  label={lst:bindgenexample},
  xleftmargin=0.5cm, % Set left margin
  basicstyle=\scriptsize\ttfamily, % Font size and style
  numbers=left, % Line numbers on the left
  breaklines=true, % Break long lines
  escapeinside=&&,captionpos=b,
]
let bindings = bindgen::Builder::default()
    // Specify the header file to generate bindings
    .header("wrapper.h")
    ....
    // Add any additional configuration options
    // Use the `core` crate instead of `std`
    .use_core() 
    // Exclude specific functions from the bindings
    .blocklist_function("function_name_1") 
    .parse_callbacks(Box::new(bindgen::CargoCallbacks))
    .generate()
    .expect("Unable to generate bindings");
\end{lstlisting}

\begin{lstlisting}[
  language=Rust,
  caption={FreeRTOS ListInitialise function in \rust{}}, 
  label={lst:RwC_list},
  xleftmargin=0.5cm, % Set left margin
  basicstyle=\scriptsize\ttfamily, % Font size and style
  numbers=left, % Line numbers on the left
  breaklines=true, % Break long lines
  captionpos=b,
]

/* generated by bindgen
pub struct xLIST {
    pub uxNumberOfItems: UBaseType_t,
    pub pxIndex: *mut ListItem_t,
    pub xListEnd: MiniListItem_t,
}
pub type List_t = xLIST;
*/
#[no_mangle]
fn vListInitialise(pxList: *mut List_t) {
    unsafe{
    (*pxList).pxIndex = transmute::<*mut xMINI_LIST_ITEM,
    *mut ListItem_t>(&mut(*pxList).xListEnd);
    (*pxList).xListEnd.xItemValue = portMAX_DELAY;
    (*pxList).xListEnd.pxNext = transmute::<*mut xMINI_LIST_ITEM,
    *mut ListItem_t>(&mut(*pxList).xListEnd);
    (*pxList).xListEnd.pxPrevious = transmute::<*mut xMINI_LIST_ITEM,
    *mut ListItem_t>(&mut (*pxList).xListEnd);
    (*pxList).uxNumberOfItems = 0;
    }
}
\end{lstlisting}

\begin{table}[]
\caption{Summary of running~\ctorust{} on various embedded~\acp{RTOS} (results discussed in~\sect{subsubsec:ctorustresults}).}
\label{tab:c2rustblinky}
\scriptsize
\centering
\begin{tabular}{cccc}
\toprule
\textbf{RTOS Name}             & \textbf{\begin{tabular}[c]{@{}c@{}}c2rust\\ transpilation?\end{tabular}} & \textbf{\begin{tabular}[c]{@{}c@{}}Generated\\ Valid Rust\\ Code?\end{tabular}} & \textbf{\begin{tabular}[c]{@{}c@{}}Overall\\ Conversion\end{tabular}} \\ \midrule
zephyrproject-rtos/zephyr      & \multirow{6}{*}{\textcolor{red}{Failed} (\faUser)}                                                  & \multirow{6}{*}{N/A}                                                            & \multirow{6}{*}{\textcolor{red}{Failed}}                                               \\ \cmidrule(lr){1-1}
RIOT-OS/RIOT                   &                                                                          &                                                                                 &                                                                       \\ \cmidrule(lr){1-1}
seL4/seL4                      &                                                                          &                                                                                 &                                                                       \\ \cmidrule(lr){1-1}
embox/embox                    &                                                                          &                                                                                 &                                                                       \\ \cmidrule(lr){1-1}
jameswalmsley/bitthunder       &                                                                          &                                                                                 &                                                                       \\ \cmidrule(lr){1-1}
stateos/StateOS                &                                                                          &                                                                                 &                                                                       \\ \midrule
RT-Thread/rt-thread            & \multirow{7}{*}{Success (\faUser)}                                                 & \multirow{7}{*}{No}                                                             & \multirow{7}{*}{\textcolor{red}{Failed}}                                               \\ \cmidrule(lr){1-1}
FreeRTOS/FreeRTOS              &                                                                          &                                                                                 &                                                                       \\ \cmidrule(lr){1-1}
azure-rtos/threadx             &                                                                          &                                                                                 &                                                                       \\ \cmidrule(lr){1-1}
apache/nuttx                   &                                                                          &                                                                                 &                                                                       \\ \cmidrule(lr){1-1}
contiki-ng/contiki-ng          &                                                                          &                                                                                 &                                                                       \\ \cmidrule(lr){1-1}
TrampolineRTOS/trampoline      &                                                                          &                                                                                 &                                                                       \\ \cmidrule(lr){1-1}
insane-adding-machines/frosted &                                                                          &                                                                                 &                                                                       \\ \midrule
kmilo17pet/QuarkTS             & Success (\faUser)                                                                  & Yes                                                                             & Success                                                               \\ \midrule
kelvinlawson/atomthreads       & \multirow{2}{*}{Success}                                                 & \multirow{2}{*}{No}                                                             & \multirow{2}{*}{\textcolor{red}{Failed}}                                               \\ \cmidrule(lr){1-1}
echronos/echronos              &                                                                          &                                                                                 &                                                                       \\ \bottomrule
\end{tabular}
\end{table}

\subsection{\texttt{std} to \texttt{no\_std} conversion}
\label{apdx:nostdconversionexample}

\begin{table*}[t!]
  \begin{tabular}{cc}
    \begin{minipage}{.5\textwidth}
    \begin{lstlisting}[
      language=Rust,
      basicstyle=\scriptsize\ttfamily,
      numbers=left,
      breaklines=true,
      xleftmargin=0.5cm
    ]
use std::fs;
use std::io;

struct Foo {
    map: std::collections::HashMap<i32, Vec<i32>>,
}

fn bar() -> Foo {
    println!("hello world");

    let _ = read_data("some message");

    Foo{ map: Default::default() }
}

fn read_data() -> io::Result<Vec<u8>> {
    let data = fs::read("somefile.txt")?;
    Ok(data)
}
    \end{lstlisting}
    \end{minipage} &
    \begin{minipage}{.5\textwidth}
    \begin{lstlisting}[
      language=Rust,
      basicstyle=\scriptsize\ttfamily,
      numbers=left,
      breaklines=true,
      xleftmargin=0.5cm
    ]
#![cfg_attr(not(feature = "std"), no_std)]

#[cfg(any(feature = "std", feature = "alloc"))]
extern crate alloc;

#[cfg(feature = "std")]
use std::fs;
#[cfg(feature = "std")]
use std::io;

struct Foo {
    #[cfg(feature = "std")]
    map: std::collections::HashMap<i32, alloc::vec::Vec<i32>>,

    #[cfg(not(feature = "std"))]
    map: hashbrown::HashMap<i32, alloc::vec::Vec<i32>>
}

fn bar() -> Foo {
    #[cfg(feature = "std")]
    println!("hello world");

    #[cfg(feature = "std")]
    let _ = read_data("some message");

    Foo{ map: Default::default() }
}

#[cfg(feature = "std")]
fn read_data() -> io::Result<Vec<u8>> {
    let data = fs::read("somefile.txt")?;
    Ok(data)
}
    \end{lstlisting}
    \end{minipage} 
    \\
 {Existing \texttt{std} compatible code}   & {After converting to be \texttt{no\_std} compatible} 
  \\
  \end{tabular}
\caption{Example demonstrating some of the changes involved in converting \texttt{std} code to \texttt{no\_std} in Rust.}
\label{lst:nostdconversion}
\end{table*}

~\lst{lst:nostdconversion} shows a small example demonstrating some of the changes that are required in order to convert an existing ~\rustcode{std} compatible \rust{} code to be \rustcode{no_std} compatible, while maintaining backwards compatibility, to the extent possible.
This particular example, assumes that, in order to make the crate suitable for both \rustcode{std} and \rustcode{no_std}, we would add a couple of features to the manifest file (\texttt{Cargo.toml}).
Firstly we add a feature named \texttt{std} to indicated that the build is targeting one of the \rustcode{std} targets.
We also add another feature named \texttt{alloc} in the manifest, which  would indicate the use of an allocator.
We also make the \rustcode{std} feature default in the manifest.
Please note that the names of these features do not matter and can be anything.
Next we briefly describe the changes required to the given code along-with the reasons for the same (the line numbers indicated in the following text are for the code listing after coversion).
\begin{itemize}
    \item Firstly, we need to conditionally enable \rustcode{no_std}, so that the crate can be built for both \rustcode{std} and \rustcode{nostd}. To achieve this, we use \rustcode{cfg_attr} and conditionally enable \rustcode{no_std} if the \rustcode{std} feature is not enabled (line 1).
    \item Next, we include \rustcode{alloc}, if \rustcode{std} feature is not enabled, but the \rustcode{alloc} feature is enabled. When building for \rustcode{std}, we do not need to specifically include \rustcode{alloc} (lines 3-4).
    \item Consider the definition of the \rustcode{struct Foo} on line 11. While \rustcode{std::collections::HashMap} is not \rustcode{no_std} compatible, we can use a functional equivalent replacement (\texttt{hashbrown::HashMap} from the \textit{hashbrown} crate), which is \rustcode{no_std} compatible.
    \item We also use the \rustcode{std} feature to conditionally enable items that are only available when building for a \rustcode{std} target. Specifically the filesystem and I/O related functionality (like printing to \texttt{stdout}, for example) is only available on \rustcode{std} (lines 6-9, 20-21).
    We also disable functionality that has no direct equivalent in the \rustcode{no_std} settings (like baremetnal/embedded domains). For example the function \rustcode{read_data} (line 30), has a return type that is not directly available on baremetal settings. Hence we enable this function only while building for \rustcode{std} (line 29).
    Moreover the call to \rustcode{read_data} (line 24) is enabled conditionally as well.
    This enables to use the converted crate in \rustcode{no_std} settings, albeit with reduced functionality.
\end{itemize}

This example demonstrated the minimum changes required in order to make a given existing code \rustcode{no_std} compatible.

\subsection{\texorpdfstring{\rust{} Interoperability}{Rust Interoperability}}

\begin{lstlisting}[
  language=Rust, % Set the language to Rust
  caption={Rust code snippet to call FreeRTOS scheduler}, 
  label={lst:rustapponcrots},
  xleftmargin=0.5cm, % Set left margin
  basicstyle=\scriptsize\ttfamily, % Font size and typewriter style
  numbers=left, % Line numbers on the left
  breaklines=true, % Break long lines
  captionpos=b,
]
extern "C" {
pub fn xTaskCreate(
    pxTaskCode: TaskFunction_t,
    pcName: *const cty::c_char,
    ... //omitted for brevity
    pxCreatedTask: *mut TaskHandle_t,
) -> BaseType_t;
}
extern "C" {
pub fn vTaskStartScheduler();
}
#[no_mangle]
pub extern "C" fn entry() {
    let task1: core::option::Option<unsafe extern "C" fn(arg1: *mut core::ffi::c_void)> =   Some(vTask1);
    let tn1 = core::str::from_utf8(b"Task 1\0").unwrap().as_ptr() as u8;
    unsafe {
        xTaskCreate(task1, &tn1, ..., ::core::ptr::null_mut());
        vTaskStartScheduler();
    ...
\end{lstlisting}

\begin{lstlisting}[
  language=C++,  % Set the language for C++
  caption={Simple C function}, 
  label={lst:ccodeadd},
  xleftmargin=0.5cm, % Set left margin
  basicstyle=\scriptsize\ttfamily, % Font size and typewriter style
  numbers=left, % Line numbers on the left
  breaklines=true, % Break long lines
  escapeinside=&&,captionpos=b,
]
// add.h
int add(int a, int b);

// add.c
#include "add.h"
int add(int a, int b) {
  return a+b;
}
\end{lstlisting}

\begin{lstlisting}[
  language=Rust,
  escapeinside=||, captionpos=b,
  caption={Calling \texttt{add} function in \ref{lst:ccodeadd} from Rust.}, 
  label={lst:callingcfromrust},
  basicstyle=\scriptsize\ttfamily,
  breaklines=true, % Break long lines
]
// bindings.rs
// FFI signature for add function in |Listing \ref{lst:ccodeadd}|
extern "C" {
    pub fn add(a: i32, b: i32) -> i32;
}

// In rust code
include!("bindings.rs");
fn main() {
  ...
  let x = add(1, 2);
  ...
}
\end{lstlisting}

\begin{lstlisting}[
  language=Rust, % Set the language to Rust
  captionpos=b,
  caption={Rust function callable from C.}, 
  label={lst:rustfunctioncallableinc},
  xleftmargin=0.5cm, % Set left margin
  basicstyle=\scriptsize\ttfamily, % Font size and typewriter style
  numbers=left, % Line numbers on the left
  breaklines=true % Break long lines
]
#[no_mangle]
pub unsafe extern "C" 
    fn sub(a: i32, b: i32) -> i32 {
  return a - b;
}
\end{lstlisting}

\subsubsection{Incompatible \texorpdfstring{~\ac{FFI}}{FFI} Types}
\label{apdx:incompatibleffitypes}
The superior~\rust{} type system has several types that are not supported in C. For instance,~\rustcode{Vec}~\cite{rustvectype}, one of the most commonly used~\rust{} types, is not supported in C.
Consequently, functions (\eg{}~\rustcode{sum_vec} in~\lst{lst:incompatiblerustfunction}) accepting parameters of these unsupported C types cannot be invoked from C.
Although ~\code{cbindgen} automates the creation of C declarations, it cannot handle these unsupported C types.
Developers need to write certain wrapper functions that convert C-supported types into appropriate~\rust{} types (\eg{}~\rustcode{wrapper_sum_vec} in~\lst{lst:incompatiblerustfunction}).

\begin{lstlisting}[
  language=Rust, % Set the language to Rust
  captionpos=b,
  caption={Example of Rust function (\texttt{sum\_vec}) using C incompatible types.}, 
  label={lst:incompatiblerustfunction},
  xleftmargin=0.5cm, % Set left margin
  basicstyle=\scriptsize\ttfamily, % Font size and typewriter style
  numbers=left, % Line numbers on the left
  breaklines=true % Break long lines
]
#[no_mangle]
pub unsafe extern "C" fn 
    sum_vec(numv: &Vec<i32>) -> i32 {
  let mut sum = 0;
  for i in 0..numv.len() {
    sum += numv[i];
  }
  sum
}

#[no_mangle]
pub unsafe extern "C" fn
    wrapper_sum_vec(p: *const i32, 
                    len: i32) -> i32 {
    let slice = unsafe {
        assert!(!p.is_null());
        slice::from_raw_parts(p, len as usize)
    };
    return sum_vec(&slice.to_vec());
}
\end{lstlisting}

\begin{lstlisting}[
  language=Rust, % Set the language to Rust
  caption={FFI incompatible type in nrf52840-pac crate}, 
  captionpos=b,
  label={lst:rocruststruct},
  xleftmargin=0.5cm, % Set left margin
  basicstyle=\scriptsize\ttfamily, % Font size and typewriter style
  numbers=left, % Line numbers on the left
  breaklines=true % Break long lines
]
#[repr(C)]
pub struct RegisterBlock {
    pub out: generic::Reg<out::OUT_SPEC>,
    pub outset: nrf52840_pac::generic::Reg<outset::OUTSET_SPEC>,
    pub outclr: nrf52840_pac::generic::Reg<outclr::OUTCLR_SPEC>,
    pub in_: nrf52840_pac::generic::Reg<in_::IN_SPEC>,
    pub dir: nrf52840_pac::generic::Reg<dir::DIR_SPEC>,
    pub dirset: nrf52840_pac::generic::Reg<dirset::DIRSET_SPEC>,
    pub dirclr: nrf52840_pac::generic::Reg<dirclr::DIRCLR_SPEC>,
    pub latch: nrf52840_pac::generic::Reg<latch::LATCH_SPEC>,
    pub detectmode: nrf52840_pac::generic::Reg<detectmode::DETECTMODE_SPEC>,
    pub pin_cnf: [nrf52840_pac::generic::Reg<pin_cnf::PIN_CNF_SPEC>; 32]
}
\end{lstlisting}

\begin{figure}[ht]
    \includegraphics[width=\linewidth]{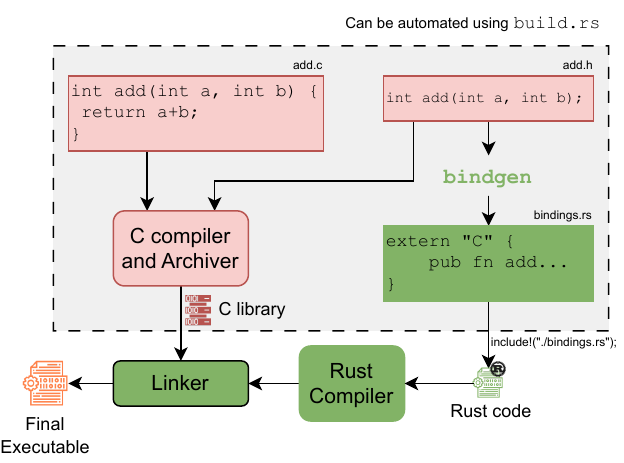}
    \caption{Process to enable invocation of a C function from~\rust{}.}
    \label{fig:bindgenfig}
    \Description[Process to enable invocation of a C function from~\rust{}.]{Process to enable invocation of a C function from~\rust{}.}
\end{figure}


\subsubsection{\texorpdfstring{Calling C function from~\rust{} (\rust{} $\rightarrow$ C)}{Calling C function from Rust (Rust to C)}}
\label{apdx:callingcfromrust}
Consider the simple C function~\code{add} as shown in~\lst{lst:ccodeadd}.
We need to follow the below steps to invoke~\code{add} from~\rust{} code:
\begin{itemize}[leftmargin=*]
\item First, we need to provide the~\ac {FFI} signature of the function~\code{add} (as shown in~\lst{lst:callingcfromrust}), so that~\rust{} compiler knows that the function is implemented in a foreign language. Developers can use~\code{bindgen}~\cite{rustbindgen} tool, which automatically generates~\ac{FFI} signatures from C header files.
\item Second, we should link with the library containing the function's implementation to get the final executable.
\end{itemize}

The~\fig{fig:bindgenfig} illustrates these steps.
One of the main tasks here is to generate~\ac{FFI} bindings for the C functions.
It is relatively straightforward to create these bindings as the~\rust{}'s type system~\cite{matsakis2014rust} is a superset of C's,~\ie{} every builtin C type has a corresponding type in~\rust{}.
To create an~\ac{FFI} declaration for a C function, developers must replace C types with corresponding~\rust{} types and use appropriate syntax.
The~\code{bindgen} tool automates most of these tasks.
Furthermore, it also provides options (\eg{}~\rustcode{use_core}~\cite{rustbindgenusecore}) to use~\rustcode{no_std} compatible types and generate embedded system-friendly bindings.
Developers can automate running the~\code{bindgen} tool by specifying the relevant commands in~\code{build.rs} (\lst{lst:bindgenexample} shows an example) and list it as build-dependency in~\code{Cargo.toml}.

\begin{figure}[ht]
    \hspace*{-1cm}
    \includegraphics[scale=0.9]{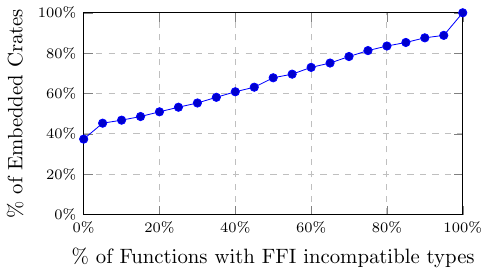}
\captionof{figure}{CDF of the percentages of crates and functions with incompatible types.}
    \label{fig:cdfincompatibletypes}
    \Description[CDF of the percentages of crates and functions with incompatible types.]{CDF of the percentages of crates and functions with incompatible types.}
\end{figure}

\subsubsection{\texorpdfstring{Calling~\rust{} function from C (C $\rightarrow$~\rust{})}{Calling Rust function from C (C to Rust)}}
\label{apdx:callingrustfromc}
Developers need to follow the below steps to call a~\rust{} function from C:
\begin{itemize}[leftmargin=*]
\item The~\rust{} function should be annotated to allow external calls.
Specifically, as shown in~\lst{lst:rustfunctioncallableinc}, it should be annotated with \rustcode{pub unsafe extern "C"} and~\rustcode{no_mangle} (to avoid name mangling and enable linking).
Also, developers need to compile the~\rust{} code into a static library (\ie{}~\rustcode{static-lib}).

\begin{lstlisting}[
  language=C++, % Set the language to C++
  captionpos=b,
  caption={C code calling Rust function in Listing \ref{lst:rustfunctioncallableinc}}, 
  label={lst:ccodewhichrust},
  xleftmargin=0.5cm, % Set left margin
  basicstyle=\scriptsize\ttfamily, % Font size and typewriter style
  numbers=left, % Line numbers on the left
  breaklines=true % Break long lines
]
// Declaration for external Rust function
extern int sub(int a, int b);

int main() {
  ...
  int c = sub(a, b);
  ...
}
\end{lstlisting}

\item In the C domain, the~\rust{} function should be declared as~\code{extern} (\lst{lst:ccodewhichrust}), and the~\rust{} library should be linked to get the final executable.
Developers can use~\code{cbindgen}~\cite{generatingcbindgen} to generate C declarations from external~\rust{} functions (\ie~\rustcode{pub extern "C"}).
\end{itemize}

\subsubsection{Non-FFI compatible~\rust{} functions}
\label{apdx:nonfficompat}

We performed a type compatibility analysis to assess the extent to which external functions in~\rust{} crates use advanced~\rust{} types,~\ie{} library functions for which developers need to engineer corresponding type wrapper functions manually.
The~\fig{fig:cdfincompatibletypes} shows the CDF of the percentage of embedded crates v/s percentage of external (or public) functions that use parameters with incompatible types (\eg{}~\texttt{Vec}).
A point $(x, y)$ on a line indicates $y$\% of crates has $x$\% (or less) of its public functions using incompatible types.
70\% of crates have at least one function with an~\ac{FFI} incompatible parameter types.
20\% of crates have 90\% of their functions with incompatible type.

\else
\fi

\end{document}